\documentclass[12pt]{article}
\usepackage{amsfonts,amsmath,amssymb,epsf}
\usepackage{amsmath,braket}
\usepackage{graphicx,color}
\newcommand{\de}{\partial}
\newcommand{\be}{\begin{equation}}
\newcommand{\ba}{\begin{eqnarray}}
\newcommand{\ea}{\end{eqnarray}}
\newcommand{\ee}{\end{equation}}

\newcommand{\f}{\frac}
\newcommand{\s}{\sqrt}
\newcommand{\vp}{\varphi}

\newcommand{\ti}{\tilde}

\newcommand{\ddd}{\cdot\cdot\cdot}
\newcommand{\no}{\nonumber \\}
\newcommand{\la}{\langle}
\newcommand{\lb}{\rangle}
\newcommand{\bea}{\begin{eqnarray}}
\newcommand{\eea}{\end{eqnarray}}
\newcommand{\bes}{\begin{equation*}}
\newcommand{\beas}{\begin{eqnarray*}}
\newcommand{\eeas}{\end{eqnarray*}}
\newcommand{\bas}{\begin{array*}}
\newcommand{\eas}{\end{array*}}
\newcommand{\ees}{\end{equation*}}

\newcommand{\ep}{\epsilon}

\def \xmap#1{\s{\f{\sin\f{\pi}{ L} (s+ #1)}{\sin\f{\pi}{L} #1 }}}
\def \ixmap#1{\s{\f{#1+s}{#1}}}

\begin{document}

\begin{titlepage}
\thispagestyle{empty}

\begin{flushright}
YITP-16-47
\\
IPMU16-0043
\\
\end{flushright}


\begin{center}
\noindent{{\textbf{EPR Pairs, Local Projections and Quantum Teleportation in Holography}}}\\
\vspace{2cm}
Tokiro Numasawa$^{a}$, Noburo Shiba$^{a}$, Tadashi Takayanagi$^{a,b}$ and Kento Watanabe$^{a}$
\vspace{1cm}

{\it
$^{a}$Center for Gravitational Physics, Yukawa Institute for Theoretical Physics,\\
Kyoto University, Kyoto 606-8502, Japan\\
$^{b}$Kavli Institute for the Physics and Mathematics of the Universe,\\
University of Tokyo, Kashiwa, Chiba 277-8582, Japan\\
}

\vskip 2em
\end{center}

\begin{abstract}
In this paper we analyze three quantum operations in two dimensional conformal field theories (CFTs): local projection measurements, creations of partial entanglement between two CFTs, and swapping of subsystems between two CFTs. We also give their holographic duals and study time evolutions of entanglement entropy. By combining these operations, we present an analogue of quantum teleportation between two CFTs and give its holographic realization. We introduce a new quantity to probe tripartite entanglement by using local projection measurement.

 \end{abstract}

\end{titlepage}

\newpage

\tableofcontents

\section{Introduction}

The holographic principle \cite{Hol}, especially the AdS/CFT \cite{Ma}, relates the structures of gravitational spacetimes to those of quantum entanglement as in \cite{Mal,RT,HRT,VR,Sw,MaSu,Sk,Skk}. This motivates us to study gravitational counterparts of quantum information theoretic properties.

In quantum information theory, operational methods are very important (see e.g. text books
\cite{Book,Geo}). Consider a bipartite system which consists of $A$ and $B$, which are far apart. We write the density matrix for the total system $AB$ as $\rho_{AB}$. We also define the reduced density matrix $\rho_A$ by tracing $\rho_{AB}$ out with respect to $B$. A basic class of physical manipulations is called LOCC (local operations and classical communications) and is defined as follows. Local operation (LO) is defined by the map
\be
\rho_{AB}\to \sum_{i,j}(A_i\otimes B_j)\rho_{AB}(A^\dagger_i\otimes B^\dagger_j),
\ee
with the trace preserving condition $\sum_{i}A_i A^\dagger_i=\sum_{j}B_j B^\dagger_j=1$. This includes the projection measurements and unitary transformations which act either $A$ or $B$ at the same time.
Moreover, we allow classical communications (CC) so that we can send a classical information from $A$ to $B$. These are called LOCC and are considered to describe possible physical operations.

For example, consider the entanglement entropy (EE) $S_A=-\mbox{Tr}[\rho_A\log\rho_A]$, which is the best measure of quantum entanglement when the total system is pure. It has the important property that it does not increase under LOCC on average. Moreover, the entanglement entropy (divided by $\log 2$) is known to be equal to the averaged maximum number of EPR pairs which we can extract from $A$ and $B$ by LOCC. In this way, the entanglement entropy has a definite operational meaning.

The quantum operations play a crucial role in the quantum teleportation \cite{Bennett}.
In this process, it is important that $A$ and $B$ are strongly entangled. Owing to this entanglement,  we can send the information of a given state from $A$ to $B$ by LOCC.

In this paper we would like to formulate several important quantum operations in the language of quantum field theories. We especially focus on two dimensional conformal field theories (2d CFTs) so that we can apply the powerful technique of conformal maps. Our quantum operations include local projection measurements and partial entangling of two CFTs as well as swapping of two CFTs. Local projection measurements mean that we perform projection measurements for all points in a region $P$ assuming a lattice regularization. Therefore, the state just after the projection has no real space entanglement in $P$.  A class of such states with no real space entanglement are described by boundary states (or Cardy states \cite{CS}) as argued in \cite{MRTW}. Therefore we can identify a class of states after the local projection measurement with boundary states, as recently pointed out by Rajabpour  \cite{Raj,Rajj,Rajjj}.

Partial entangling is defined by adding maximal entanglement between two CFTs in a particular region. Swapping is to exchange two intervals in two CFTs. We will also give holographic duals of these operations and compute the holographic entanglement entropy (HEE) \cite{RT,HRT} (also refer to the reviews \cite{Review}) in various setups with time evolutions. Finally we will combine our quantum operations to give an analogue of quantum teleportation between two CFTs. We present its holographic realization by considering an AdS black hole. This holographic model of quantum teleportation is closely related to the one by Susskind \cite{Sk,Skk} as in both setups the information is teleported through the Einstein-Rosen bridge.

This paper is organized as follows: In section two, we explain how to realize local projection measurements in CFTs. We also compute the evolution of entanglement entropy after the measurement
in a free fermion CFT. In section three, we introduce two more quantum operations: partially entangling and swapping of two CFTs.  In section four, we present holographic dual of our quantum operations in CFTs. In section five, we compute the time evolution of entanglement entropy after these operations by using gravity duals. In section six, we will present an analogue of quantum teleportation between two CFTs as well as its holographic realization. In section seven, we summarize our conclusions and discuss future problems. In appendix A, we summarized our conventions of theta functions. In appendix B, we present a toy analytical model of partially entangling two CFTs. In appendix C, we summarized our result for holographic entanglement entropy for two symmetric intervals under local projection measurements.

\section{Local Projection Measurements in CFTs}\label{sec:Proj}

Consider a two dimensional CFT defined on an infinite line $-\infty<x<\infty$. This is described by a
path-integral on a complex plane, whose coordinate is expressed as $(w,\bar{w})$ such that $w=\f{x+iy}{\s{2}}$.

Then we would like to describe an operation of projection measurement along an interval $P$, given by
$x\in [-q,q]$. We especially focus on a local projection measurement, which means that in a discretized description as a lattice theory, we specify a specific quantum state for each site by the projection. In other words, we consider the following projection operator
\be
{\cal P}=\left(\prod_{x\in P}|\psi_x\lb \la \psi_x|\right)\otimes\left(\prod_{x\in P^c} I_{x}\right),
\ee
where $I_x$ is the identity operator at the site $x$. The total quantum state we are interested in is
given by ${\cal P}|\Psi_0\lb$, where $|\Psi_0\lb$ is the ground state of the CFT.
By a local unitary transformation we can choose $|\psi_x\lb$ to be a canonical one $|0_x\lb$. The state $\prod_x |\psi_x\lb$ has no real space entanglement as it is a direct product state on the interval. In \cite{MRTW}, a class of such states, which are translationally invariant, is give by boundary state (Cardy state) \cite{CS} in the boundary conformal field theories (BCFTs).

\subsection{General Prescription at $t=0$}

\begin{figure}
  \centering
  \includegraphics[width=8cm]{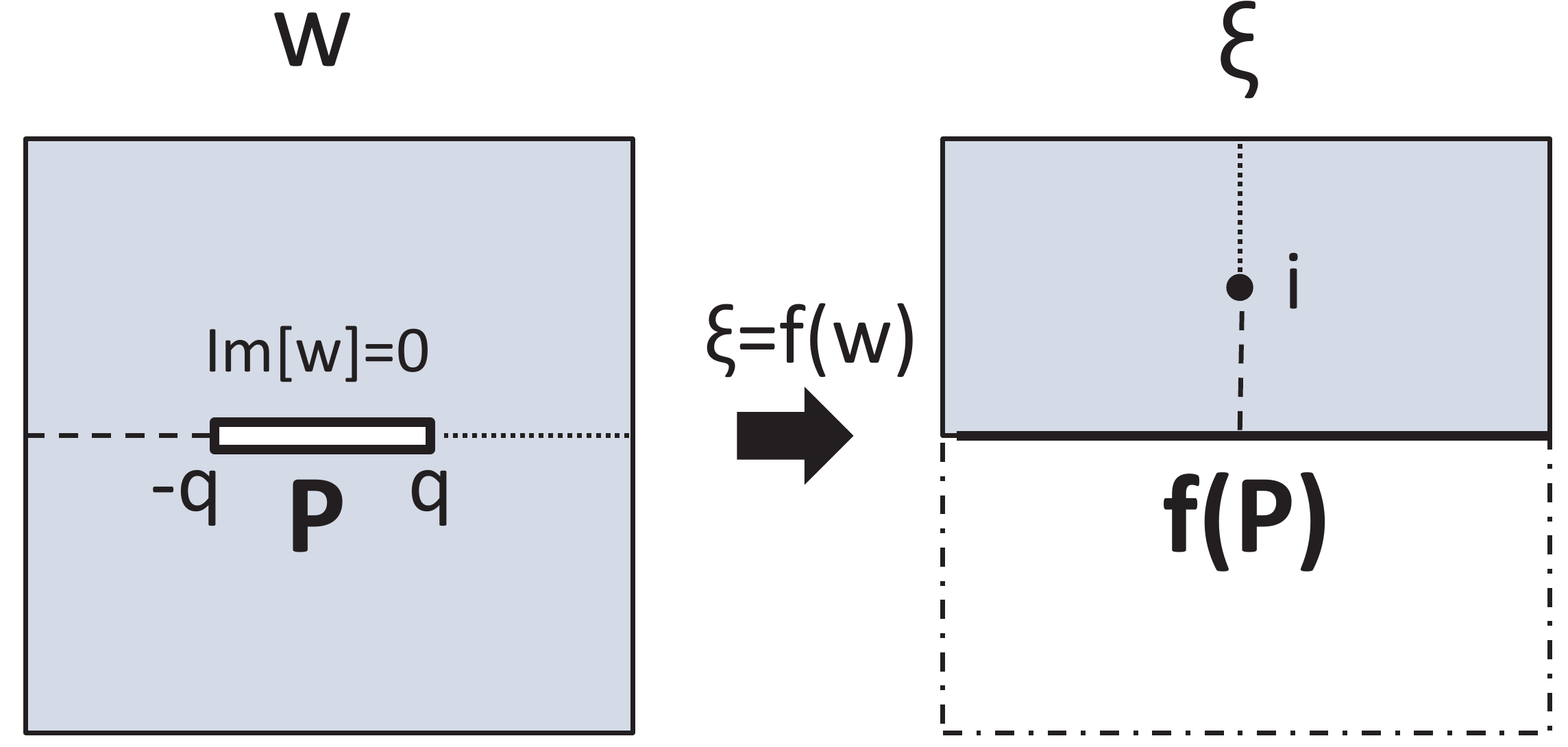}
  \caption{The conformal transformation from the coordinate $(x,y)$ into an upper half plane $(\xi_1,\xi_2)$,
  where $w=(x+iy)/\s{2}$ and $\xi=(\xi_1+i\xi_2)/\s{2}$.}
\label{fig:onecutmap}
  \end{figure}

The recent papers \cite{Raj,Rajj,Rajjj} by Rajabpour argue that such a projection measurement is realized by inserting a slit along the interval $P$ in the Euclidean path-integral description as depicted in the left picture of Fig.\ref{fig:onecutmap}. The upper edge and lower one each give the state $\prod_{x\in P}|0_x\lb$ and $\prod_{x\in P}\la 0_x|$, respectively and thus they are equivalent to the projection
operation ${\cal P}$.
Calculations of various physical quantities can be done by performing
the following conformal map\footnote{The factor $\s{2}$ in $q/\s{2}$ is correlated with that in our coordinate
definition $w=(x+iy)/\s{2}$ which is consistent with our convention of holographic description.}
\be
\xi=\s{\f{q/\s{2}+w}{q/\s{2}-w}}, \label{scmap}
\ee
which is sketched in the right picture of Fig.\ref{fig:onecutmap}. This maps our one slit geometry into an upper half plane.

A quantity which we can calculate immediately is the energy stress tensor $T$. Since $T$ is vanishing on the upper half plane, its contribution after the conformal mapping purely comes from  the Schwarzian derivative term
\be
T(w)=-\f{c}{6}\left(\f{3(f'')^2-2f'f'''}{4(f')^2}\right), \label{emtw}
\ee
where $c$ is the central charge of the 2d CFT.
In our example (\ref{scmap}) we find explicitly
\be
T(w)=\f{cq^2}{16(q^2/2-w^2)^2}. \label{lww}
\ee
We are focusing on the quantum state at the Euclidean time $t_E=0$, equally Im$w=0$.
The position where we measure the energy stress tensor is specified by the coordinate Re$w$.
The result (\ref{lww}) shows that the energy density gets divergent at $w=\pm q$, i.e. the two edges of
the projected interval $P$. In the next subsection we will introduce a UV cut off and resolve this singular behavior.

Now let us compute the entanglement entropy $S_A$ when the subsystem $A$ is defined as an interval
$[q,q+l]$ next to the projected region $P$.\footnote{For more general choices of subsystem $A$, results are not universal and will be discussed later in the case of free fermion CFTs and holographic CFTs.} We can compute $S_A$ in the replica method. We introduce the twist operator $\sigma_{n}$ which produces an end point of the cut for a $n$-sheeted Riemann surface as in the standard treatment \cite{CC}. The chiral conformal dimension of $\sigma_n$ is $\f{c}{24}(n-1/n)$ for a central charge $c$.
Then the trace $\mbox{Tr}(\rho_A)^n$ corresponds to the one point function $\la \sigma(w_1,\bar{w}_1)\lb_w$ on the $w$-plane, with $w_1=\bar{w_1}=\f{q+l}{\s{2}}$.  By the map
(\ref{scmap}) into the upper half plane (UHP), this is evaluated as
\be
\la\sigma(w_1,\bar{w}_1)\lb_w=\left|\f{\de \xi_1}{\de w_1} \right|^{\f{c}{12}(n-1/n)} \cdot
\la\sigma(\xi_1,\bar{\xi}_1)\lb_{UHP}\propto \left(\f{qa}{l(2q+l)}\right)^{\f{c}{12}(n-1/n)},
\ee
where $a$ is the UV cut off (lattice spacing) and we defined $\xi_1=i\s{(2q+l)/l}=-\bar{\xi}_1$.

Therefore by taking a derivative with respect to $n$, setting $n=1$, we find
\be
S_A=\f{c}{6}\log\f{2l(l+2q)}{qa}+\gamma_b,  \label{cftpr}
\ee
where $\gamma_b$ represents an additive constant.
This agrees with the decompactifying limit of the result in \cite{Raj} for a two dimensional CFT on a circle. Note also that as is obvious from the above analysis, the constant $\gamma_b$ depends on the boundary condition of the boundary state and is given by the boundary entropy \cite{bdye} plus a numerical constant which depends on the choice of the UV cut off $a$. In this paper we will simply set $\gamma_b = 0$, which does not change the outline of our results.

\subsection{UV Regularized Description and Time Evolution}

\begin{figure}
  \centering
  \includegraphics[width=6cm]{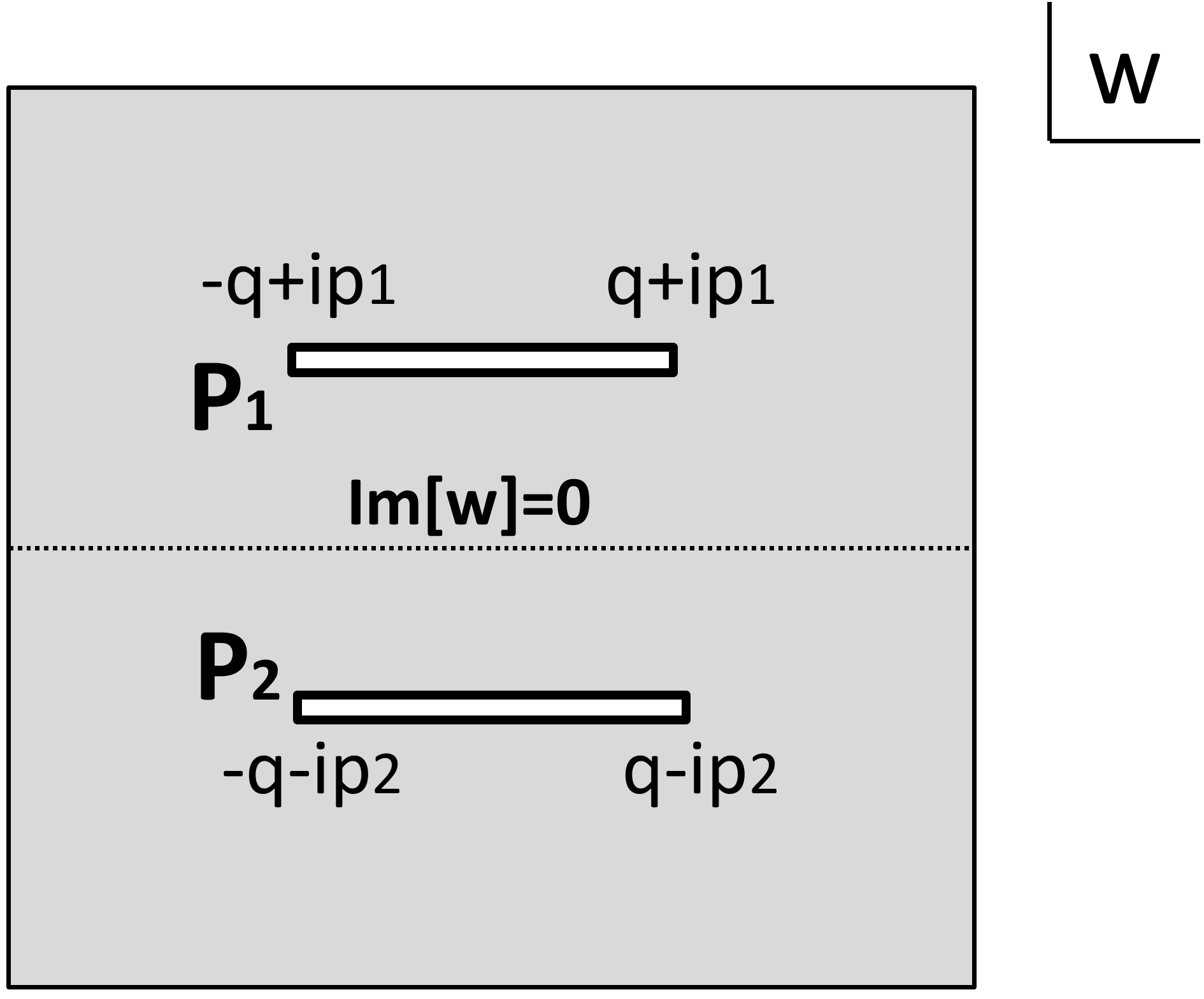}
  \hspace{1cm}
   \includegraphics[width=7cm]{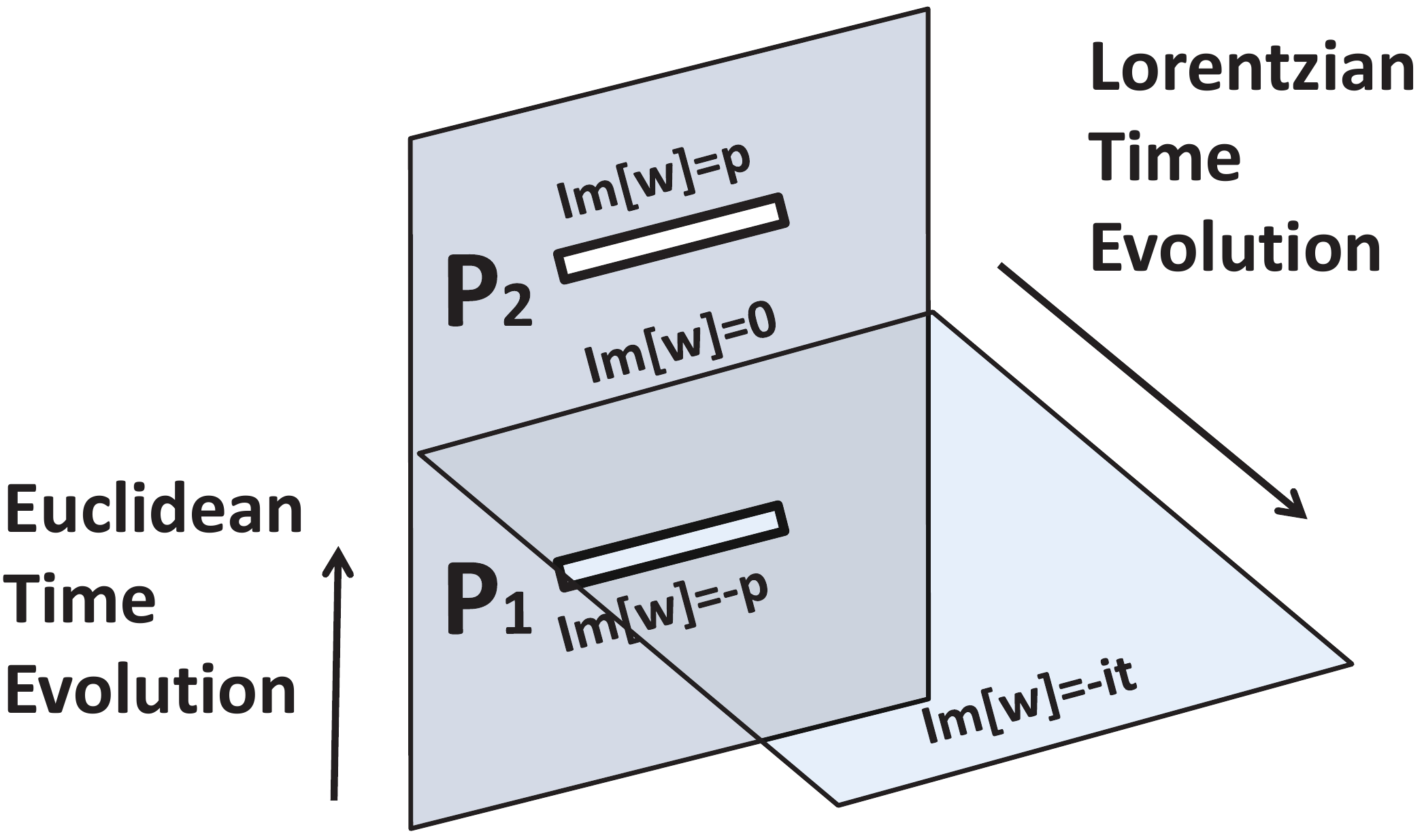}
    \caption{The left picture describes the Euclidean path-integral expression of the quantum state, evolved by an Euclidean time after the projection measurement. The right picture describes
    the real-time evolution after the projection measurement.}
\label{fig:twocutmap}
  \end{figure}

Actually, the previous description with a single cut and the conformal transformation (\ref{scmap}) leads to a physically singular setup as the energy density gets divergent near the two endpoints of the interval $P$ as we saw in (\ref{lww}). This is because we projected the state on $P$ even for high energetic modes. To resolve this issue, we add a small Euclidean time evolution of the projected state in the path-integration. This leads to the two cuts geometry explained in the left picture of Fig.\ref{fig:twocutmap}, where we choose the length of the interval $P$ to be $2q$. This regularized description has an advantage that we can study the time
evolution in a systematic way.

For the real time evolution we set
\be
p_1=p-it,\ \ \ p_2=p+it,
\ee
via an analytical continuation of Euclidean time as in the right picture of Fig.\ref{fig:twocutmap}.  This describes the time evolved state
\be
e^{-itH}e^{-pH}\cdot {\cal P}|\Psi_0\lb.  \label{totop}
\ee

\begin{figure}
  \centering
  \includegraphics[width=9cm]{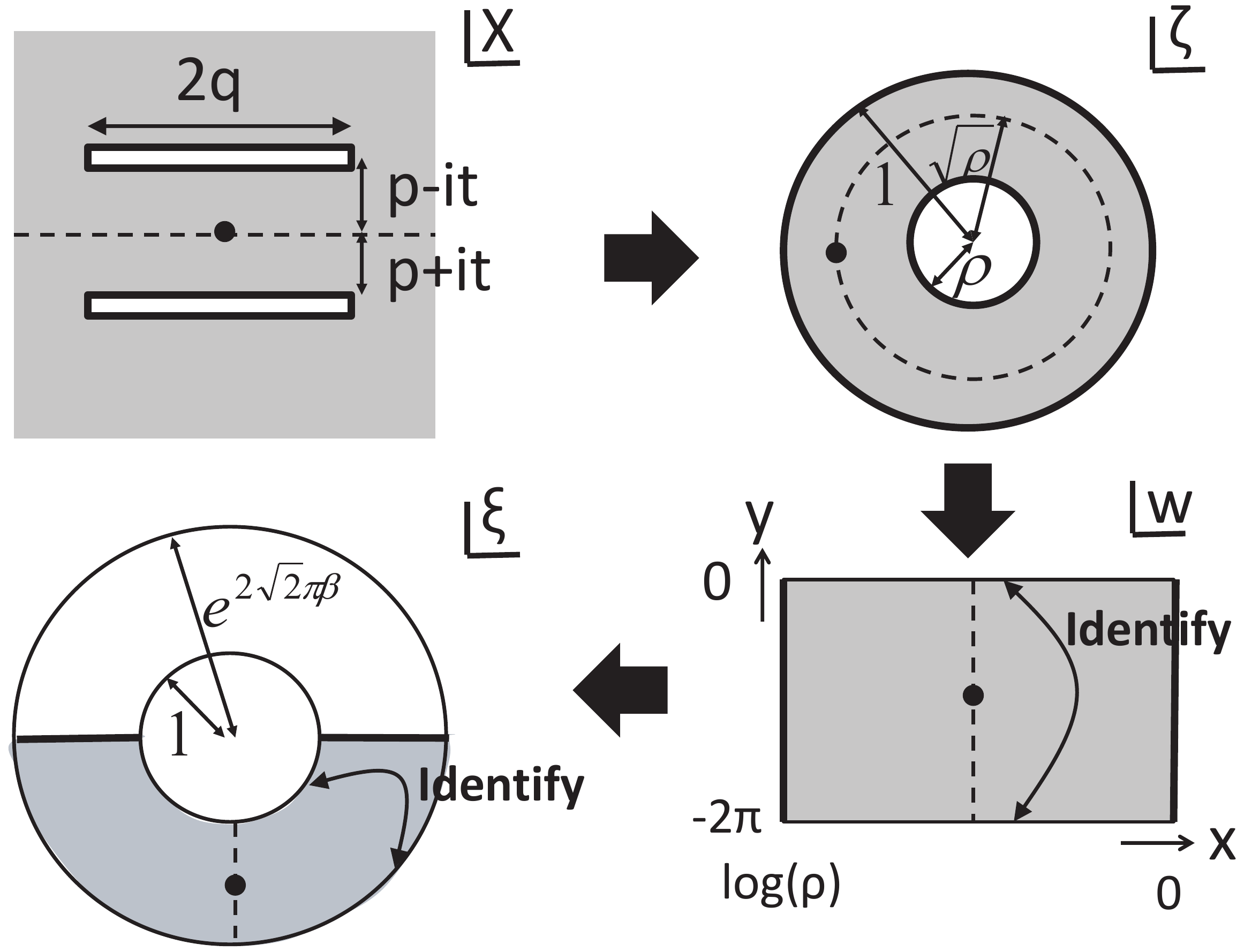}
  \caption{The conformal map between the two cut geometry and the cylinder}
\label{fig:twocutsmap}
  \end{figure}

 In order to analyze this setup, we can employ a chain of conformal maps from the two cut geometry into an annulus or cylinder, sketches in the upper half of Fig.\ref{fig:twocutsmap}. The map $X=X(\zeta)$ from the annulus $\rho\leq |\zeta|\leq 1$ to our two cut geometry is found in \cite{anmap} (see also
the general analysis \cite{DM}) as follows\footnote{
Here we set $A=2ip$ and $\chi=\f{\pi}{2}$ in eq.(2.1) of \cite{anmap} and used the fact $K(-1)=\f{1}{2}$ and
$K(-\rho)=0$.}:
\be
X(\zeta)=2ip\left(K\left(\zeta/\s{\rho}\right)+K\left(\zeta\s{\rho}\right)-\f{1}{2}\right)-t, \label{xcx}
\ee
where $K(\zeta)$ is defined by
\ba
&& K(\zeta)\equiv\zeta\f{d\log P(\zeta)}{d\zeta},  \no
&& P(\zeta)\equiv(1-\zeta)\prod_{k=1}^\infty (1-\rho^{2k}\zeta)(1-\rho^{2k}\zeta^{-1}).
\ea
Note that its complex conjugate is given by
\be
\bar{X}(\bar{\zeta})=-2ip\left(K\left(\bar{\zeta}/\s{\rho}\right)+K\left(\bar{\zeta}\s{\rho}\right)-\f{1}{2}\right) + t,
\ee
as we need to regard $it$ as a real valued Euclidean time.

We can express this function as follows
\be
K(\zeta)=\f{\zeta}{\zeta-1}+\sum_{k=1}^\infty \left(\f{\rho^{2k}\zeta^{-1}}{1-\rho^{2k}\zeta^{-1}}
-\f{\rho^{2k}\zeta}{1-\rho^{2k}\zeta}\right).
\ee
We can easily prove the relation $K(1/\zeta)=1-K(\zeta)$. Note also $K(1)=\infty$.

We can show that this map (\ref{xcx}), the radius $1$ and $\rho$ circle, which are boundaries of the annulus, are mapped into the cuts $[ip-q,ip+q]$ and $[-ip-q,-ip+q]$, respectively. Especially, the points
$\zeta=1$ and $\zeta=-1$ are both mapped into the point $X = ip - t$.  The points
$\zeta=\s{\rho}$ and $\zeta=-\s{\rho}$ inside the annulus are mapped into the $X=\infty$ and $X= - t$.

The value of $q$ can also be found from the transformation (\ref{xcx}) as we will see below. For this, we would like to study the mapping of the circle $|\zeta|=1$. The point
$\zeta=e^{i\theta}$ is transformed as
\be
X(e^{i\theta})=ip-t+2p\cdot \f{d\log F(\theta)}{d\theta},  \label{fxt}
\ee
where $F(\theta)$ is defined by (our theta-function convention is summarized in appendix \ref{thetaf})
\ba
F(\theta)=\prod_{n=1}^\infty (1-\rho^{n-1/2}e^{-i\theta})(1-\rho^{n-1/2}e^{i\theta})
=\rho^{\f{1}{24}}\cdot \f{\theta_4(\nu,is)}{\eta(is)}.
\ea
Here we defined $\theta=2\pi\nu$ ($0\leq \nu<1$) and $\rho=e^{-2\pi s}$.

\begin{figure}
  \centering
  \includegraphics[width=6cm]{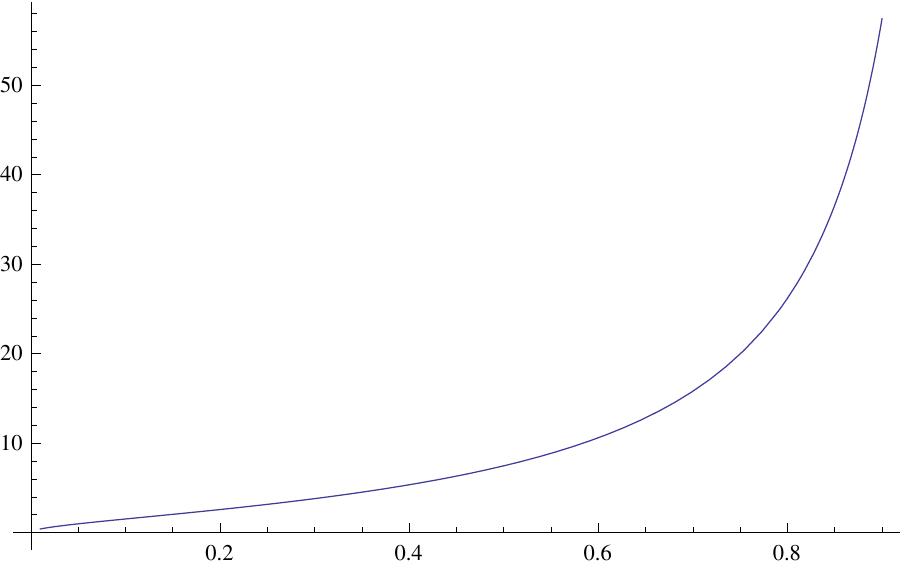}
  \caption{The ratio $\f{q}{p}$ as a function of $\rho$.}
\label{fig:qprho}
  \end{figure}

 Note that since $F(\theta)$ is real valued function and thus Im$[X (e^{i\theta})]= p$. The modular transformation leads to
\be
F(\theta)=\rho^{\f{1}{24}}\cdot e^{-\f{\pi \nu^2}{s}}\cdot \f{\theta_2\left(i\f{\nu}{s},\f{i}{s}\right)}{\eta(i/s)}.
\ee
We are interested in the limit $\rho\to 1$ or equally $\tau = i/s \to \infty$. In this limit we find
\be
F(\theta)\sim e^{-\pi \f{\nu^2}{s}}e^{-\f{\pi}{6s}}e^{\f{\pi\nu}{s}}.
\ee
Thus we find the following estimation in this limit
\be
X(e^{i\theta})\simeq ip-t+\left(1-2\nu\right)\f{p}{s}.
\ee
Since this takes the maximal value\footnote{Here we mean that $\nu$ is chosen to be infinitesimally small in the limit $\rho\to 0$. The strict value $\nu=0$ actually leads to $X= ip - t$ as we already mentioned.}
 at $\nu=0$, we obtain
\be
\f{q}{p}\simeq \f{1}{s},\ \ \ (q\gg p).  \label{qps}
\ee
The full behavior of $q/p$ on $\rho$ is shown in Fig.\ref{fig:qprho}.

For later purposes, it is useful to perform two further conformal transformations (sketched in
the lower half of Fig.\ref{fig:twocutsmap}).
We can map the annulus in the $\zeta$ coordinate into a cylinder in $w$ coordinate as follows:
\be
\zeta=\rho\cdot e^{-\s{2}w}.  \label{zw}
\ee
The coordinate $w=\f{x+iy}{\s{2}}$ takes values in the range
\be
\log\rho<x<0,\ \ \ -2\pi< y<0.  \label{rhoxy}
\ee

Moreover, we perform the following transformation into an annulus in $\xi$ coordinate:
\be
\xi=e^{2i \beta w}, \label{xw}
\ee
where we set
\be
\beta=\f{1}{2\s{2}s}.  \label{betas}
\ee

In the appendix B, we presented a toy analogous examples which allow much simple analytical calculations,
where we replaced the two slit with two disks.

\subsection{Explicit Example: 2d Free Fermion CFT}

Now we would like to apply the previous formulation of local projection measurement in 2d CFTs
to a 2d free fermion CFT as an explicit example. Especially, our goal is to compute the time evolution of entanglement entropy after the local projection measurement. For this purpose, first we briefly explain the expression of the twist operator in 2d free massless fermion, which is necessary for the replica method computation of entanglement entropy \cite{CC,CaHu,ANT}.
First we consider the Dirac fermion on $n$-sheeted manifold which has a branch point at $z=0$ on a plane $\mathbb{C}$. This can be seen as a collection of $n$ Dirac fermions $(\psi^{(a)},\bar{\psi}^{(a)}), a = 0,1,\cdots,n-1 \in Z_N$ on a plane $\mathbb{C}$ with the twisted boundary condition
\be
\psi^{(a)}_L(e^{2 \pi i} z ) = \psi_L^{(a+1)}(z),  \ \ \  \psi^{(a)}_R(e^{-2\pi i }\bar{z}) = \psi_R^{(a+1)}(\bar{z}),
\ee
where the $\psi^{(a)}_L(z)$ is the chiral part of Dirac fermion and $\psi^{(a)}_R(\bar{z})$ is the anti-chiral part.
Since the Lagrangian is gaussian, action remains unchanged after the discrete Fourier transformation
\be
\psi^{(a)} \to \f{1}{\s{n}} \sum_{b = 0}^{n-1} e^{\f{2\pi i a b}{n}}\psi^{(b)}.
\ee
After this discrete Fourier transformation, the twisted boundary condition is diagonalized:
\be
\psi^{(a)}_L(e^{2 \pi i} z ) =e^{\f{2 \pi i a}{n}} \psi_L^{(a)}(z), \label{bdydiagonal}
\ee
Thus the theory factorizes into $n$ Dirac fermions which decouple from each other with different boundary conditions given by (\ref{bdydiagonal}).
Next we consider the bosonization of free Dirac fermion. The free massless Dirac fermion is mapped to the free scalar $X(z,\bar{z})$ via (we follow the convention in \cite{ANT,TU})
\be
\psi_L(z) = e^{i X_L(z)}, \ \ \bar{\psi}_L(z) = e^{- i X_L(z)}, \ \ \psi_R(\bar{z}) = e^{i X_R(\bar{z})}, \ \ \bar{\psi}_R(\bar{z}) = e^{- i X_R(\bar{z})}
\ee
The twist operator for Dirichlet boundary condition is explicitly given by \cite{TU}
\be
\sigma^{(a)}(z,\bar{z}) = e^{i\f{a}{ n}(X_L(z) + X_R(\bar{z}))}, \label{Dtwist}
\ee
and for Neumann boundary condition, twist operator is given by
\be
\sigma^{(a)}(z,\bar{z}) = e ^{i \f{a}{n}(X_L(z) - X_R(\bar{z}))}.\label{Ntwist}
\ee
For example, we can confirm that the OPE with (\ref{Dtwist}) or (\ref{Ntwist}) reproduces the correct boundary condition (\ref{bdydiagonal}).  Then, the full twist operator is given by the product of each twist operator $\sigma^{(a)}$:\footnote{Here we only consider the case of odd $n$. For even $n$, we need to consider the choice of spin structure carefully, but the final result is not changed. }
\be
\sigma_n (z,\bar{z})= \prod_{n=-\f{n-1}{2}}^{\f{n-1}{2}}\sigma^{(a)}(z,\bar{z}).
\ee
Using this expression, we can calculate the correlation function of twist operators from the correlation functions of vertex operators.

\subsubsection{Entanglement Entropy at $t=0$}

We consider the (R\'enyi) entanglement entropy with one interval. The two point functions of twist operators on the upper half plane are given by
\ba
\braket{\sigma_n(\xi_1,\bar{\xi}_1) \sigma _{-n}(\xi_2,\bar{\xi}_2)}_{UHP} = \tilde{d}_n \left( \frac{a'^2 (\xi_1-\bar{\xi}_2)(\bar{\xi}_1-\xi_2)}{| \xi_1 - \bar{\xi}_1||\xi_2 -\bar{\xi}_2|(\xi_1-\xi_2)(\bar{\xi}_1-\bar{\xi}_2) } \right)^{\frac{1}{12}(n - \frac{1}{n})}, \label{twopt}
\ea
where $\ep$ is an UV cutoff and $\tilde{d}_n$ is the normalization factor of two point function on UHP. In $\text{Im} z$ and $\text{Im} w \to 0$ limit, the two point function should factorize to the product of one point functions on UHP:
\ba
\braket{\sigma_n(\xi_1,\bar{\xi}_1) \sigma_{-n}(\xi_2,\bar{\xi}_2)}_{UHP} &\sim& \braket{\sigma_n(\xi_1,\bar{\xi}_1) }_{UHP} \braket{\sigma_{-n}(\xi_2,\bar{\xi}_2)}_{UHP} \notag \\
&=& \tilde{c}_n^2\left(\f{a'}{|\xi_1-\bar{\xi}_1|} \f{a'}{|\xi_2-\bar{\xi}_2|}\right)^{\f{1}{12}(n-\f{1}{n})},
\ea
where $\tilde{c}_n$ is the normalization factor of the one point function and we use the same notation in \cite{CC}.
Also we define the cutoff $a'=a/\s{2}$ to match the notation of cutoff in \cite{Raj,Rajj,Rajjj}\footnote{Because we choose the notation of $w = \f{x+iy}{\s{2}}$, we need $\s{2}$ factor to match the notation with \cite{Raj,Rajj,Rajjj}.}.
On the other hand, in this limit the explicit two point function (\ref{twopt}) becomes
\be
\braket{\sigma_n(\xi_1,\bar{\xi}_1) \sigma _{-n}(\xi_2,\bar{\xi}_2)}_{UHP} \sim \tilde{d}_n  \left(\f{a'}{|\xi_1-\bar{\xi}_1|} \f{a'}{|\xi_2-\bar{\xi}_2|}\right)^{\f{1}{12}(n-\f{1}{n})}.
\ee
Thus we find that $\tilde{d}_n = \tilde{c}_n^2$.
The map from the cylinder with slit to UHP is given by
\be
\xi(w) = \s{\f{\sin\f{\pi}{L} (q+\s{2}w)}{\sin\f{\pi}{L}(q - \s{2}w)}},
\ee
and the differential is given by
\be
\frac{d\xi}{dw} = \f{\pi}{\s{2}L} \f{\sin\f{2\pi}{L}q}{\sqrt{\sin^3\f{\pi}{L}(q-\s{2}w) \sin\f{\pi}{L}(q+\s{2}w)}}.
\ee
From this, the correlation function of twist operators on cylinder is given by
\ba
\braket{\sigma_n(w_1,\bar{w}_1) \sigma_{-n}(w_2,\bar{w}_2)}_{cyl/slit} &=&  \tilde{c}_n^2 \Bigg( \s{\frac{d\xi}{dw}\Big|_{w=w_1 } \frac{d\bar{\xi}}{d\bar{w}}\Big|_{\bar{w}=\bar{w}_1 } \frac{d\xi}{dw}\Big|_{w=w_2 } \frac{d\bar{\xi}}{d\bar{w}}\Big|_{\bar{w}=\bar{w}_2 } }\notag \\
&&\frac{a'^2 (\xi_1-\bar{\xi}_2)(\bar{\xi}_1-\xi_2)}{ |\xi_1 - \bar{\xi}_1||\xi_2 -\bar{\xi}_2|(\xi_1-\xi_2)(\bar{\xi}_1-\bar{\xi}_2) } \Bigg)^{\frac{1}{12}(n - \frac{1}{n})}.
\ea
By taking the derivative with respect to $n$ and set $n=1$, the entanglement entropy is given by
\ba
S_A=\frac{1}{6}\log\left( \f{4L}{\pi} \f{\sin\f{\pi}{L}(l_1+2q)\sin\f{\pi}{L}l_1}{a \sin \f{2\pi}{L}q}\right) + \frac{1}{6}\log\left( \f{4L}{\pi} \f{\sin\f{\pi}{L}(l_2+2q)\sin\f{\pi}{L}l_2}{a \sin \f{2\pi}{L}q}\right) + 2 \tilde{c}_1'
\notag \\
+\f{1}{3} \log \f{|\xmap{l_1}-\xmap{l_2}|}{|\xmap{l_1}+\xmap{l_2}|}.
\ea
Note that $\ti{c}'_1$ is related to the boundary entropy $\log g$ \cite{bdye,CC} up to a UV regularization dependent constant.

When we take $L\to \infty$ , then we get
\be
S_A = \f{1}{6}\log \f{2(l_1+2q)l_1}{aq} + \f{1}{6}\log \f{2(l_2+2q)l_2}{aq} + 2 \tilde{c}_1' + \f{1}{3}\log \f{|\ixmap{l_1}-\ixmap{l_2}|}{|\ixmap{l_1}+\ixmap{l_2}|}.
\ee
When the region $[q+l_1,q+l_2]$ is far from the end of slit ($\f{q}{l_1} << 1$, $\f{|l_2-l_1|}{l_1} << 1$ ), entanglement entropy becomes
\be
S_A \sim \f{1}{3}\log \f{|l_1-l_2|}{a} + 2 \tilde{c}_1' ,
\ee
which is the same behavior with the entanglement entropy with no projection measurement up to the constant term.
Generically the two point function of twist operators on UHP is given by
\be
\braket{\sigma_n(w_1,\bar{w}_1)\sigma_{-n}(w_2,\bar{w}_2)}_{UHP} = \tilde{c}_n^2 \left(\frac{a'^2 }{| w_1 - \bar{w}_1||w_2 -\bar{w}_2|} \right) ^{\f{c}{12}(n-\f{1}{n})} \mathcal{F}(\eta).
\ee
Here $\eta = \f{(w_1-\bar{w}_2)(\bar{w}_1-w_2)}{(w_1-w_2)(\bar{w}_1-\bar{w}_2)}$ is the cross ratio and $\mathcal{F}(\eta)$ is the function that only depends on the cross ratio $\eta$ and satisfies $\mathcal{F}(1) = 1$ and $\mathcal{F}(\eta) \sim f_ n \eta ^{\f{c}{12}(n-\f{1}{n})} $ with some constant $f_n$.  We have $f_1 = 1$ in the limit $\eta \to 0$. From this, in the limit of $\f{q}{l_1} << 1$ and  $\f{|l_2-l_1|}{l_1} << 1$ , the entanglement entropy behaves
\be
S_A = \f{c}{3}\log \f{l_2-l_1}{a} + 2 \tilde{c}_1' + f_1'.
\ee
 In this way, we can confirm the usual logarithmic behavior of entanglement entropy for general CFTs. On the other hand, there is not only the term $\tilde{c}_1'$, related to the boundary entropy but also another constant term $f_1'$ which comes from the non universal term $\mathcal{F}(\eta)$.

\subsubsection{UV Regularized Description and Time Evolution}

\begin{figure}
  \centering
  \includegraphics[width=6cm]{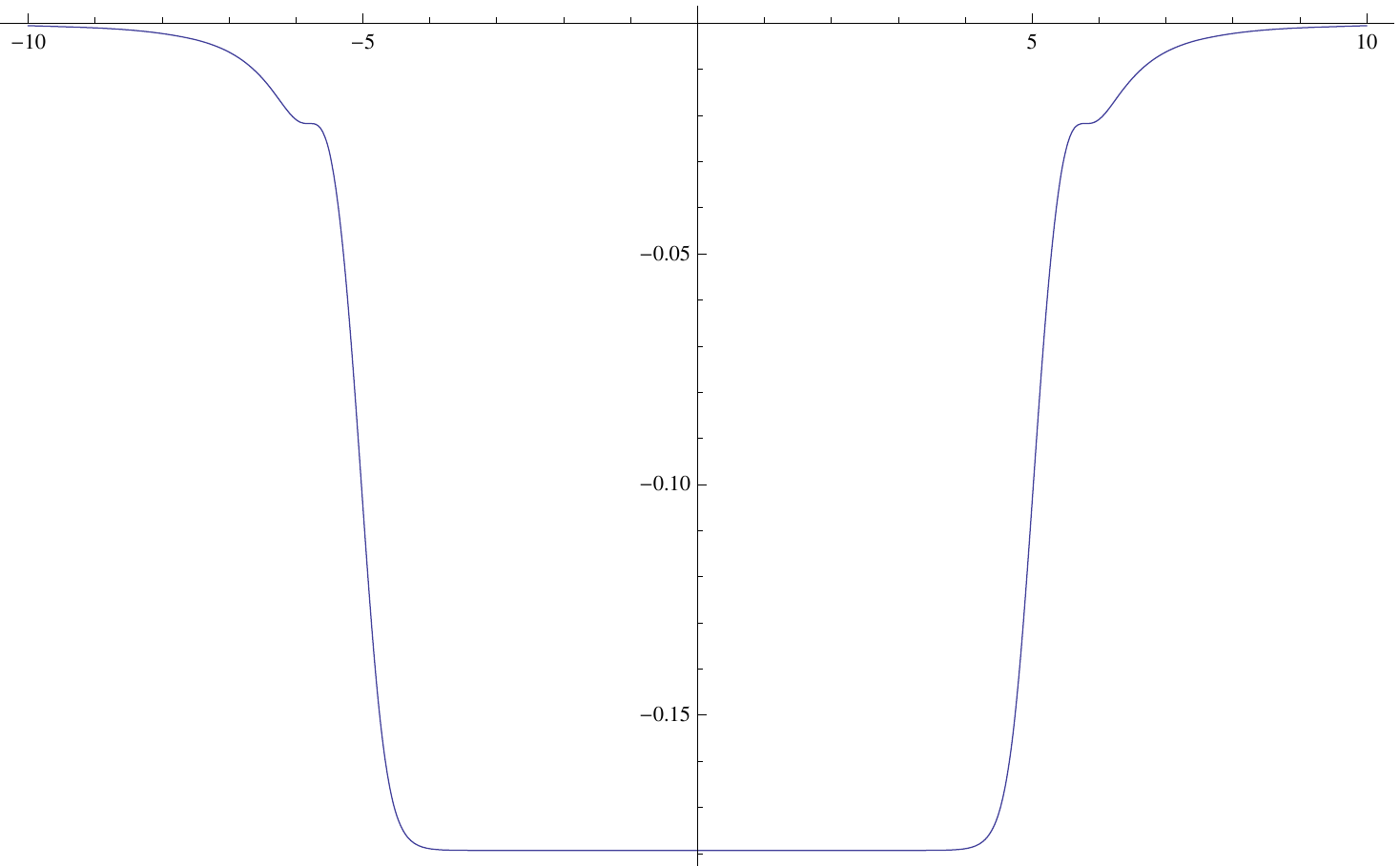}
  \hspace{1cm}
  \includegraphics[width=6cm]{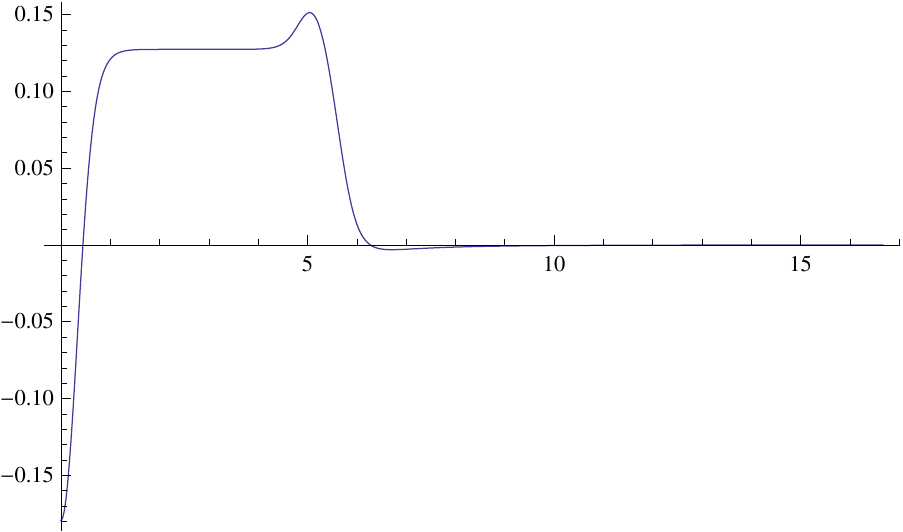}
  \caption{The behavior of the growth of entanglement entropy $\Delta S_A$ in the free fermion CFT after the local projection measurement. The left graph describes $\Delta S_A$ as a function $x$, where the subsystem $A$ is chosen to be the interval $[-0.5+x,0.5+x]$. The right one shows the time evolution of $\Delta S_A$ for the fixed subsystem $A$ given by $[-0.5,0.5]$.}
\label{fig:free}
  \end{figure}

Next we consider the projection measurement with cutoff. The map from the plane with two cuts to the annulus is given by
\ba
X(\zeta) &=& 2ip(K(\zeta\s{\rho}) + K(\zeta/\s{\rho}) -\f{1}{2}) - t ,\notag \\
\bar{X}(\bar{\zeta}) &=& -2ip(K(\bar{\zeta}\s{\rho}) + K(\bar{\zeta}/\s{\rho}) -\f{1}{2}) + t.
\ea
By putting $\zeta = \rho e ^y  (\bar{\zeta} = \rho e^{\bar{y}})$, we can map the annulus to the cylinder given by $0 \le \text{Re} y \le -\log \rho $ and $  0 \le \text{Im} y \le 2 \pi$.
Using this map, the correlation function of vertex operators $\sigma^{(a)}(x_1) $ and $\sigma^{(-a)}(x_2)$ is given by
\be
\braket{\sigma^{(a)}(x_1) \sigma^{(-a)}(x_2)}_{two cut} = \Big( \f{dy}{dX} \Big) ^{\f{a^2}{2 n^2}}\Big( \f{d\bar{y}}{d\bar{X}} \Big) ^{\f{a^2}{2 n^2}}\braket{\sigma^{(a)}(y_1,\bar{y}_1) \sigma^{(-a)}(y_2,\bar{y}_2)}_{cylinder},
\ee
where the coordinate $y_i $ and $\bar{y}_i$ on cylinder is given by
\be
y_i = -\f{1}{2} \log \rho + 2 \pi i \nu_i, \ \ \ \bar{y}_i = -\f{1}{2} \log \rho - 2 \pi i \bar{\nu}_i \label{point1and2}
\ee
$\nu_i\ (\bar{\nu}_i)$ is the solution of $X(\sqrt{\rho}e^{2\pi i \nu_i}) = x_i \ (\bar{X}(\sqrt{\rho}e^{-2\pi i \bar{\nu}_i}) = x_i) $.
The correlation function of vertex operators on cylinder with Neumann boundary condition  is considered in \cite{TU} and given by
\ba
&&\braket{V_{(k_R,k_L)}(y_1,\bar{y}_1)V_{(-k_R,-k_L)}(y_2,\bar{y}_2)} \notag \\
&=&\f{\braket{B|e^{- 2\pi s H} V_{(k_L,k_R)}(y_1,\bar{y}_1)V_{(-k_L,-k_R)}(y_2,\bar{y}_2)|B}_N }{\braket{B|e^{-2\pi s H }|B}_N}\notag \\
&=&\f{ \sum_{w=-\infty} ^{\infty} e^{- \f{R^2 w ^2 \pi s }{2}} e^{\f{R}{2}(k_Lw(y_1-y_2)-k_Rw (\bar{y}_1-\bar{y}_2))}}{\sum_{w=-\infty}^{\infty} e^{-\f{R^2 w ^2 \pi s }{2}}}
 \notag \\
&&\cdot \left(  \f{\eta(2is )^3}{\theta _1(\f{y_2-y_1}{2 \pi i}|2is)}  \right)^{k_L^2 }
\cdot \left(  \f{\eta(2is )^3}{\theta _1(\f{\bar{y}_2-\bar{y}_1}{2 \pi i}|2is)}  \right)^{k_R^2 } \cdot \left( \f{\theta _1(\f{y_1+\bar{y}_1}{2 \pi i}|2is) \theta _1(\f{y_2+\bar{y}_2}{2 \pi i}|2is)}{\theta _1(\f{y_1+\bar{y}_2}{2 \pi i}|2is) \theta _1(\f{y_2+\bar{y}_1}{2 \pi i}|2is)} \right)  ^{k_Lk_R}. \label{correlationcylinder}
\ea
We consider the case of symmetric interval i.e. $x_2 = -x_1$.
Then, we find that $\nu_1  -\nu_2= \bar{\nu}_1-\bar{\nu}_2 $ for any time $t$.
We also find that for any interval $[x_1,x_2]$ at $t = 0$, $\nu_1  -\nu_2= \bar{\nu}_1-\bar{\nu}_2 $ holds.
From this observation, after substituting the value of $k_R = \f{a}{n}, k_L = -\f{a}{n}$ and (\ref{point1and2}), we find that the numerator and the denominator of the third line of (\ref{correlationcylinder}) cancels.
Finally the correlation function on cylinder we need is given by
\ba
&&\braket{\sigma^{(a)}(y_1,\bar{y}_1)\sigma^{(-a)}(y_2,\bar{y}_2)}_{cylinder}  \notag \\
&=&\left(  \f{\eta(2is )^3}{\theta _1(\nu_2-\nu_1|2is)}  \right)^{\f{a^2}{n^2} }
\cdot \left(  \f{\eta(2is )^3}{\theta _1(\bar{\nu}_2-\bar{\nu}_1|2is)}  \right)^{\f{a^2}{n^2} } \cdot \left( \f{\theta _1(\nu_1 - \bar{\nu}_2 - is|2is) \theta _1(\nu_2 - \bar{\nu}_1 - is|2is)}{\theta _1(\nu_1 - \bar{\nu}_1 - is|2is) \theta _1(\nu_2- \bar{\nu}_2 - is|2is)} \right)  ^{\f{a^2}{n^2}}. \notag \\
\ea
Here $s = -\f{1}{2\pi}\log \rho$. The final expression of correlation function of twist operators on the plane with two cuts is given by
\ba
&&\braket{\sigma_n(y_1,\bar{y}_1)\sigma_{-n}(y_2,\bar{y}_2)}_{twocut} \notag \\
&=&\Bigg[ \s{\f{dy}{dX}\Big|_{X=x_1}\f{d\bar{y}}{d\bar{X}}\Big|_{\bar{X}=x_1}\f{dy}{dX}\Big|_{X=x_2}\f{d\bar{y}}{d\bar{X}}\Big|_{\bar{X}=x_2}} \notag \\
&& \cdot\left(  \f{\eta(2is )^3}{\theta _1(\nu_2-\nu_1|2is)}  \right)
\cdot \left(  \f{\eta(2is )^3}{\theta _1(\bar{\nu}_2-\bar{\nu}_1|2is)}  \right) \cdot \left( \f{\theta _1(\nu_1 - \bar{\nu}_2 - is|2is) \theta _1(\nu_2 - \bar{\nu}_1 - is|2is)}{\theta _1(\nu_1 - \bar{\nu}_1 - is|2is) \theta _1(\nu_2- \bar{\nu}_2 - is|2is)} \right)  \Bigg]^{\f{1}{12}(n-\f{1}{n})}. \notag \\
\ea
After taking the differential with regard to $n$ and put $n=1$, we get entanglement entropy:
\ba
S_A &=& \frac{1}{6} \log \Bigg[ \s{\f{dy}{dX}\Big|_{X=x_1}\f{d\bar{y}}{d\bar{X}}\Big|_{\bar{X}=x_1}\f{dy}{dX}\Big|_{X=x_2}\f{d\bar{y}}{d\bar{X}}\Big|_{\bar{X}=x_2}} \notag \\
&& \cdot\left(  \f{\eta(2is )^3}{\theta _1(\nu_2-\nu_1|2is)}  \right)
\cdot \left(  \f{\eta(2is )^3}{\theta _1(\bar{\nu}_2-\bar{\nu}_1|2is)}  \right) \cdot \left( \f{\theta _1(\nu_1 - \bar{\nu}_2 - is|2is) \theta _1(\nu_2 - \bar{\nu}_1 - is|2is)}{\theta _1(\nu_1 - \bar{\nu}_1 - is|2is) \theta _1(\nu_2- \bar{\nu}_2 - is|2is)} \right)  \Bigg]. \notag \\
\ea
To extract the effect of projection measurement, we consider the difference of entanglement entropy from that of the ground state $S_A^{ground}$:
\be
\Delta S_A \equiv S_A - S_A^{ground}.
\ee
We plotted $\Delta S_A$ in the case of $p=\f{1}{2}$ and $\rho = 0.6$ (corresponding to $q\simeq 5.3$) in Fig.\ref{fig:free}. The left graph shows that the quantum entanglement is reduced around the region $P$
$-q\leq x\leq q$ as expected. The right graph shows the time evolution of entanglement entropy for a fixed interval. Just after the local projection at $t=0$, the entropy grows linearly, whose mechanism is very similar to the global quenches \cite{GQ}. It starts saturated around the time given by the half of the length of interval and gets constant for a time period $q$. After that it rapidly goes to zero.  This is because the excitations, which are originally produced in the region $P$ at $t=0$, simply go a way from the interval $A$ for the late time region $t>q$, as they propagate at the speed of light.

\section{Partially Entangling and Swapping of Two CFTs}

Next we would like to consider two different quantum operations which act on two identical 2d CFTs, called CFT$_1$ and CFT$_2$. They are (a) partially entangling and (b) partially swapping, of the two CFTs.

\begin{figure}
  \centering
  \includegraphics[width=7cm]{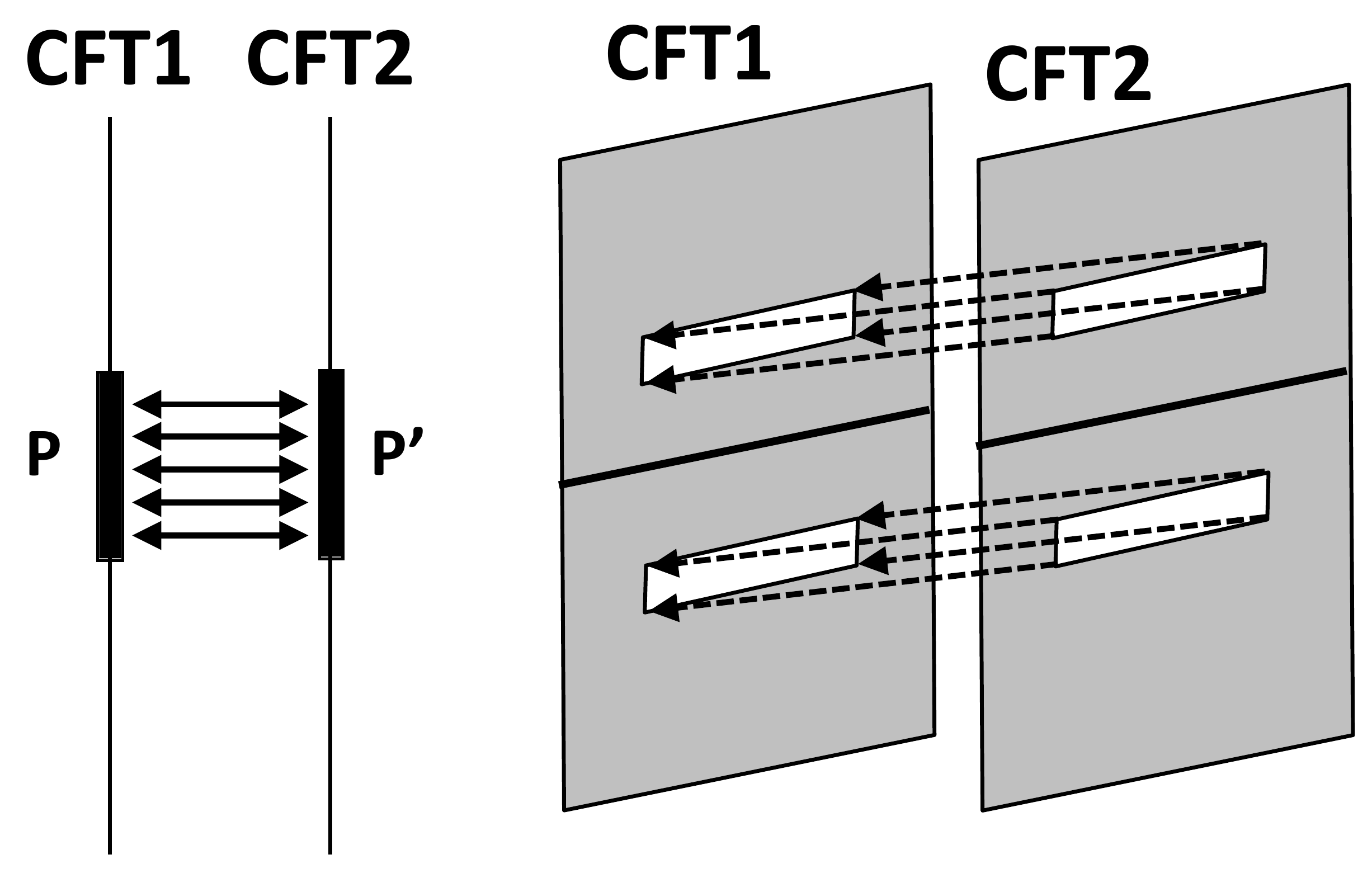}
  \hspace{1cm}
  \includegraphics[width=7cm]{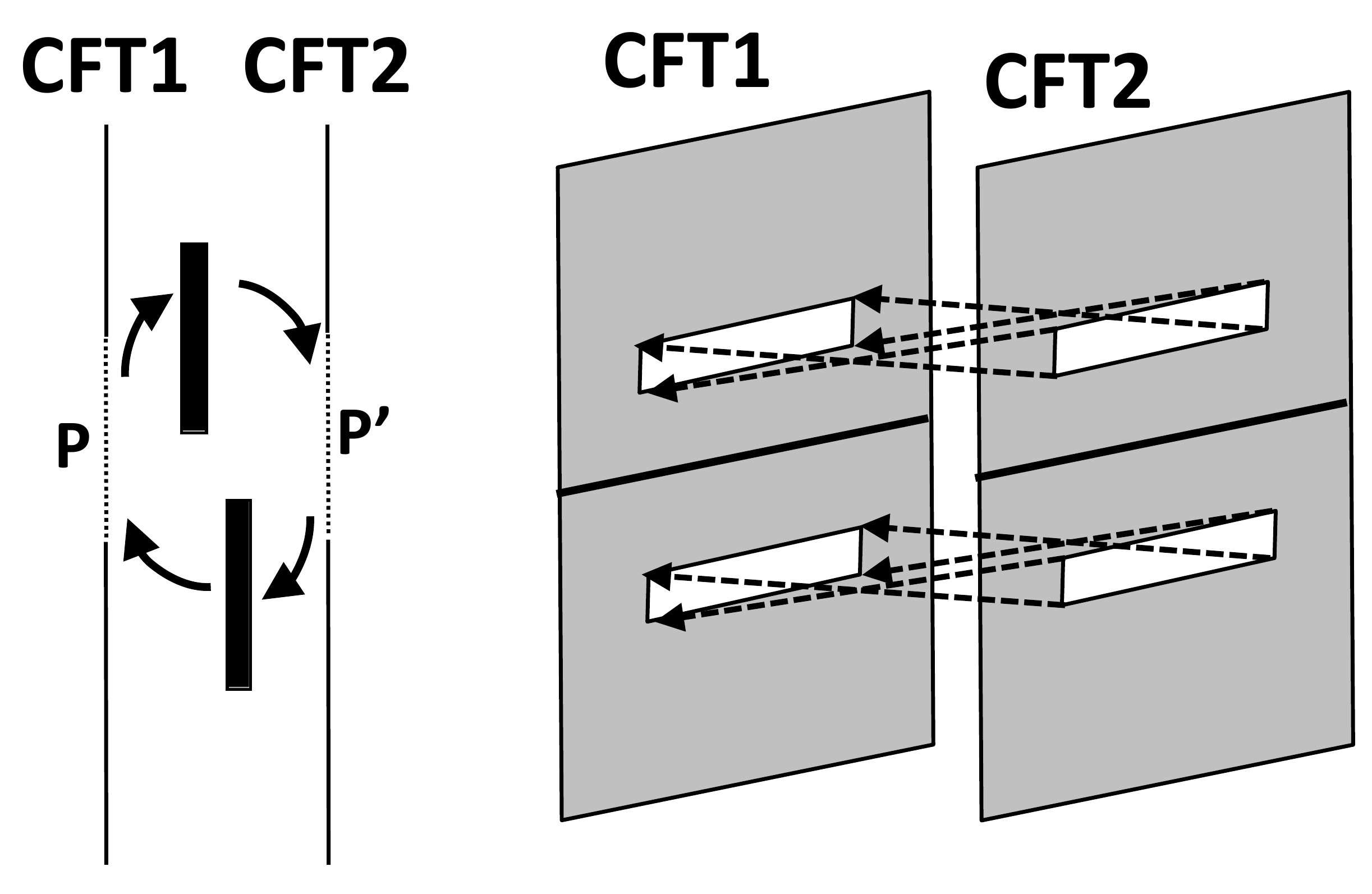}
  \caption{The description of the partial entangling of two CFTs (left) and partial swapping between two CFTs (right).}
\label{fig:swap}
  \end{figure}

The former (a) partially entangling, is defined as the simplest local projection described by gluing an interval $P$ with length $2q$ in CFT$_1$ and an interval $P'$ with the same size in CFT$_2$. This procedure and its Euclidean path-integral with a UV regularization are sketched in the left half of Fig.\ref{fig:swap}. Equally this is obtained by attaching a pair of the sheet with two slits shown in
Fig.\ref{fig:twocutmap}. This projection ${\cal P}$ is explicitly written as
\be
{\cal P}_e=\prod_{x\in P}\left(\sum_{n_x}|n_x\lb_{1}|n_x\lb_{2}\right)\left(\sum_{m_x}\la m_x|_{1}\la m_x|_{2}\right) \otimes \prod_{x\in P^c}\left(I^1_{x}\otimes I^2_{x}\right), \label{pento}
\ee
where $|n_x\lb$ and $|m_x\lb$ denote arbitrary states in the 2d CFT; $I_x$ is the identity operator; $P^c$ is the complement of the region $P$.
This corresponds to a projection onto a maximally entangled state (i.e. analogue of EPR state) between the two CFTs on the interval $P$.  We can insert the real time evolution by a small period $p$ and UV regularization, which leads to the same expression of total quantum state (\ref{totop}). Note that this procedure also introduce some minor entanglement between $P^c$ in CFT$_1$ and that in CFT$_2$ at the same time.

Another interesting operation for the identical two CFTs is (b) partially swapping.  This is defined as a swapping of the two intervals $P$ and $P'$ as depicted in the right half of Fig.\ref{fig:swap}. At time $t=0$ we cut out the intervals $P$ in CFT$_1$ and $P'$ in CFT$_2$ and glue them again by exchanging them. This is described by the operation :
\be
\prod_{x\in P}\left(\sum_{n_x,m_x}|n_x\lb_{1}|m_x\lb_{2}\la m_x|_{1}\la n_x|_{2}\right) \otimes \prod_{x\in P^c}\left(I^1_{x}\otimes I^2_{x}\right). \label{pentosw}
\ee

These two quantum operations have simple descriptions in the Euclidean path-integral description as also shown in Fig.\ref{fig:swap}. We first prepare the vacuum states of the two identical CFTs by path-integrals from the past infinity of the Euclidean time $t_E=-\infty$. In the case of (a) partial entangling of the two CFTs, we glue them with each other along the intervals $P$ and $P'$ at $t_E=-p$. Soon after that we open up new two sheets just above this and propagates by a period $p$. This defines the wave function of total quantum state $e^{-pH}{\cal P}_e|\Psi_0\lb$ at $t_E=0$. On the other hand, in the case of (b) partial swapping, we exchange the intervals $P$ and $P'$ at $t_E=-p$. After that we perform the Euclidean time evolution until $t_E=0$ to obtain the regularized wave function.

It is useful to note that the topology of the Euclidean path-integral for the whole time $-\infty<t_E<\infty$
is given by a torus in both cases as is clear from
Fig.\ref{fig:swap}. Since we are working on CFTs, an important quantity is the period $\tau=\tau_1+i\tau_2$ of this torus ($\tau_1$ is vanishing in all of examples in this paper).

If we assume $\tau_2\gg 1$ i.e. high temperature limit, the thermal entropy for a CFT on a circle is given by the universal formula $S=\f{\pi c}{3}\tau_2$. By the conformal map we explained, this entropy coincides with the entanglement entropy $S_1$ when we trace out the whole CFT$_2$. In the AdS/CFT setup which we will study in the next section, we find
\ba
&& S_1=\f{\pi c}{3}\tau_2\ \  \ (\tau_2>1),\no
&& S_1=0\ \ \ (\tau_2<1),  \label{entot}
\ea
where the former is computed as the entropy of BTZ black hole.
Note that when $\tau_1=0$, $\tau_2$ is given by the ratio of lengths of two cycles
$C_x$ and $C_y$ of the torus as
\be
\tau_2=\f{|C_y|}{|C_x|}. \label{cxy}
\ee

In the appendix B, we presented a toy analogous examples which allow much simple analytical calculations,
where we replaced the two slit with two disks.

\subsection{Case (a): Partial Entangling of Two CFTs}\label{sec:PEnt}

In the case of partial entangling of the two identical CFTs,
we need to glue the upper edge (and lower one) of a slit in the first sheet to the upper one (and lower one) of a cut in the second sheet.\footnote{Though this geometry is topologically a torus, we cannot describe it by the curve (\ref{elp}), which will be used to describe the case (b). If we consider a sheet with two cuts which are both along a line e.g. the real line, the two different setups (swapping and entangling) are described by the same torus, because we can simply rotate  the sheet by the angle $\pi$ along the real axis.
However, if we act the $SL(2,C)$ transformation such that the four points on the real axis are mapped into the four points $-ip-q,-ip+q,ip-q,$and $ip+q$, the two cuts are mapped into two parts of a radius
$\s{p^2+q^2}$  circle in an obvious way. Therefore, the final setup is different from the entangling one, though it is equivalent in the swapping case.} We can employ the explicit conformal map between the two slit geometry and annulus (\ref{xcx}). The two sheet are now mapped into two annuli with the same size and we simply need to glue the two boundary circles in each annulus with those in the other annulus to obtain the torus.

\begin{figure}
  \centering
  \includegraphics[width=6cm]{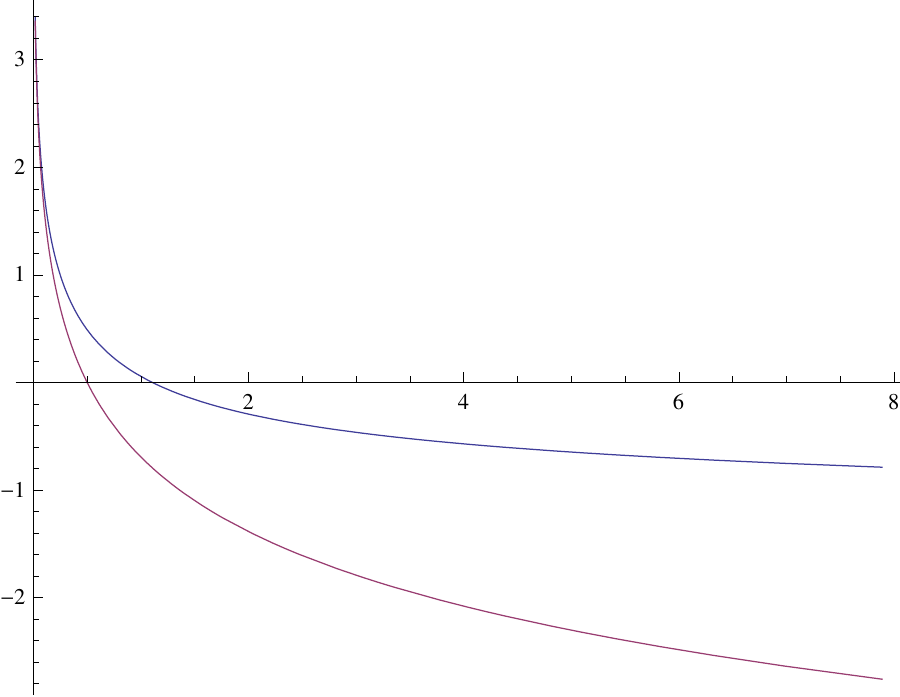}
  \caption{We showed the plot of entanglement entropy between the two CFTs as a function of (Euclidean time $p$ )/(projection length $q$) in the case of partial entangling. The vertical and horizontal axis correspond to $\log|\tau|=\log\f{C_y}{C_x}$ and $\gamma=p/q$. The blue curve describes the behavior of the entanglement entropy, while the red one
  corresponds to the approximation (\ref{aplq}).
  When the blue curve crosses the real axis, there is a phase transition and the entropy becomes vanishing for larger values of $\gamma$.}
\label{fig:twocutperiod}
  \end{figure}

 The period $\tau_2=\f{|C_y|}{|C_x|}=\f{1}{2s}$ (note $\tau_1=0$) can be found from an analysis of (\ref{fxt}) and we plotted in Fig.\ref{fig:twocutperiod}. When $q\gg p$, we find from (\ref{qps}) that the period of the torus is estimated by
 \be
\tau_2=\f{1}{2s}\simeq \f{q}{2p}.  \label{aplq}
\ee
Finally, the entanglement entropy between CFT$_1$ and CFT$_2$ can be obtained from (\ref{entot}) as follows:
\be
S_1\simeq \f{\pi c}{6}\cdot \f{q}{p}, \ \ \ (q\gg p). \label{entvol}
\ee
This behavior matches with our expectation. After the partial entangling along the interval $P$,
we expect that the state becomes maximally entangled on $P$. Since the length is given by $2q$ and the UV cut off (or lattice spacing of entangled pairs) is given by $p$, we naturally obtain the above estimation, which is extensive.

\subsection{Case (b): Partial Swapping of Two CFTs}

\begin{figure}
  \centering
  \includegraphics[width=6cm]{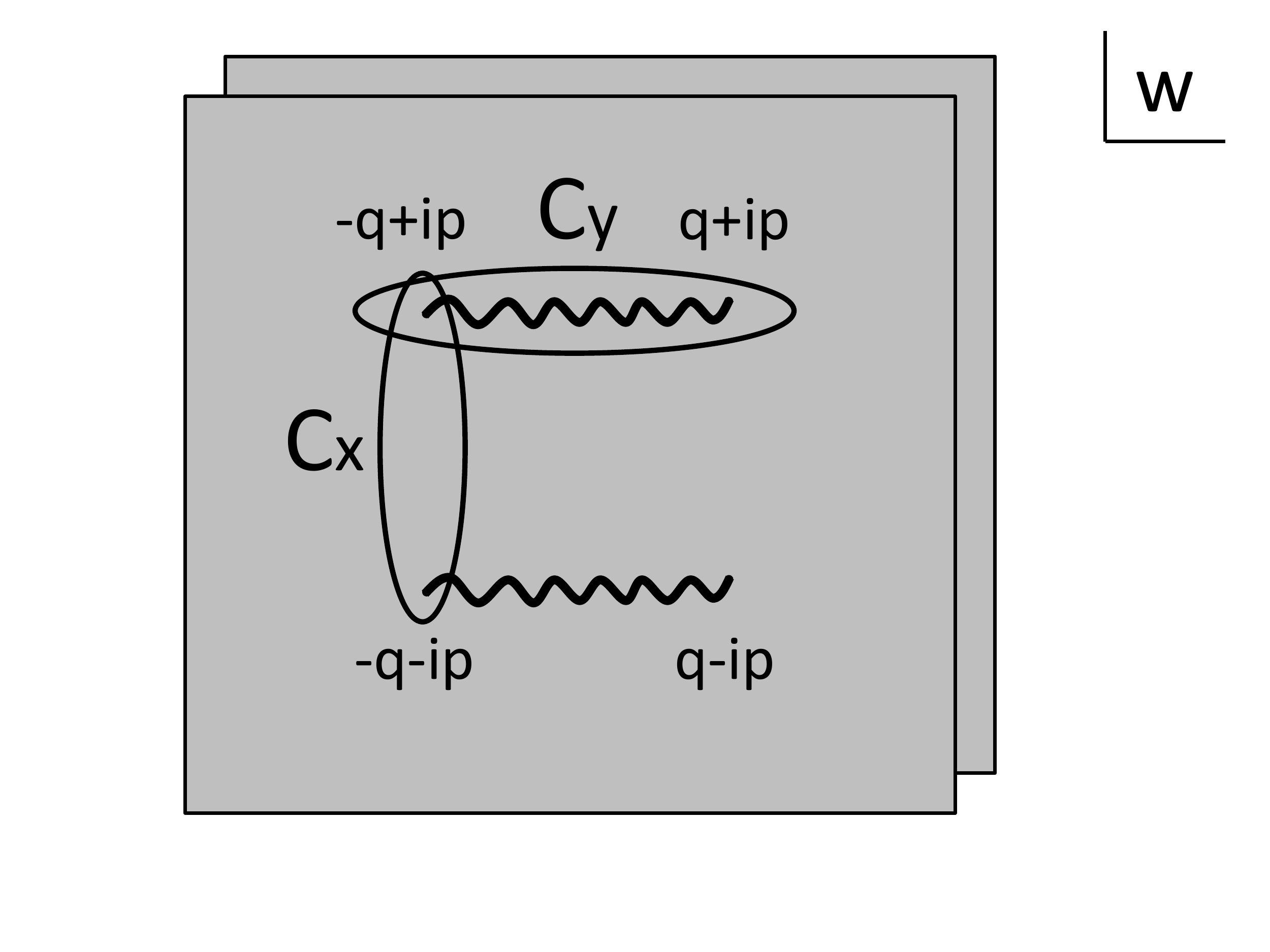}
  \hspace{1cm}
  \includegraphics[width=6cm]{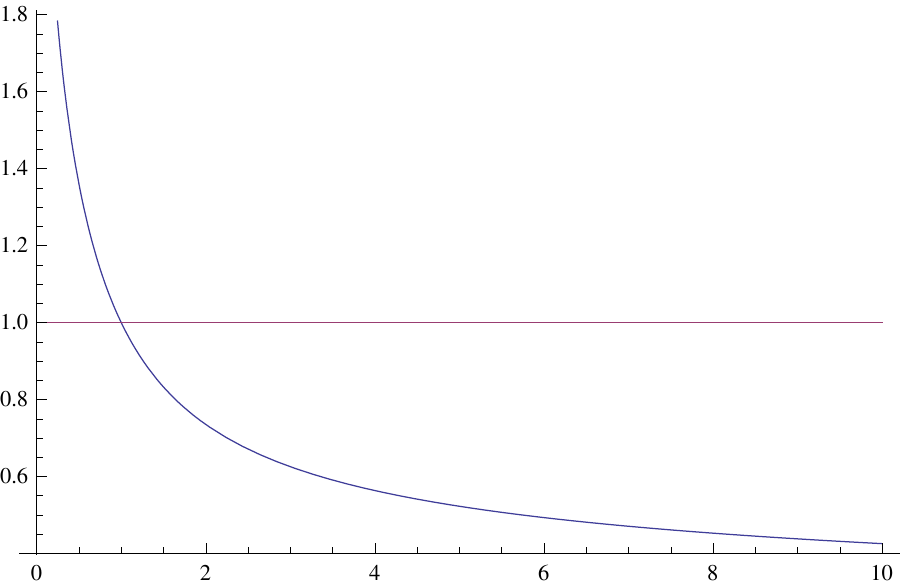}
  \caption{The left picture describes the double sheeted geometry and cycles of a torus. The right plot
  shows the entanglement entropy between the two CFTs as a function of (Euclidean time $p$ )/(projection length $q$) in the case of partial swapping. The vertical and horizontal axis correspond to $\f{C_y}{C_x}$ and
  $\gamma=p/q$. The blue curve for $0<\gamma<1$ describes the behavior of the entanglement entropy.
  At $\gamma=1$ there is a phase transition and the entropy becomes vanishing for $\gamma>1$.}
\label{fig:TwoCutr}
  \end{figure}

To describe the partial swapping, we paste two planes along the two cuts
$[-ip-q,-ip+q]$ and $[ip-q,ip+q]$ (refer to the left picture of Fig.\ref{fig:twocutmap} for each plane). The lower part of each cut in the first sheet continues to the upper part of the cut in the second sheet, as depicted the left picture of Fig.\ref{fig:TwoCutr}. This geometry is described by the elliptic curve
\be
y^2=(x-ip-q)(x-ip+q)(x+ip-q)(x+ip+q). \label{elp}
\ee
The doubled planes are mapped into a torus with a period $\tau$. We denote two cycles by $C_x$ and $C_y$, respectively (see the left picture of Fig.\ref{fig:TwoCutr}). The period $\tau_2$ of torus is explicitly evaluated from the ratio (\ref{cxy}) of the integrals of the holomorphic one form around each cycle:
\ba
&& |C_x|=\f{2}{p}\int^1_{-1}\f{dz}{\s{(1-z^2)(4\gamma^{-2}+1-z^2-4i\gamma^{-1}z)}},\no
&& |C_y|=\f{2}{q}\int^1_{-1}\f{dz}{\s{(1-z^2)(4\gamma^2+1-z^2-4i\gamma z)}},
\ea
where we set $\gamma=p/q$. The entanglement entropy is found from (\ref{entot})  as follows:
\be
S_1=\f{\pi c}{3}\cdot \f{|C_y|}{|C_x|},  \label{entrocut}
\ee
assuming $|C_y|>|C_x|$. We numerically plotted the ratio in the right picture of Fig.\ref{fig:TwoCutr}.

When the size of the projected region is much larger than the UV cut off scale i.e. $q\gg p$ we find the following behavior:\footnote{We can derive this result from the standard expression of elliptic curve
$y^2=(x-1)(x+1)(x-u)$, (we use the notation of \cite{SW}), because we can map our curve (\ref{elp}) into this form by a $SL(2,C)$ transformation. Note that under this map, the shape of cut is deformed into a circle. However, the geometry does not depend on the shape of the cut. We find the cross ratio of $[-1,1,u,\infty]$ is given by $\eta=\f{z_{12}z_{34}}{z_{13}z_{24}}=\f{2}{u+1}$. On the other hand, in our example we find
$\eta=\f{q^2}{p^2+q^2}$ Thus we find $u=1+\f{2p^2}{q^2}$. When $u\simeq 1$, the period of the torus
behaves like $\tau\sim -\f{i}{\pi}\log(u-1)\sim \f{i}{\pi}\log\f{2p^2}{q^2}$. Thus we find the result (\ref{perods}).}
\be
S_1 \simeq \f{2c}{3}\log\left(\f{q}{p}\right). \label{perods}
\ee
Interestingly, this result can be easily understood from the partial swap procedure.
Before the swapping, the interval $P$ with the length $2q$ is entangled with the other part of CFT$_1$. After the swap, this entanglement transferred into that between CFT$_1$ and CFT$_2$, leading to the entanglement entropy $S_P=\f{c}{3}\log \f{q}{p}$, where we remembered that $p$ represents the UV cut off. Since there is the same contribution from the interval $P'$ in the CFT$_2$, totally we reproduce (\ref{perods}).

At $\gamma=1$ there is a phase transition from the BTZ black hole to the thermal AdS.
Thus the entropy becomes vanishing for $\gamma>1$ in the large $c$ limit of holographic
CFTs.

Finally it is intriguing to note that the two types attachments of the two sheets lead to two different tori and thus lead to the two different behaviors of entropy (\ref{entvol}) and (\ref{perods}).

\section{Holographic Local Projection Measurement}
 In this section we explain how we construct gravity duals of local projection measurements explained in section \ref{sec:Proj} in the AdS$_3/$CFT$_2$ setups. We will study the behavior of entanglement entropy. Here we set the UV regularization parameter $p$, introduced in (\ref{totop}), to be zero for simplicity and focus on the state at $t=0$. We will analyze a gravity dual for non-zero $p$ and its time evolution in the next section.

\subsection{Conformal Transformation and AdS$_3/$CFT$_2$}
First we start with an Euclidean version of the holographic dual of general conformal map for AdS$_3/$CFT$_2$
in \cite{Ro}. Let us start with the Poincare AdS$_3$
\be
ds^2=R^2\left(\f{d\eta^2+2d\xi d\bar{\xi}}{\eta^2}\right), \label{poi}
\ee
where $(\xi,\bar{\xi})$ corresponds to the coordinate of complex plane at the AdS boundary.
This is dual to the vacuum state of a two dimensional (2d) CFT on $R^2$. Now we would like to perform the standard conformal map in 2d CFT (or holomorphic map): $\xi\to w$ as follows
\be
\xi=f(w),\ \ \ \bar{\xi}=\bar{f}(\bar{w}).
\ee

In the bulk AdS, this is extended to the following coordinate transformation:
\ba
&& \xi=f(w)-\f{2z^2(f')^2\bar{f}''}{8(f')(\bar{f}')+z^2f''\bar{f}''}, \no
&& \bar{\xi}=\bar{f}(\bar{w})-\f{2z^2(\bar{f}')^2f''}{8(f')(\bar{f}')+z^2f''\bar{f}''}, \no
&& \eta=\f{8z(f'\bar{f}')^{3/2}}{8(f')(\bar{f}')+z^2f''\bar{f}''}. \label{mapz}
\ea

By this coordinate transformation, the Poincare AdS metric (\ref{poi}) is mapped into
\be
ds^2=R^2\left(L(w)dw^2+\bar{L}(\bar{w})d\bar{w}^2
+\left(\f{2}{z^2}+\f{z^2}{2}L(w)\bar{L}(\bar{w})\right)dwd\bar{w}+\f{dz^2}{z^2}\right).
\label{metm}
\ee
Here we defined
\be
L(w)=\f{3(f'')^2-2f'f'''}{4(f')^2},
\ee
which is proportional to the energy momentum tensor (\ref{emtw}) induced by the conformal anomaly (i.e. the Schwarzian derivative terms). Note that in the AdS boundary limit $z\to 0$, the boundary metric becomes flat $ds^2\simeq \f{2R^2}{z^2}dwd\bar{w}$. Moreover, if we want to consider a Lorentzian metric, we can simply set $(w,\bar{w})\to (w^+,w^-)$.

For example, if we perform the conformal transformation $f(w)=e^{2i\beta w}$, the bulk coordinate transformation is given by
\ba
&& \xi=e^{2i\beta w}\cdot\left(\f{2-z^2\beta^2}{2+z^2\beta^2}\right), \no
&& \eta=e^{i\beta(w-\bar{w})}\cdot \left(\f{4\beta z}{2+\beta^2 z^2}\right). \label{holexp}
\ea
We find $L(w)=-\beta^2$ and the final metric (\ref{metm}) looks like
\be
ds^2=\f{R^2}{z^2}\left[dz^2+\left(1-\f{\beta^2}{2}z^2\right)^2dx^2
+\left(1+\f{\beta^2}{2}z^2\right)^2dy^2\right], \label{btz}
\ee
where $w=\f{x+iy}{\s{2}}$. For a smooth geometry we need to require the periodicity
$x\sim x+\f{\pi\s{2}}{\beta}$. If we regard $x$ as an Euclidean time, this metric describes the BTZ
black hole \cite{BTZ}, which is topologically a solid torus. The boundary of this geometry describes a torus and the two independent cycles can be chosen to be
\be
C_x: x\in \left[-\f{\pi}{\s{2}\beta},\f{\pi}{\s{2}\beta}\right],\ \ \ \ C_y: y\in \left[0,|C_y|\right].
\ee
Note that the cycle $C_x$ is contractible in the solid torus geometry (\ref{btz}), while $C_y$ is not, representing the black hole horizon. This black hole solution is thermodynamically favored when
$|C_y|>|C_x|$. When $|C_y|<|C_x|$ the thermal AdS solution is favored which is obtained from
(\ref{btz}) by a simple renaming $(x,y)\to (y,x)$ of the torus coordinates.

\subsection{Holographic Dual of Local Projection Measurement}

Consider an Euclidean 2d CFT on $R^2$: $ds^2=2dwd\bar{w}$ with the coordinate $(w,\bar{w})=\left((x+iy)/\s{2},(x-iy)/\s{2}\right)$.
The region $P$ is defined by the interval $-q\leq x\leq q$ and at $t=0$ we do the local projection measurement for all points in $P$. As we explained in section two, we can employ the Euclidean path-integral formulation and construct the state at $t=0$ by imposing a BCFT boundary condition around the slit $P$.

\begin{figure}
  \centering
  \includegraphics[width=6cm]{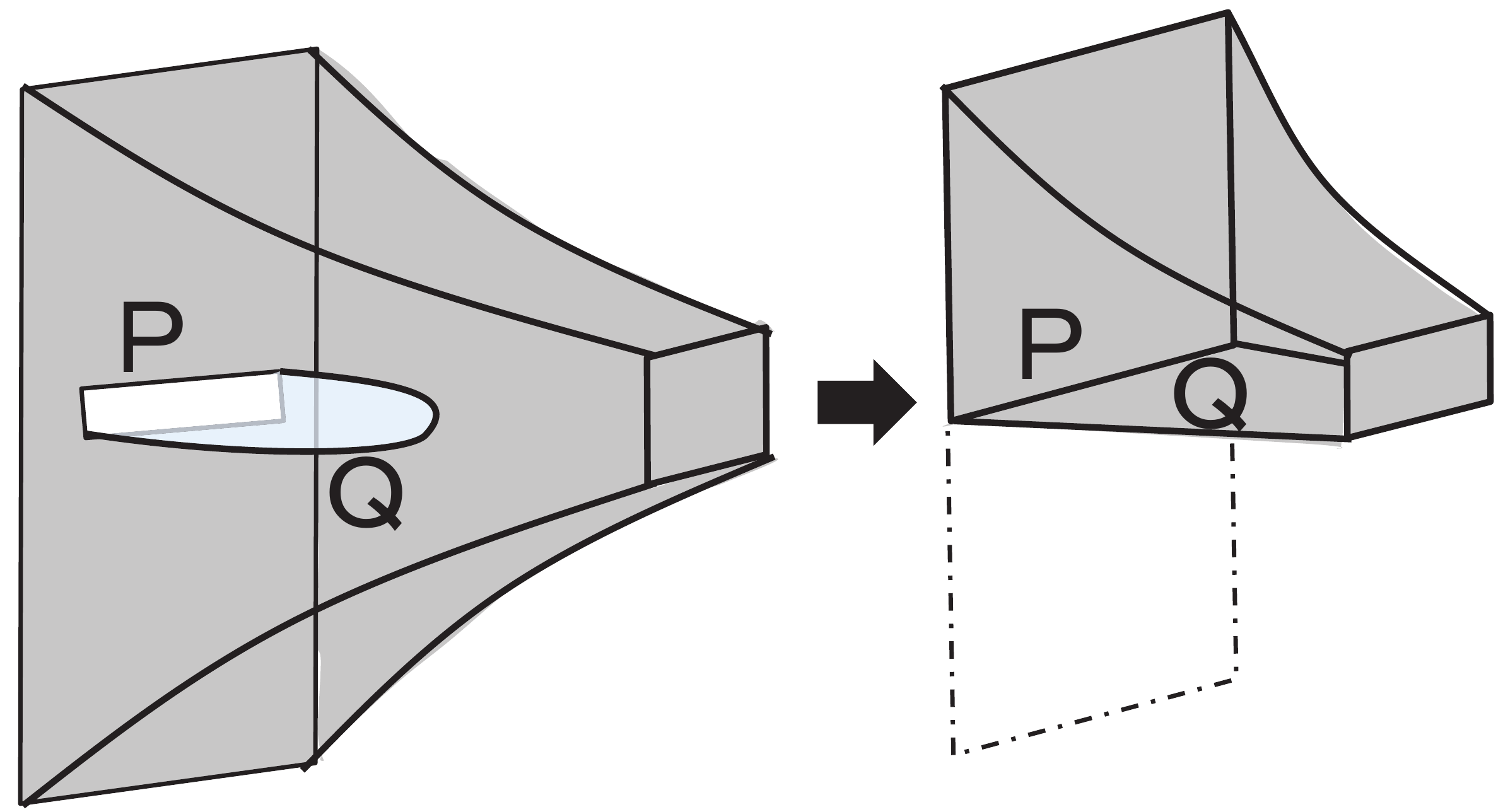}
  \caption{The setup of AdS/BCFT and its conformal transformation into a half of Poincare AdS.}
\label{fig:adsmap}
  \end{figure}

A holographic dual of such a BCFT can be found by using the prescription of AdS/BCFT
\cite{AdSBCFT} (refer to Fig.\ref{fig:adsmap}), which gives a bottom up model for such problems. An upshot is that we can construct the gravity dual by extending the boundary in a holographic CFT toward the three dimensional bulk such that this extended two dimensional surface $Q$
satisfies the following condition:
\be
K_{ab}-Kh_{ab}+Th_{ab}=0, \label{abcft}
\ee
where $h_{ab}$ and $K_{ab}$ are the induced metric and the extrinsic curvature of the surface $Q$;
$K$ is the trace of $K_{ab}$. The parameter $T$ corresponds to the tension when we regard $Q$ as a brane.
For example, the quantity called boundary entropy \cite{bdye} is a monotonically increasing function of $T$. In this paper, we simply set $T=0$, for which the surface $Q$ becomes a totally geodesic surface.

Note that the condition (\ref{abcft}) does not have solutions in general if we ignore the back-reaction
by the surface $Q$ and thus we need to solve again the Einstein equation with the boundary condition (\ref{abcft}) as explicitly done in \cite{NTU} in a concrete example. However, in our three dimensional pure gravity, we know that all solutions to Einstein equation with a negative cosmological constant is locally given by a pure AdS$_3$. Indeed we can employ the coordinate transformation (\ref{mapz}) to map a generic solution into the Poincare AdS$_3$.

For our problem, let us perform the conformal map (\ref{scmap}), sketched in Fig.\ref{fig:onecutmap}.
This transformation maps a plane with the cut along $P$ into an upper half-plane. Therefore we can easily identify a gravity dual of the latter i.e. a BCFT defined on an upper half-plane using the known result in \cite{AdSBCFT} (refer to Fig.\ref{fig:adsmap}). This gravity solution is given by a half of Poincare AdS$_3$ defined by the metric (\ref{poi}) with the restriction Im$\xi>0$. The function $L(w)$ in this metric
is found to be $L(w)=-\f{3q^2}{8 (q^2/2-w^2)^2}$.

\subsection{Computation of Holographic Entanglement Entropy: A Single Interval} \label{sipr}

Now we define the subsystem $A$ to be the interval $q\leq x\leq q+l$ and compute the HEE $S_A$.
The conformal map (\ref{scmap}) maps the CFT geometry into an upper half plane. The two end points of projected region $P$: $(x,y)=(-q,0)$ and $(x,y)=(q,0)$, are mapped into $\xi=0$ and $\xi = \infty$, respectively. The end point of $A$ given by $(x,y)=(q+l,0)$ is mapped to $\xi = i\s{\f{2q+l}{l}}$.

The gravity dual is obtained by performing the extension (\ref{mapz}) of the specific map (\ref{scmap}) to the gravity dual of upper half plane with a BCFT boundary condition at the boundary. The CFT on upper half place is dual to a half of the Poincare AdS i.e. the space defined by the metric (\ref{poi}) restricted to $\mbox{Im} (\xi)>0$. As follows from a standard calculation \cite{RT,AdSBCFT}, the holographic entanglement entropy for the interval $A$ can be found from the length of geodesic in the Poincare AdS as follows\footnote{As same as the CFT part (in Fig.\ref{fig:onecutmap}), we set $\xi = (\xi_1 + i \xi_2)/\sqrt{2}$. In this convention, the end point of $A$ is mapped to $(\xi_1, \xi_2) = (0, \sqrt{2 (2q+l)/l})$.}:
\be
S_A=\f{R}{4G_N}\log \f{2\s{\f{2(2q+l)}{l}}}{\ep},
\ee
where $\ep$ is the UV cut off at $\eta=z$ in the metric (\ref{poi}). We can relate $\ep$ to the UV cut off of the original space (\ref{metm}) by the map (\ref{mapz})
\be
\f{\ep}{a}=|f'(w)|=\f{\s{2}q}{l^{3/2}(2q+l)^{1/2}}.
\ee
This leads to the final expression:
\be
S_A=\f{c}{6}\log\f{2(l+2q)l}{aq}, \label{holdger}
\ee
which agrees with the previous CFT result (\ref{cftpr}) up to the finite constant contribution $\gamma_b$. The finite constant, which is essentially the boundary entropy, is related to the tension $T$ in (\ref{abcft}). Note that we set $T=0$ in our holographic computations (\ref{holdger}).

Note that (\ref{holdger}) is a monotonically decreasing function of $q$. This agrees with our expectation that the local projection reduces quantum entanglement and this effect gets larger when the size of projected region $P$ is large. The smallest size limit of the projected region $P$ corresponds to $q\sim a$. Then we reproduce the familiar result $S_P=\f{c}{3}\log\f{l}{a}$ \cite{HLW}.\\

Now we calculate $S_A$ when $A$ is a general interval defined by $q+l_1\leq x\leq  q+l_2$.
We expect some phase transition depending on the value of $l_1$ and $l_2$.

\begin{figure}
\begin{minipage}{0.3\hsize}
  \centering
  \includegraphics[width=3.5cm]{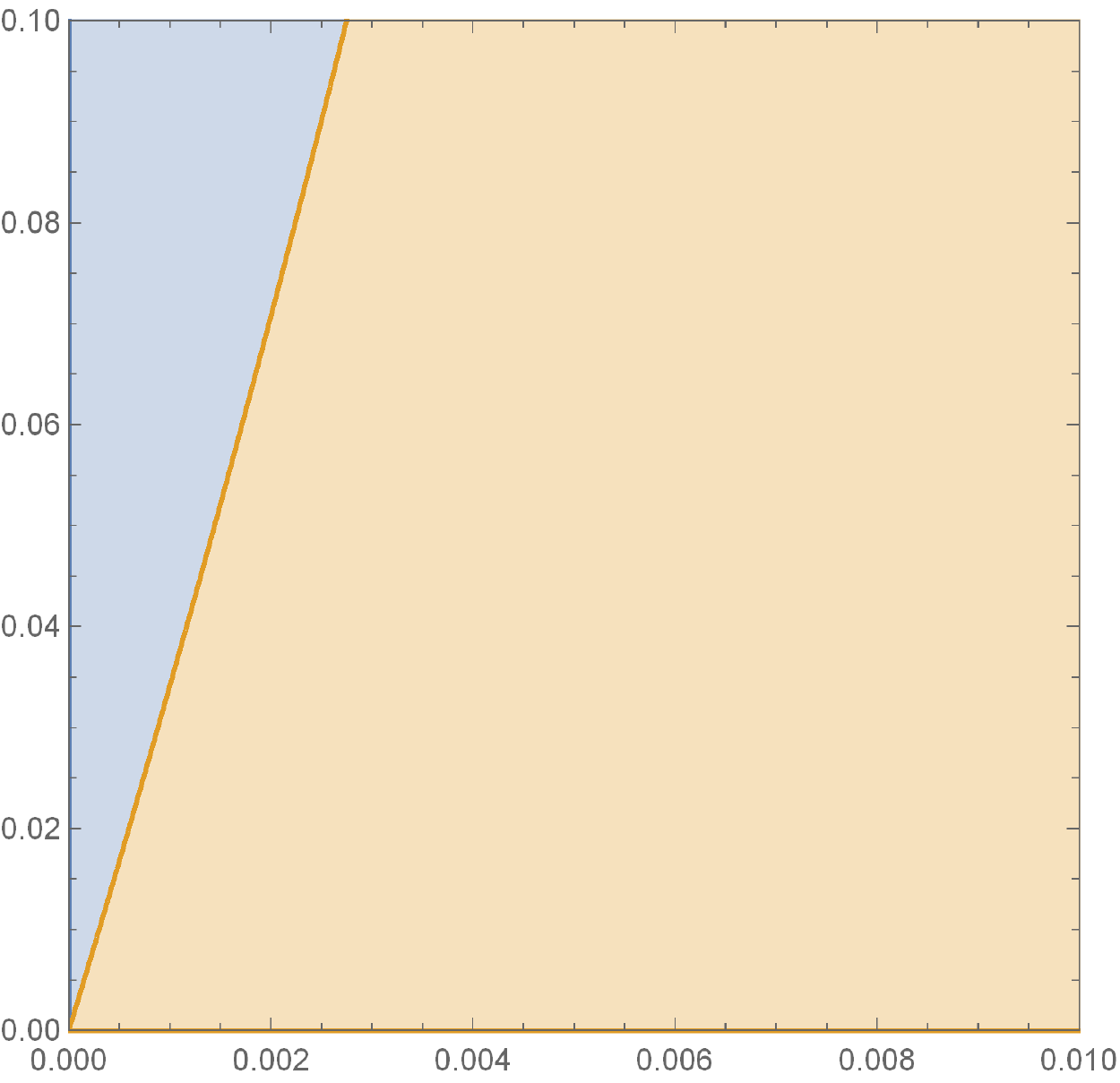}
\end{minipage}
\begin{minipage}{0.3\hsize}
  \centering
  \includegraphics[width=3.5cm]{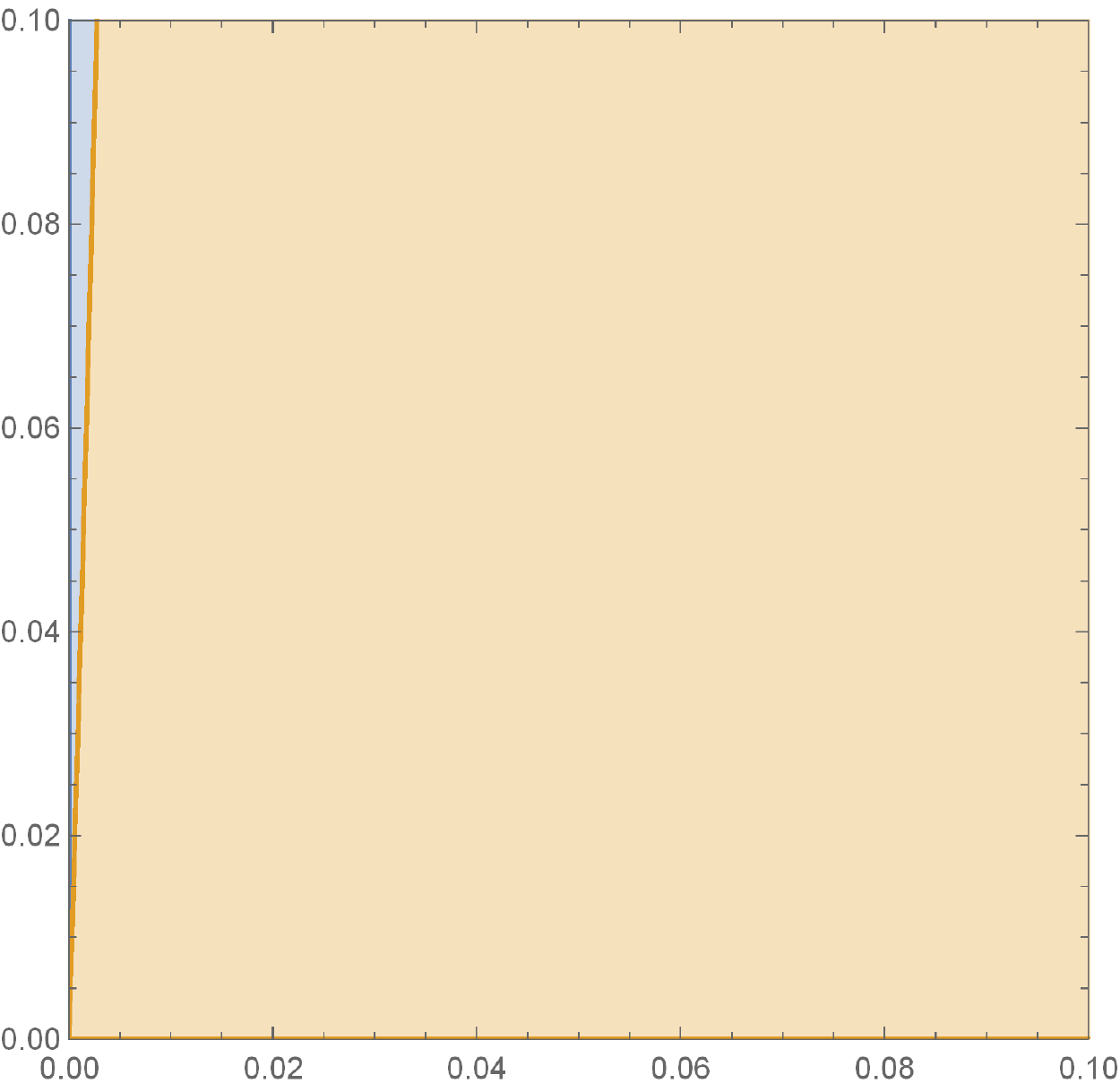}
\end{minipage}
\begin{minipage}{0.3\hsize}
  \centering
  \includegraphics[width=3.5cm]{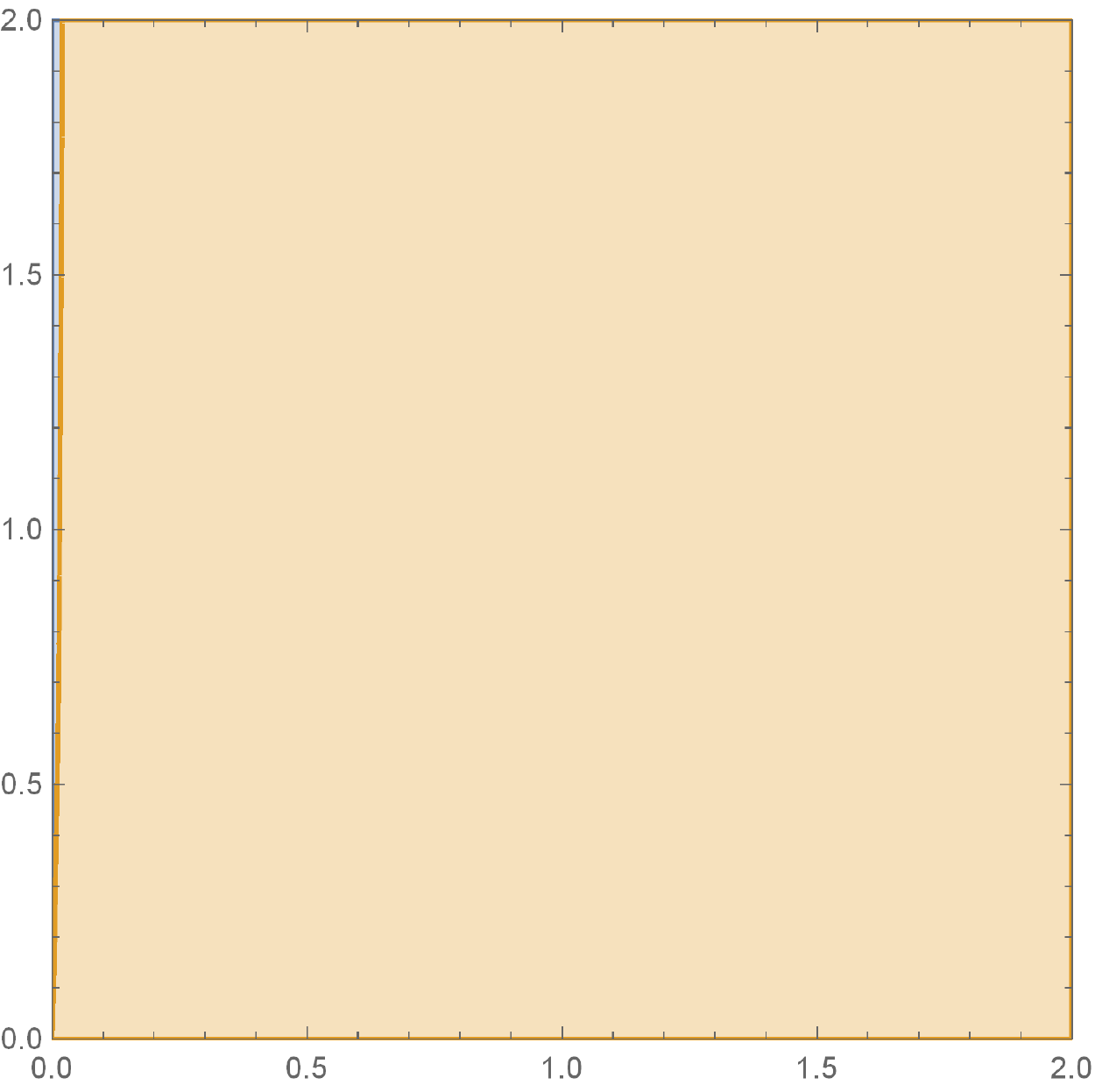}
\end{minipage}
  \caption{The phase diagrams of the HEE $S_A$. The horizontal and vertical coordinate correspond to $\f{l_1}{2q}$ and $\f{l_2-l_1}{2q}$ (zooming out from the left picture to the right picture).
In the (tiny) blue region, $S_A (1) > S_A (2)$ and thus the disconnected solution is chosen. In the (huge) orange region, the situation is opposite.}
\label{fig:1intHEEdia}
\end{figure}

\begin{figure}
  \centering
  \includegraphics[width=7cm]{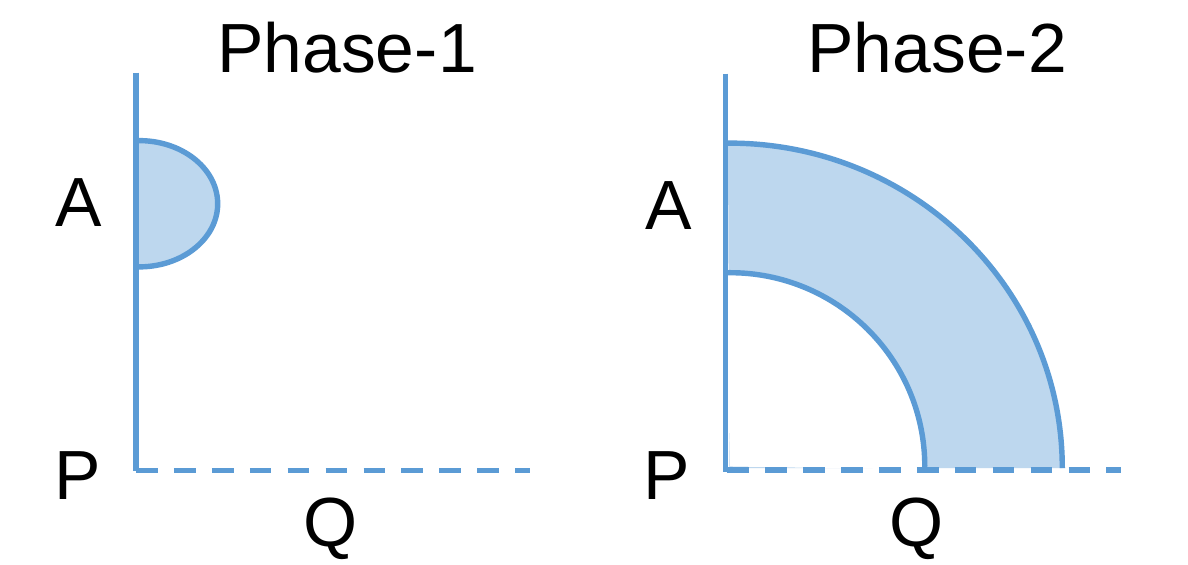}
  \hspace{1cm}
  \includegraphics[width=6cm]{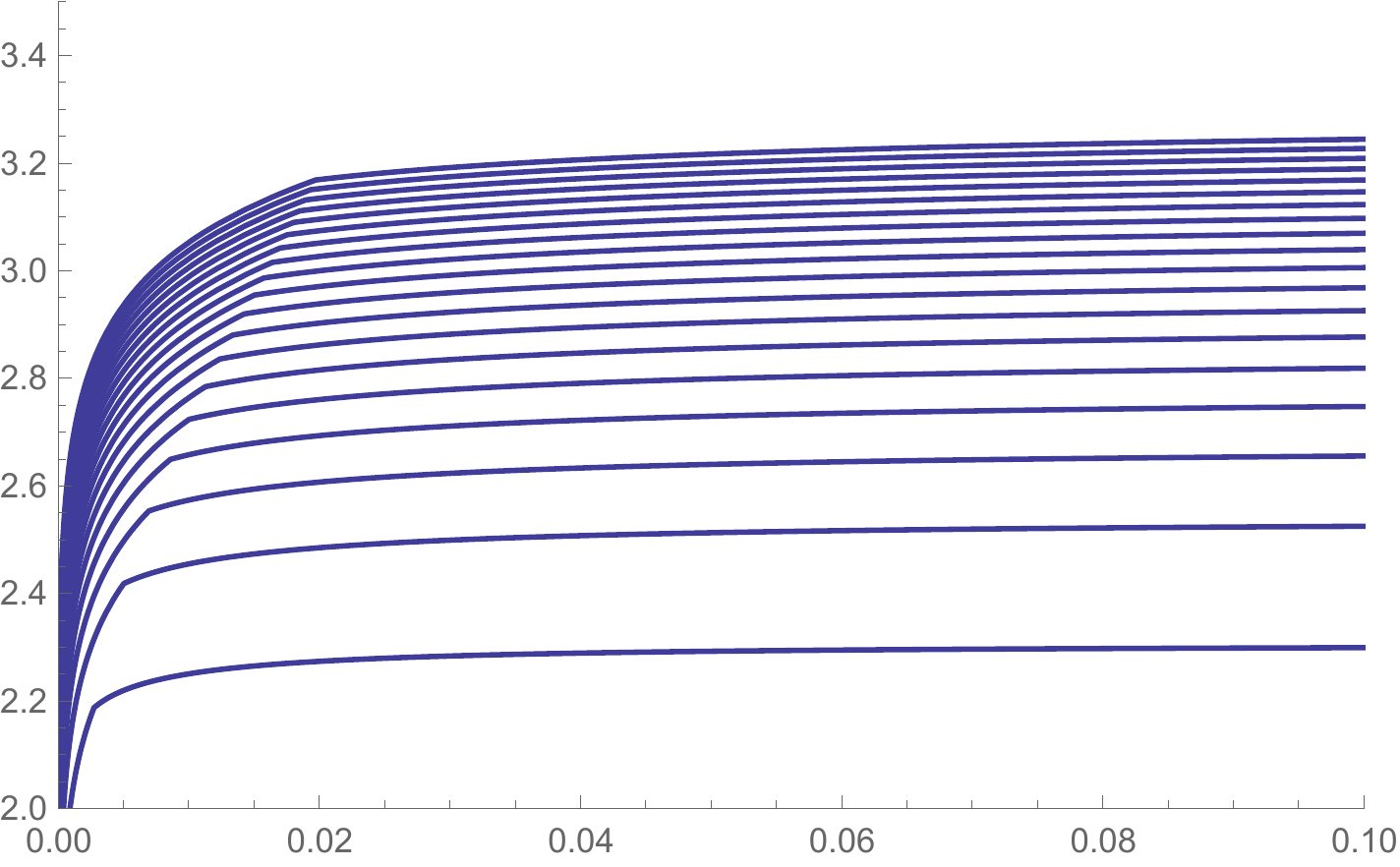}
  \caption{The left picture sketches two phases of the geodesics responsible to the holographic entanglement entropy. The right plot shows the HEE $S_A$ as the function of $\f{l_1}{2q}$ with the different values of $\f{l_2-l_1}{2q}$.
$S_A$ has a kink when the configurations of the minimal surfaces change. (In this plot, we set the UV cutoff $\f{a}{2q} = 0.0001$.)}
  \label{fig:1int_HEE2D}
\end{figure}

In fact, $S_A$ has two phases in the $\f{l_1}{2q}$ -- $\f{l_2-l_2}{2q}$ plane (Figure \ref{fig:1intHEEdia}) :
the phase where the minimal surface is connected between the edges of the interval $A$ (Phase-1)
and the phase where the minimal surface consists of two disconnected geodesics which end on the bulk boundary Im$\xi=0$ (Phase-2). Refer to the left picture of Fig.\ref{fig:1int_HEE2D}.

We find the following expressions of $S_A$:
\begin{align}
S_A (1)
&= \f{c}{6} \log \f{ 2\left( \s{ \f{2q+l_1}{l_1}} - \s{ \f{2q+l_2}{l_2}}  \right)^2}{ \ep_1 \ep_2 } \notag \\
&= \f{c}{6} \log \left[\f{2(2q+l_1)l_1}{a q} \f{2(2q+l_2)l_2}{a q} \f{ \left( \s{ \f{2q+l_1}{l_1}} - \s{ \f{2q+l_2}{l_2}}  \right)^2 }{ 4 \s{\f{2q+l_1}{l_1}} \s{\f{2q+l_2}{l_2}} }\right], \label{saiprw}
\end{align}
in the Phase-1 and
\be
S_A (2)
= \f{c}{6} \log \f{ 2 \s{ \f{2(2q+l_1)}{l_1}} }{ \ep_1 } \f{ 2 \s{ \f{2(2q+l_2)}{l_2}} }{ \ep_2 }
= \f{c}{6}\log \left[\f{2(2q+l_1)l_1}{a q} \f{2(2q+l_2)l_2}{a q}\right],  \label{saiprww}
\ee
in the Phase-2.  According to the basic rule of holographic entanglement entropy \cite{RT,HRT}, we always choose the smaller value from $S_A(1)$ and $S_A(2)$. This leads to
\begin{align}
S_A
= & \ \mathrm{min} [S_A (1) , S_A (2)] \notag \\
= &
\renewcommand{\arraystretch}{1.8}
\left\{
\begin{array}{l}
S_A (1) \ \ \
\left( \f{ \f{l_1}{2q} \left(1+\f{l_1}{2q} \right) }{ \alpha -\f{l_1}{2q} } > \f{l_2-l_1}{2q} >0 ,\  \alpha > \f{l_1}{2q} > 0 \right) \ , \
\left( \f{l_2-l_1}{2q} > 0 , \f{l_1}{2q} \geq \alpha > 0  \right) \\
S_A (2) \ \ \
\left( \f{l_2-l_1}{2q} > \f{ \f{l_1}{2q} \left(1+\f{l_1}{2q} \right) }{ \alpha - \f{l_1}{2q} } >0 ,\  \alpha > \f{l_1}{2q} > 0 \right)
\end{array}
\right.
\renewcommand{\arraystretch}{1}
\end{align}
where $\alpha$ is a positive constant: $\alpha = \f{- 4+3 \sqrt{2}}{8}$.
The phase diagram of $S_A$ in the $\f{l_1}{2q}$ -- $\f{l_2-l_2}{2q}$ plane are plotted
in Figure \ref{fig:1intHEEdia}.
As mentioned before, $S_A$ has two phases and
at the transition point it has a kink as in the right graph of Fig.\ref{fig:1int_HEE2D}. Finally we would like to mention that in our holographic computation, we assumed $T=0$ in (\ref{abcft}). If we choose other values of $T$, the phase boundary should change.

We can generalize our analysis to multi-intervals. Refer to appendix C for the analysis in the case of two symmetric intervals.

\subsection{Towards New Multi-Partite Entanglement Measure}

The entanglement entropy $S_A$ measures the amount of bipartite (=two body) quantum entanglememt between
$A$ and $B$ when the combined system $AB$ is a pure state. As a next step, it is very intriguing to
explore a measure of tri-partite (=three body) quantum entanglement between $A$, $B$ and $C$ assuming that
the system $ABC$ is pure. For recent discussions of tri-partite entanglement in the light of holography refer to \cite{Ba,EN,Sk,Skk}.

Note that the tripartite mutual information
defined by $I(A:B:C)=S_A+S_B+S_C-S_{AB}-S_{BC}-S_{CA}+S_{ABC}$ \cite{HHM} is simply vanishing in this setup,
because $S_{ABC}=0$, $S_A=S_{BC}$, etc. Since the number of independent values of entanglement entropy, which is three, coincides with that of two body entanglement in this system, it is clear that we cannot estimate any tripartite entanglement from them. However, this may change
if we take into account projection measurements.\footnote{Projection measurements are employed to define
a quantity called quantum discord \cite{QD}, which is considered to be a measure of two body entanglement even in mixed states.} For example, let us introduce the following new quantity:
\be
\delta_{A}^{B}=2(S_A-S^{\Pi_B}_A)-I(A:B),  \label{ww}
\ee
where $I(A:B)=S_{A}+S_B-S_{AB}$ is the mutual information. The quantity $S^{\Pi_B}_A$ denotes the
entanglement entropy for $A$ when we perform a projection measurement of $B$. Especially it is natural to take the minimum value when we allow any projection to any state in B. We would like to argue that $\delta_{A}^{B}$ can probe tripartite (or more generally multi-partite) entanglement.

First assume there is only bipartite entanglement in our three body system $ABC$.
We write the amount of the entanglement between $A$ and $B$ as $e_{AB}$ etc. In this case we obtain
$I(A:B)=2e_{AB}$. $S^{\Pi_B}_A$ becomes minimum when the projection $\Pi_B$ removes all entanglement between
$A$ and $B$. Thus we find $S^{\Pi_B}_A=e_{AC}$. In this way we obtain
\be
\delta_{A}^{B}=0,
\ee
for two body entanglement.

On the other hand, if we consider a GHZ state $\f{1}{\s{2}}\left(|000\lb+|111\lb\right)$ for three qubit system,
which is known as a state with maximum tripartite entanglement, we find the non-trivial result:
\be
\delta_{A}^{B}=\log 2.
\ee
These suggest that $\delta_{A}^{B}$ can probe the tripartite entanglement. However notice that this quantity is not always positive definite as we will see below and is not a standard measure.

Now let us estimate this quantity $\delta_{A}^{B}$ when $A$ and $B$ are finite size intervals in a two dimensional CFT by using the results in section \ref{sipr} (we replace the interval $P$ with $B$). It is natural to choose the projection $\Pi_B$ to be a local projection measurement. $S_{AB}$ can be computed as the entanglement entropy for two intervals as follows:
\begin{align}
S_{AB}
= & \mathrm{min} [ S_{AB} (1) , S_{AB} (2) ] \notag \\
= &
\renewcommand{\arraystretch}{1.8}
\left\{
\begin{array}{l}
S_{AB} (1) \ \ \
\left( \f{ \f{l_1}{2q} \left(1+\f{l_1}{2q} \right) }{ 1-\f{l_1}{2q} } > \f{l_2-l_1}{2q} >0 ,\  1 > \f{l_1}{2q} > 0 \right) \ , \
\left( \f{l_2 - l_1}{2q} > 0,  \f{l_1}{2q} \geq 1  \right) \\
 S_{AB} (2) \ \ \
\left( \f{l_2-l_1}{2q} >\f{ \f{l_1}{2q} \left(1+\f{l_1}{2q} \right) }{ 1-\f{l_1}{2q} } > 0 ,\  1 > \f{l_1}{2q} > 0 \right)
\end{array}
\right.
\renewcommand{\arraystretch}{1}
\end{align}
where
\begin{align}
S_{AB} (1)
&= \f{c}{3} \log \f{2q (l_2-l_1)}{a^2}, \label{wqere1} \\
S_{AB} (2)
&= \f{c}{3} \log \f{l_1 (2q+l_2)}{a^2}. \label{wqere2}
\end{align}
Note that $S_{AB} (1)$ corresponds to the disconnected phase and $S_{AB} (2)$ to the connected phase.

Now our quantity $\delta_{A}^{B}$ (\ref{ww}) is expressed as follows:
\be
\delta_{A}^{B}
= S_{AB} (1) - 2 S_B +\mathrm{min}[S_{AB} (1) , S_{AB} (2)] -2 \cdot \mathrm{min}[S_{A}  (1) , S_{A} (2)] ,
\ee
where $S_B$ is given by the familiar formula \cite{HLW}
\begin{align}
S_{B}
&= \f{c}{3} \log \f{2q}{a} .
\end{align}
Also $S_{AB}(1,2)$ and $S_A(1,2)$ are defined in (\ref{wqere1}), (\ref{wqere2}), (\ref{saiprw}) and (\ref{saiprww}).

\begin{figure}
\begin{minipage}{0.3\hsize}
  \centering
  \includegraphics[width=5cm]{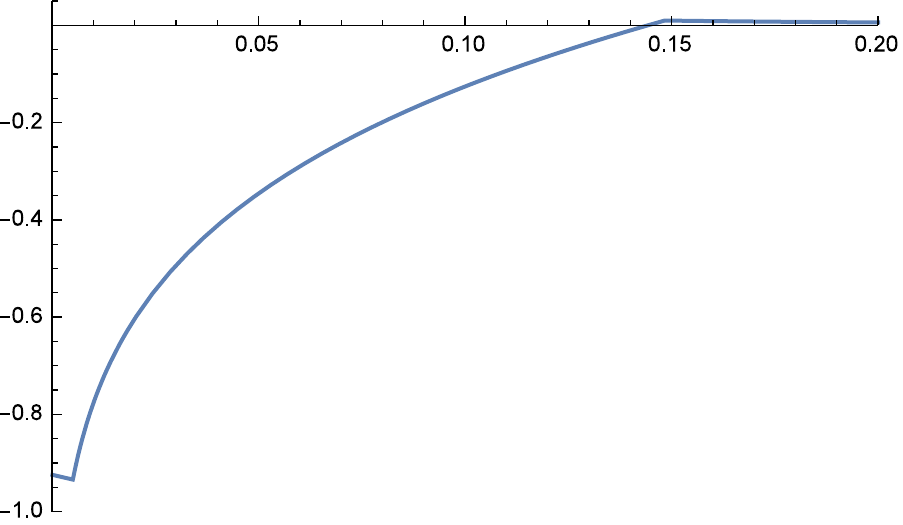}
\end{minipage}
\begin{minipage}{0.3\hsize}
  \centering
  \includegraphics[width=5cm]{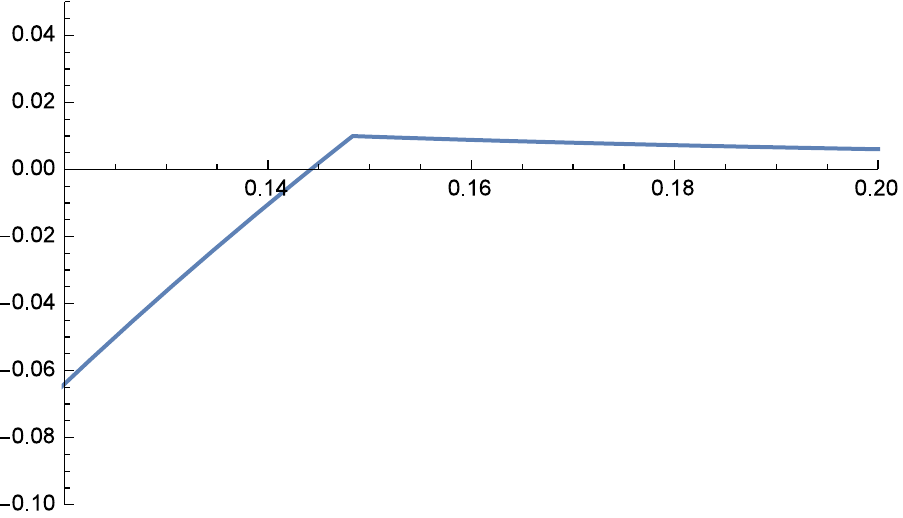}
\end{minipage}
\begin{minipage}{0.3\hsize}
  \centering
  \includegraphics[width=5cm]{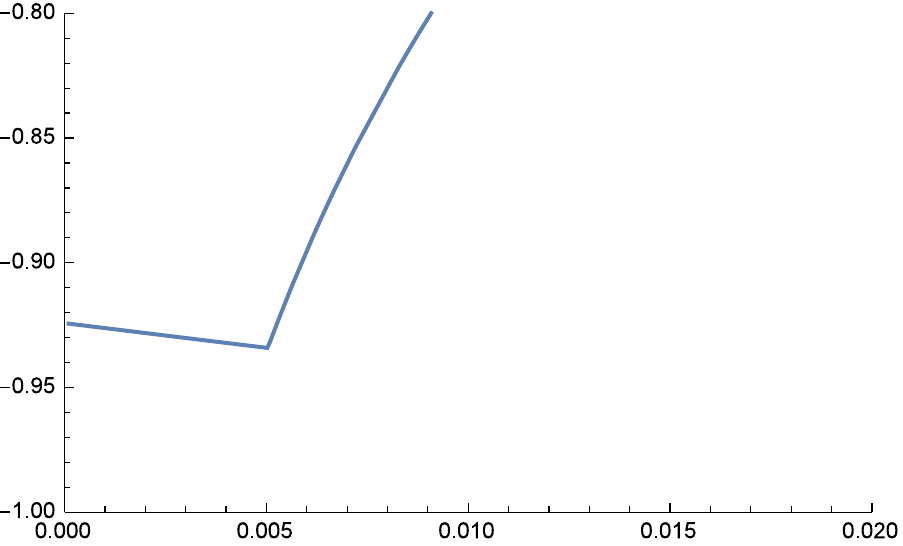}
\end{minipage}
  \caption{The plot of  $\delta_{B}^{A}$ as the function of $\f{l_1}{2q}$ ( $\f{l_2-l_1}{2q} = 0.2,  c=1$). In general, $\delta_{B}^{A}$ has 2 kinks, decreases slightly from the finite negative value $\sim -\f{4}{3} \log 2 \sim - 0.924$, then increases to the small positive value ($\sim 0.010$) and finally approaches to $0$.}
  \label{fig:1int_deltaAB2D2}
\end{figure}

More explicitly, we have
\begin{align}
\delta_{B}^{A}& =   \notag \\
&\left\{
\renewcommand{\arraystretch}{1.8}
\begin{array}{l}
S_{AB} (1) -2 S_B + S_{AB} (2) -2 S_{A} (2)
= -\f{c}{3} \log \f{16 (2q+l_1) l_2 }{2q( l_2 - l_1) }  \\
\ \ \ \ \ \ \ \ \ \ \left( 0<\f{l_1}{2q}<\alpha, \  \f{l_2-l_1}{2q} >  \f{ \f{l_1}{2q} \left(1+\f{l_1}{2q} \right) }{\alpha - \f{l_1}{2q}}   \right) \\
S_{AB} (1) -2 S_B + S_{AB} (2) -2 S_{A} (1)
= -\f{c}{3} \log \f{16(2q+l_1) l_2}{2q  (l_2-l_1)} \f{ \left( \s{ \f{2q+l_1}{l_1}} - \s{ \f{2q+l_2}{l_2}}  \right)^2 }{ 4 \s{\f{2q+l_1}{l_1}} \s{\f{2q+l_2}{l_2}} }  \\
\ \ \ \ \ \ \ \ \ \ \left( 0<\f{l_1}{2q}<\alpha , \  \f{\f{l_1}{2q} \left(1+\f{l_1}{2q} \right)}{1-\f{l_1}{2q}} < \f{l_2-l_1}{2q} < \f{ \f{l_1}{2q} \left(1+\f{l_1}{2q} \right) }{\alpha - \f{l_1}{2q}}  \right) ,\ \left(\alpha \leq \f{l_1}{2q}<1 \ , \  \f{\f{l_1}{2q} \left(1+\f{l_1}{2q} \right)}{1-\f{l_1}{2q}} < \f{l_2-l_1}{2q} \right) \\
2 (S_{AB} (1)- S_B - S_{A} (1) )
= -\f{c}{3} \log \f{16(2q+l_1) l_1 (2q+l_2) l_2}{ (2q (l_2-l_1))^2} \f{ \left( \s{ \f{2q+l_1}{l_1}} - \s{ \f{2q+l_2}{l_2}}  \right)^2 }{ 4 \s{\f{2q+l_1}{l_1}} \s{\f{2q+l_2}{l_2}} }  \\
\ \ \ \ \ \ \ \ \ \ \left( 0 < \f{l_1}{2q}<1, \ \f{l_2-l_1}{2q} < \f{ \f{l_1}{2q} \left(1+\f{l_1}{2q} \right) }{1-\f{l_1}{2q}} \right) \ ,\ \left(  1 \leq \f{l_1}{2q} , 0 < \f{l_2-l_1}{2q} \right)
\end{array}
\renewcommand{\arraystretch}{1}
\right. \label{eq: deltaABexplicit}
\end{align}
This behavior is plotted in Fig.\ref{fig:1int_deltaAB2D2}.

We find that  $\delta_{B}^{A}$ approaches to 0
 in the large separation limit ($\f{l_1}{2q} \to \infty$ with fixed $\f{l_2 - l_1}{2q}$).
\be
\delta_{B}^{A}  \to \f{c}{48} \left(\f{l_2-l_1}{2q} \right)^2 \left(\f{2q}{l_1} \right)^4 +
\cdots \sim 0 \ \ \ \ \left(\f{l_1}{2q} \to +\infty \right).\label{wcerdsqq}
\ee
On the other hands, $\delta_{B}^{A}$ approaches to an finite value
in the small separation limit ($\f{l_1}{2q} \to +0$ with fixed $\f{l_2 - l_1}{2q}$).
\be
\delta_{B}^{A}  \to -\f{4 c}{3} \log 2 \ \ \ \ \left(\f{l_1}{2q} \to +0 \ \ \mathrm{with \ fixed} \ \ \f{l_2 - l_1}{2q} \right). \label{wcerdsq}
\ee
In contrast, the mutual information $I (A : B)$, which measures two body entanglement,  diverges in the small separation limit.\footnote{
This difference in the small separation limit seems to be true only for two dimensional QFTs which have one spatial direction.
In two dimension, only two subsystems can be attached together at a same point.
This suggests that the amount of tripartite entanglement (or multi-partite entanglement) is finite in two dimension.
However, in higher dimension, it can diverge in the small separation limit
because any number of subsystems can be attached together at a same point.}

Note that in our holographic computations we always set the tension parameter $T$ in (\ref{abcft}) to zero.
However, since we are interested in a local projection which minimize $S^{\Pi_B}_A$ we need to take a smallest possible value of $T$, which is related to the smallest value of boundary entropy as found in \cite{AdSBCFT} and thus depends on the details of holographic CFT. This gives a positive constant shift to $\delta_{B}^{A}$  if the phase includes any disconnected geodesic (which ends on $Q$). This leads to a positive contribution to (\ref{wcerdsq}) but does not change (\ref{wcerdsqq}). In this way we find that when $A$ and $B$ are closed to each other, the small $\f{l_1}{2q}$ behaviour of $\delta_{B}^{A}$ (\ref{eq: deltaABexplicit}) can detect tripartite entanglement.

\subsection{Local Projection Measurements at Finite Temperature}

Here we would like to study quantum entanglement when we perform a local projection measurement
in a 2d CFT at finite temperature. We apply our holographic method to compute the holographic entanglement
entropy. We can describe a CFT at finite temperature as a pure state (called thermo field double)
$|TFD\lb$ in doubled CFTs (called CFT$_1$ and CFT$_2$):
\be
|TFD\lb=\sum_{n}e^{-\f{\beta_H}{2}E_n}|n\lb_1|n\lb_2, \label{tfdsw}
\ee
where $|n\lb_{1,2}$ denote complete energy eigenstates $H|n\lb=E_n|n\lb$ in CFT$_1$ and CFT$_2$.
The parameter $\beta_H$ is the inverse temperature and is the same as that of the dual BTZ black hole
\cite{Mal}. If we trace out CFT$_2$, we get the density matrix at finite temperature for CFT$_1$.

In the Euclidean path-integral description, we take the complex coordinate $(w,\bar{w})$ with
$w=(x+i\tau)/\s{2}$, $-\infty < x <\infty$ and $\tau\sim \tau+i\beta_H$. We take the projected region to be
$\tau=0$ and $-q\leq x\leq q$.
Note that in this description, the CFT$_1$ and CFT$_2$ of the state (\ref{tfdsw}) live at $\tau=0$ and
$\tau=i\f{\beta_H}{2}$, respectively.

When we define the two end points of the subsystem $A$ to be $w=w_1$ and $w=w_2$,
the entanglement entropy is given by
\begin{equation}
\begin{split}
S_A=-\lim_{n\to1}\frac{\partial}{\partial n}
\ln \langle \sigma_{n} (w_1,\bar{w}_1) \bar{\sigma}_n (w_2, \bar{w}_2)
\rangle
\end{split}  \label{ee TFD1}
\end{equation}
where $\sigma_n$ is the twist operator with conformal weight $\frac{cn}{24}(1-\frac{1}{n^2})$ and $c$ is the central charge. We apply the conformal map,
\begin{equation}
\begin{split}
\xi(w)=\sqrt{\dfrac{\sinh \frac{\pi}{\beta_H}(q/\s{2}+w)}{\sinh \frac{\pi}{\beta_H}(q/\s{2}-w)}}.
\end{split}  \label{map TFD}
\end{equation}
This maps the cylinder with one slit P into an upper half plane.
Finally the holographic entanglement entropy becomes
\begin{equation}
\begin{split}
S_A=\frac{c}{6} \mbox{min} \left[ \ln \frac{(|\xi_1|-|\xi_2|)^2}{\epsilon_1 \epsilon_2}, \ln \frac{4|\xi_1| |\xi_2|}{\epsilon_1 \epsilon_2} \right]
\end{split}  \label{HEE TFD1}
\end{equation}
where we assumed $\xi_1$ and $\xi_2$ are pure imaginary because
$\tau=0$ and $\tau=\frac{i\beta_H}{2}$ are mapped into the imaginary axis in the $\xi$ plane.
$\frac{c}{6} \ln \frac{(|\xi_1|-|\xi_2|)^2}{a^2}$ and $\frac{c}{6} \ln \frac{4|\xi_1| |\xi_2|}{a^2}$ correspond to the connected geodesics and the disconnected geodesics.
From (\ref{map TFD}) and (\ref{HEE TFD1}), we obtain the entanglement entropy as
\begin{equation}
\begin{split}
S_A=&\frac{c}{6} \left[ \ln \frac{2\beta_H |\xi_1|}{\pi} \dfrac{\left|\sinh \frac{\pi}{\beta_H}(q-w_1)\right|^2 }{\sinh \frac{2q\pi}{\beta_H}}
+\ln \frac{2\beta_H |\xi_2|}{\pi} \dfrac{\left|\sinh \frac{\pi}{\beta_H}(q-w_2)\right|^2 }{\sinh \frac{2q\pi}{\beta_H}} \right.  \\
&\left. + \mbox{min} \left[ \ln \frac{(|\xi_1|-|\xi_2|)^2}{a^2}, \ln \frac{4|\xi_1| |\xi_2|}{a^2} \right]  \right].
\end{split}  \label{HEE TFD2}
\end{equation}

\subsubsection{A Single Interval in CFT$_1$ ($\tau=0$)}

\begin{figure}
 \centering
 \includegraphics[width=6.5cm, clip]{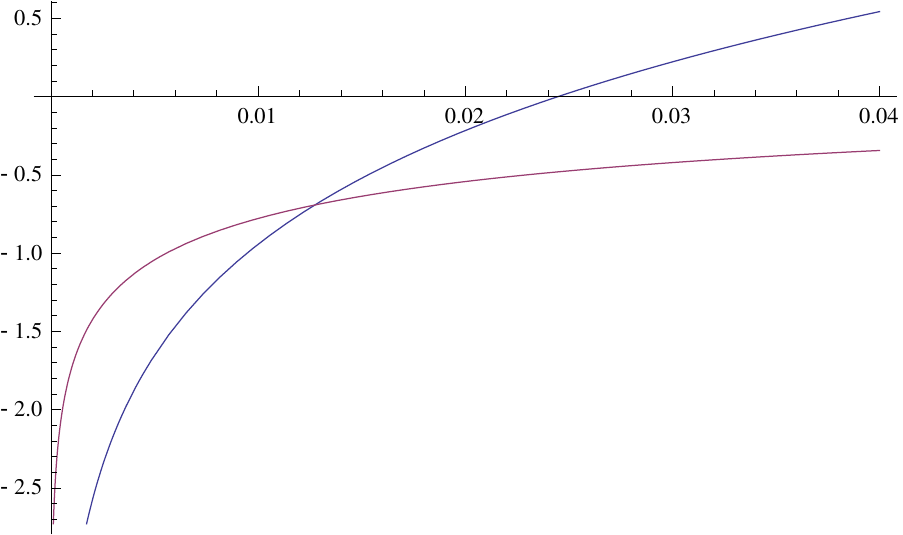}%
 \caption{We plotted $\Delta S_A$ as a function of $\frac{\pi l_1}{\beta_H}$, where $l_2=l_1+l$.
 We normalized the total value by choosing the central charge $c=6$ and assume the length parameters to be $\frac{\pi q}{\beta_H}=\frac{\pi l}{\beta_H}=1$.
 The blue curve corresponds to the HEE of disconnected geodesics and the red one corresponds to the HEE of connected geodesics.  There is a phase transition.}
 \label{eeTFDtau0}
 \end{figure}

First we take the two end points of the subsystem A to be in CFT$_1$: $w_1=q+l_1$ and $w_2=q+l_2$,  where $l_2>l_1$.
From (\ref{HEE TFD2}), we obtain
\begin{equation}
\begin{split}
S_A=&\frac{c}{6} \left[ \ln \frac{2\beta_H |\xi_1|}{\pi} \dfrac{\left|\sinh \frac{\pi}{\beta_H}l_1\right|^2 }{\sinh \frac{2q\pi}{\beta_H}}
+\ln \frac{2\beta_H |\xi_2|}{\pi} \dfrac{\left|\sinh \frac{\pi}{\beta_H}l_2\right|^2 }{\sinh \frac{2q\pi}{\beta_H}} \right.  \\
&\left. + \mbox{min} \left[ \ln \frac{(|\xi_1|-|\xi_2|)^2}{a^2}, \ln \frac{4|\xi_1| |\xi_2|}{a^2} \right]  \right],
\end{split}  \label{HEE TFD tau0}
\end{equation}
where
\begin{equation}
\begin{split}
|\xi_{1,2}|=\sqrt{\dfrac{\sinh \frac{\pi}{\beta_H}(2q+l_{1,2})}{\sinh \frac{\pi}{\beta_H}l_{1,2}}}.
\end{split}  \label{map TFD tau0}
\end{equation}
When $q/\beta_H \to 0$, $S_A$ becomes the entanglement entropy at finite temperature $\beta_H^{-1}$ \cite{CC},
\begin{equation}
\begin{split}
S_A^{(0)} \equiv \lim_{q/\beta_H \to 0}S_A=\frac{c}{3} \ln \left( \frac{\beta_H}{\pi a} \sinh \frac{\pi}{\beta_H} (l_2-l_1) \right) .
\end{split}  \label{HEE TFD tau0 q0}
\end{equation}
We plot $\Delta S_A \equiv S_A-S_A^0$ in Fig.\ref{eeTFDtau0}. We observe that the entanglement entropy is reduced as the interval $A$ gets closer to $P$ as expected.

\subsubsection{A Single Interval in CFT$_2$ ($\tau=\frac{i\beta_H}{2}$)}
\begin{figure}
 \centering
 \includegraphics[width=6.5cm, clip]{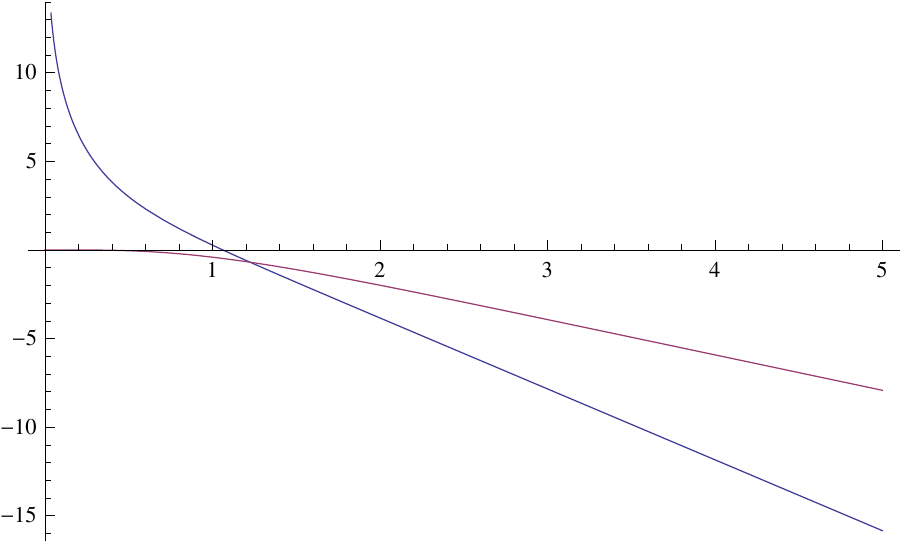}%
 \hspace{1cm}
 \includegraphics[width=6.5cm, clip]{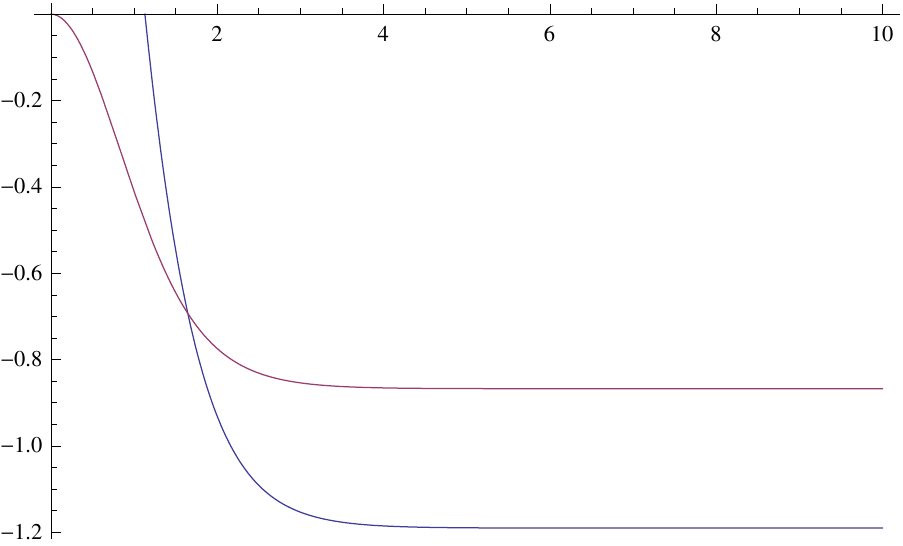}
 \caption{In the left figure, we plotted $\Delta S_A$ as a function of $\frac{\pi q}{\beta_H}$, for $x_1=-q,~ x_2=q$. $\Delta S_A$ is a linearly decreasing function of $\frac{\pi q}{\beta_H}$ when $\frac{\pi q}{\beta_H}$ is large.  In the right figure, we plotted $\Delta S_A$ as a function of $\frac{\pi x}{\beta_H}$ where $x_2=-x_1=x$ for $\frac{\pi q}{\beta_H}=1$. In this case, $\Delta S_A$ approaches a constant value when $\frac{\pi x}{\beta_H}$ is large.
 We normalized the total value by choosing the central charge $c=6$.
 The blue curves correspond to the HEE of disconnected geodesics and the red ones correspond to the HEE of connected geodesics.  There are phase transitions.}
 \label{eeTFDtaui}
 \end{figure}

 Next we take the two end points of the subsystem A to be in CFT$_2$: $w_1=x_1+\frac{i\beta_H}{2}$ and $w_2=x_2+\frac{i\beta_H}{2}$, where $x_1<x_2$.
From (\ref{HEE TFD2}), we obtain
\begin{equation}
\begin{split}
S_A=&\frac{c}{6} \left[ \ln \frac{2\beta_H |\xi_1|}{\pi} \dfrac{\left|\cosh \frac{\pi}{\beta_H}(q-x_1)\right|^2 }{\sinh \frac{2q\pi}{\beta_H}}
+\ln \frac{2\beta_H |\xi_2|}{\pi} \dfrac{\left|\cosh \frac{\pi}{\beta_H}(q-x_2)\right|^2 }{\sinh \frac{2q\pi}{\beta_H}} \right.  \\
&\left. + \mbox{min} \left[ \ln \frac{(|\xi_1|-|\xi_2|)^2}{a^2}, \ln \frac{4|\xi_1| |\xi_2|}{a^2} \right]  \right],
\end{split}  \label{HEE TFD taui}
\end{equation}
where
\begin{equation}
\begin{split}
|\xi_{1,2}|=\sqrt{\dfrac{\cosh \frac{\pi}{\beta_H}(q+x_{1,2})}{\cosh \frac{\pi}{\beta_H}(q-x_{1,2})}}.
\end{split}  \label{map TFD taui}
\end{equation}
We find $\lim_{q\to 0} S_A=S_A^0=\frac{c}{3} \ln \left( \frac{\beta_H}{\pi a} \sinh \frac{\pi}{\beta_H} (x_2-x_1) \right)$.
We plot $\Delta S_A \equiv S_A-S_A^0$ in Fig.\ref{eeTFDtaui}. We can observe that the entanglement between
CFT$_1$ and CFT$_2$ is locally reduced by the projection on the region $P$ in CFT$_1$. This amount of reduction can be estimated by the saturated value in the right graph of Fig.\ref{eeTFDtaui}.

\section{Evolution of Holographic Entanglement Entropy after Local Projections and Partial Entangling}

In this section, we introduce the UV cut off i.e. the parameter $p$ in (\ref{totop}) in the holographic
description and analyze time evolutions of gravity duals of both (i) partial projection measurement (introduced in section \ref{sec:Proj}) and (ii) partial entangling of two identical CFTs (introduced in section \ref{sec:PEnt}).\footnote{We can also treat the partial swapping of two CFTs introduced in section 3.2 as
it is again conformally equivalent to a torus, though we will not study this in detail.}

\subsection{Holographic Description with UV Cut Off}

Let us start with a gravity dual of (i) partial projection measurement. In the Euclidean path-integral formalism, we can introduce the UV cut off $p$ by considering the two slit geometry (see Fig.\ref{fig:twocutmap}). This can be conformally mapped into a cylinder or annulus by the map (\ref{xcx}) and (\ref{zw}) as in Fig.\ref{fig:twocutsmap}. Its dual geometry depends on the moduli $\rho=e^{-2\pi s}$ of the annulus, which is related to the ratio $q/p$ (refer to Fig.\ref{fig:qprho}) as follows:
\ba
&& \mbox{BTZ BH phase}: q/p>1,\ \ \  \label{btzph} \\
&& \mbox{Thermal AdS Phase}: q/p<1.\ \ \ \label{tadsph}
\ea

In the former phase, the dual metric in the coordinate $(w,\bar{w})$ is given by (\ref{btz}) by restricting the range of $x$ as
\be
-\f{\pi}{\s{2}\beta}<x<0.    \label{xrange}
\ee
Note also that $y$ is periodic as $y\sim y+2\pi$.
The parameter $\beta$ is given in terms of the moduli parameter $s$ by the relation (\ref{betas}).
From the range (\ref{xrange}), we find that the holographic geometry is given by a half of solid torus.

In the latter phase, the dual metric in the same coordinate is obtained from (\ref{btz}) by exchanging
$x$ and $y$ with the identification $\beta=\s{2}s$.

Below we focus on the former phase (\ref{btzph}) because we interpret the parameter $p$ as a UV cut off and therefore we are interested in the region $q\gg p$. By applying the holographic transformation (\ref{mapz}) to the conformal map (\ref{xw}), we find that the dual metric simply takes the form of Poincare AdS$_3$ (\ref{poi}) in the coordinate $(\xi,\bar{\xi},\eta)$. However we need to remember the restriction of $x$ and the periodicity of $y$ as shown in (\ref{rhoxy}). The
identification $y\sim y+2\pi$ is equivalent to the identification: $(\eta,\xi,\bar{\xi})\sim e^{2\s{2}\pi\beta}(\eta,\xi,\bar{\xi})$. In this way, we find that the gravity dual is described by a part of Poincare AdS restricted to the region:
\be
1\leq \s{\f{\eta^2}{2}+|\xi|^2} \leq e^{2\s{2}\pi\beta},  \label{poitwo}
\ee
with the two half sphere boundaries identified.

Now we move onto to a gravity dual of (ii) partial entangling of two identical CFTs. As in
the left picture in Fig.\ref{fig:swap}, we glue two planes with two slits together.
After the conformal transformation (\ref{xcx}) into the annulus, we perform the map
$\zeta=\rho e^{\s{2}w}$ instead of (\ref{zw}). This leads to the range $0<x<\f{\pi}{\s{2}\beta}$ and
together with (\ref{xrange}), the total geometry describes a torus with the period
\be
\tau_2=\f{|C_y|}{|C_x|}=\s{2}\beta=\f{1}{2s}.
\ee
Therefore its gravity dual is given by the BTZ black hole (\ref{btz}) for $\tau_2>1$. On the other hand
for $\tau_2<1$, it becomes the thermal AdS, obtained from (\ref{btz}) with $x$ and $y$ exchanged.
In the former phase, the Bekenstein-Hawking area law formula computes the black hole entropy of the BTZ black hole and this is clearly identified with the entanglement entropy $S_1$ when we trace out the whole of one of the two CFTs. This leads to the formula (\ref{entrocut}) as we find
\be
S_1=\f{\mbox{Horizon Length}}{4G_N}=\f{\pi R}{G_N}\cdot \f{\beta}{\s{2}}=\f{\pi c}{3}\cdot \f{|C_y|}{|C_x|},
\ee
where we employed the well-known relation $\f{R}{G_N}=\f{2c}{3}$ \cite{BrHe}.

The latter phase $\tau_2<1$, on the other hand, has no black hole entropy and thus $S_1=0$.
This is because due to the large spectrum gap in holographic CFTs, the order one energy cut off
removes the large part of degrees of freedom of order $c$. Since we are interested in the high energetic limit of UV cut off $p\ll q$, we concentrate on the former phase $\tau_2>1$ below.

\subsection{Time evolution of Holographic Entanglement Entropy}

Now we would like to compute the holographic entanglement entropy (HEE) $S_A$ for an interval $A$ defined as
$X\in [X_1,X_2]$ in the coordinate $(X,\bar{X})$ which describes the two slit geometry.
We would like to discuss the two different setups: (i) local projection measurement of a single
CFT and (ii) partial entangling of two CFTs, at the same time as the computations are similar. In the latter case (ii), we take the interval $A$ in one of the two CFTs and trace out all other parts. To compute the holographic entanglement entropy in both cases, we need to pick up the relevant geodesics and compute the shortest length \cite{RT,HRT}.

\begin{figure}
  \centering
  \includegraphics[width=7cm]{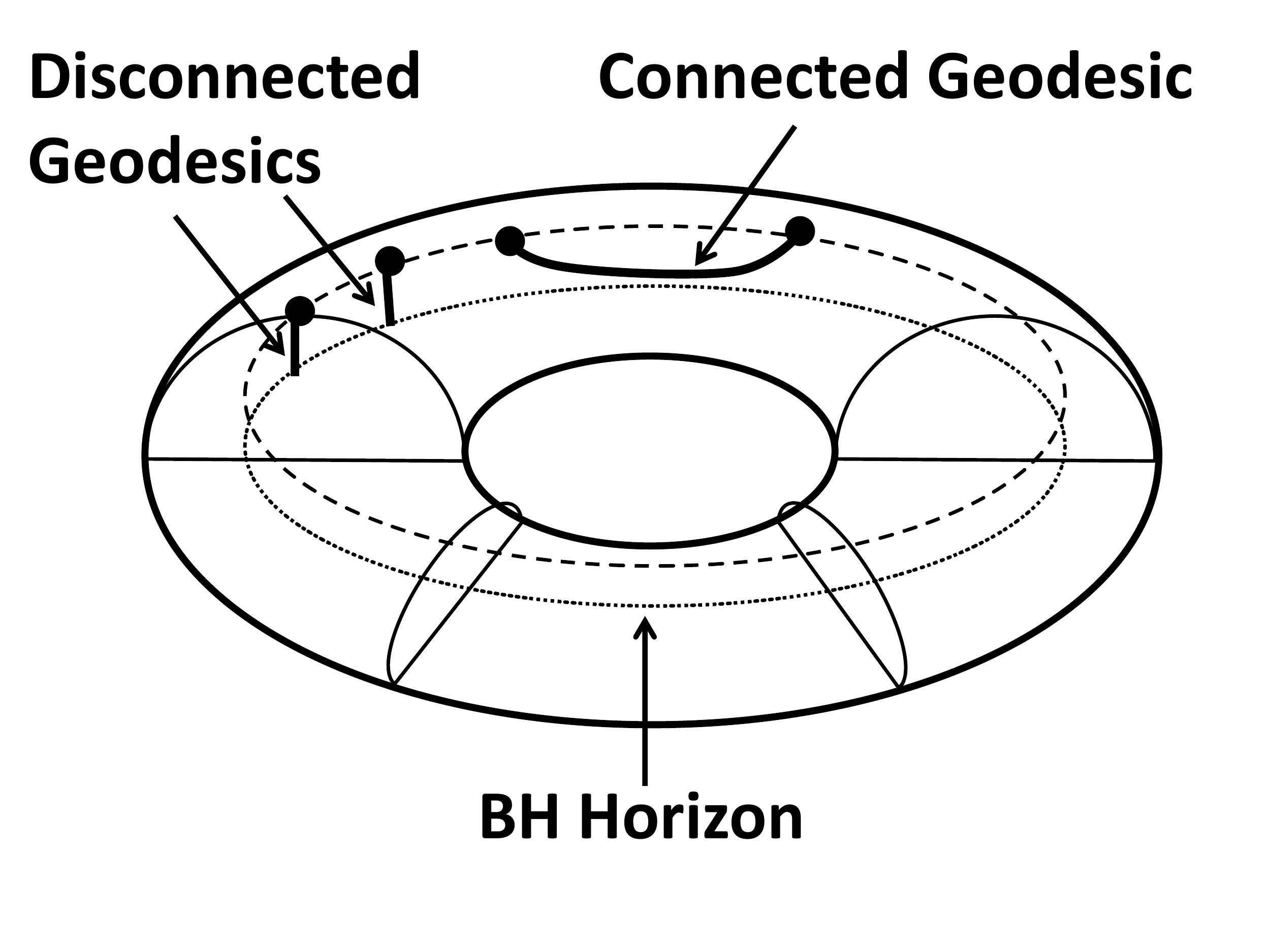}
  \caption{The computations of holographic entanglement entropy in a half of BTZ black hole geometry. In the case of gravity dual of a single CFT with the projection measurement along $P$, the geodesic can end  on the boundary, which is on the horizon in our example. In the case of gravity dual of doubles CFTs pasted with each other along $P$, only connected geodesics are allowed.}
\label{fig:BTZ}
  \end{figure}

In the black hole phase $\tau_2>1$, the gravity dual of (i) local projection measurement is given by cutting a solid torus (= Euclidean BTZ geometry (\ref{btz})) into a half as explained in the previous subsection. This space is sketched in Fig.\ref{fig:BTZ}. The boundary of this three dimensional geometry
is given by an union of an annulus (described by the coordinate $\zeta$) where the CFT is defined, and another annulus $Q$, which extends in the bulk at the bottom of the picture. Therefore, following the AdS/BCFT prescription \cite{AdSBCFT}, we need to choose the shortest one among connected or disconnected geodesics. The disconnected one is possible because the geodesic can end on the new boundary $Q$ at the bottom.

On the other hand, if we consider the gravity dual of (ii) partial entangling of two CFTs, we allow only connected geodesics in the solid torus geometry (\ref{btz}). Since the calculation of the connected geodesic length is the same as that in the case (i), we will present results together below.

First we compute the length of connected geodesic. For this, it is useful to map the two slit geometry in the $X$ coordinate into the Poincare AdS metric in the coordinate $(\xi,\bar{\xi},\eta)$. In this coordinate, the holographic entanglement entropy is computed from the geodesic distance as follows:
\be
S_A=\f{c}{6}\log \f{(\xi_1-\xi_2)(\bar{\xi}_1-\bar{\xi}_2)}{\ep_1\ep_2},  \label{heeint}
\ee
where $\xi_1$ and $\xi_2$ are the two end points corresponds to $X=X_1$ and $X=X_2$.
The UV cut off $\eta=\ep_{1,2}$ should respect the original UV cut off $z=a$ in the $X$ coordinate.

The point $X$ on the real axis is mapped into the point on the cirle $|\zeta|=\s{\rho}$ by the map (\ref{xcx}). We set $p=1/2$ in this section. More explicitly, we find the relation (we choose $0\leq \nu, \bar{\nu}<1$)
\ba
&& X\left(\zeta=\s{\rho}e^{2\pi i\nu}\right)=i\left(K(e^{2\pi i\nu})+K(\rho e^{2\pi i\nu})-\f{1}{2}\right)-t,\no
&&  \bar{X}\left(\zeta=\s{\rho}e^{-2\pi i\bar{\nu}}\right)=-i\left(K(e^{-2\pi i\bar{\nu}})
+K(\rho e^{-2\pi i\bar{\nu}})-\f{1}{2}\right)+t.
\ea
Even though $\nu$ and $\bar{\nu}$ are both real, they are not the same when $t\neq 0$. This is due to our analytical continuation of Euclidean time into Lorentzian one.

It is helpful to rewrite $K$ function as follows (remember $s=-\f{\log \rho}{2\pi}$, refer to our theta-function convention in appendix \ref{thetaf})
\ba
&& K(e^{2\pi i\nu})=\f{1}{2}+\f{1}{2\pi i}\f{\de_\nu \theta_1(\nu|2is)}{\theta_1(\nu|2is)},\no
&& K(\rho e^{2\pi i\nu})=\f{1}{2}+\f{1}{2\pi i}\f{\de_{\ti{\nu}} \theta_1(\ti{\nu}|2is)}{\theta_1(\ti{\nu}|2is)},
\ea
where we defined $\ti{\nu}=\nu+is$.

The final map into $\xi$ coordinate is given by
\be
\xi=(-i)\cdot e^{2\pi\beta\s{2}\nu},\ \ \
\bar{\xi}=i\cdot e^{2\pi\beta\s{2}\bar{\nu}}.
\ee

\begin{figure}
  \centering
  \includegraphics[width=6.5cm]{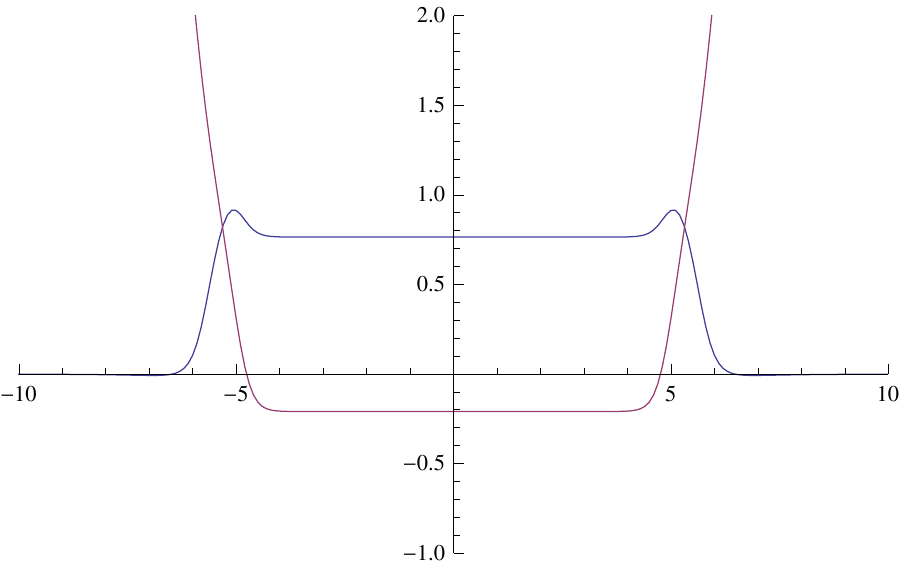}
  \hspace{1cm}
  \includegraphics[width=6.5cm]{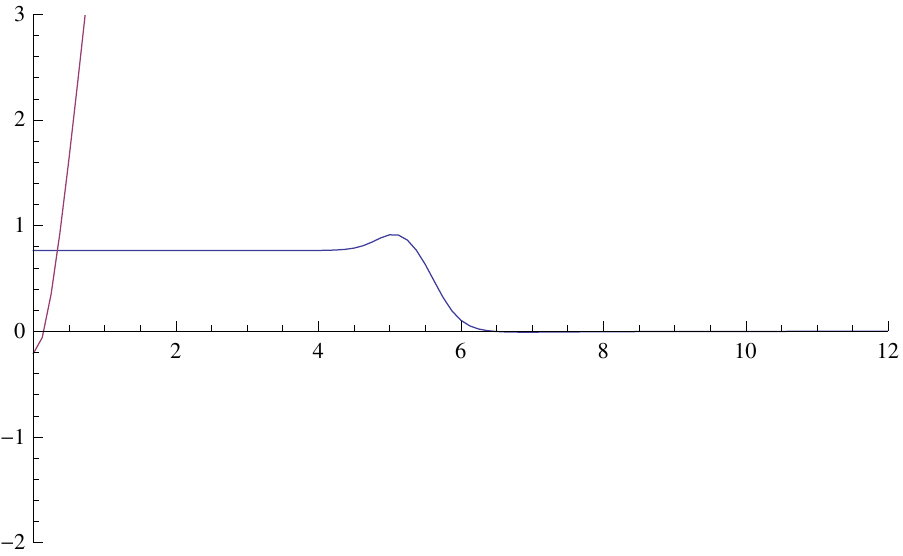}
  \caption{In the left picture, we showed $\Delta S_A$ as a function of $x$ for the interval subsystem $A$ defined by $[x-0.5,x+0.5]$, related to the connected geodesic. We chose $\rho=0.6$. The blue and red graph correspond to the result from the connected and disconnected geodesic. In the right picture we plotted $\Delta S_A$
  as a function of time $t$ when we defined $A$ to be the interval $[-0.5,0.5]$. In the case of entangled two CFTs, we always choose the blue curve (connected geodesic). On the other hand, in the case of local projection measurement of a single CFT, we choose the smaller value among the blue and red curve at each point, where we observe the phase transition behavior at the point they intersect.}
\label{fig:twocutDIS}
  \end{figure}

\begin{figure}
  \centering
  \includegraphics[width=6.5cm]{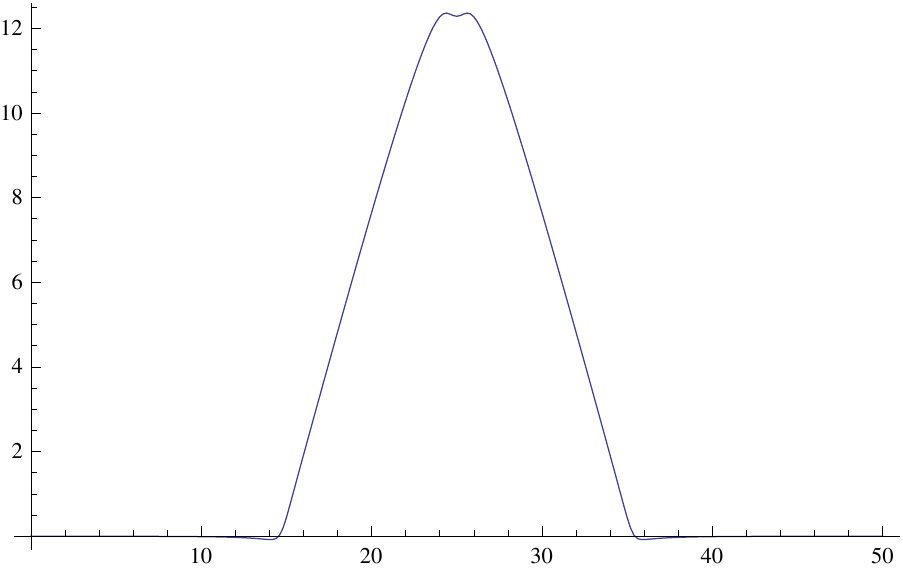}
  \hspace{1cm}
  \includegraphics[width=6.5cm]{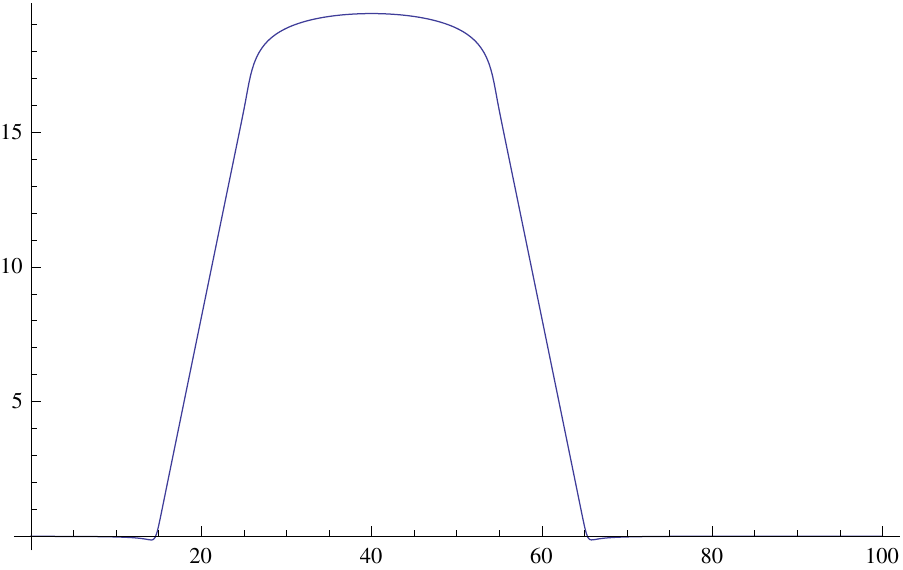}
  \caption{The behaviors of holographic entanglement in the setup dual to doubled CFT defined on the two cut geometry. We chose $\rho=0.6$. In the left picture, we showed $\Delta S_A$ as a function of time $t$ for the interval subsystem $A$ defined by $[20,30]$. In the right picture, we presented the same plot for the subsystem $A$ $[20,60]$.}
\label{fig:twocutInt}
  \end{figure}

The relation between the cut off $\ep$ in the Poincare coordinate and the original one $a$ is found as
\be
\f{\ep}{a}=\s{\f{d\xi}{dX}\f{d\bar{\xi}}{d\bar{X}}}=2\pi\s{2}\beta\cdot e^{\pi \s{2}\beta(\nu+\bar{\nu})}
\cdot\left(\f{dX}{d\nu}\f{d\bar{X}}{d\bar{\nu}}\right)^{-1/2}.  \label{eampp}
\ee
We plotted results of the connected geodesic in Fig.\ref{fig:twocutDIS} and Fig.\ref{fig:twocutInt} as blue graphs.

To calculate the disconnected geodesic, it is easier to work with the BTZ black hole (\ref{btz}) in the $w$ coordinate. If we trust the Euclidean geometry, which is correct at $t=0$, each geodesic from a boundary point ends at the horizon and thus the total length for an interval $A=[X_1,X_2]$ is computed as
\be
{\mbox{Length}}=\int^{\s{2}/\beta}_{\delta_1}\f{dz}{z}+\int^{\s{2}/\beta}_{\delta_2}\f{dz}{z}=
\log\left[\f{2}{\beta^2\delta_1\delta_2}\right],
\ee
where $\delta_{1,2}$ are the UV cut off in the metric (\ref{btz}). The relation between $\ep$ and $\delta$ can be found from the form of map (\ref{xw}) as
\be
\f{\ep}{\delta}=2\beta e^{\s{2}\pi\beta(\nu+\bar{\nu})}.
\ee
By combining this with (\ref{eampp}), we can calculate the holographic entanglement entropy for disconnected geodesics.

However, if we consider the real time evolution $t>0$, then we need to consider space-like geodesic in the Lorentzian spacetime so that it ends on a point on the boundary inside the bulk, which is Lorentzian continuation of $Q$. We choose this point by extremizing the length of geodesic.
Since our holographic spacetime is described by the Lorentzian version of BTZ black hole (\ref{btz}), setting $x=i\tau$, the identification of such geodesic can be done in the same way as done for the holographic quantum quenches \cite{HaMa}. In the end, we find
\be
{\mbox{Length}}(t)=\log \left[\cosh(\s{2}\beta \tau_1)\right]+\log \left[\cosh(\s{2}\beta \tau_2)\right]+\log\left[\f{2}{\beta^2\delta_1\delta_2}\right], \label{eet}
\ee
where $\tau_1$ and $\tau_2$ are given by $\tau_{i}=\pi(\nu_{i}-\ti{\nu}_{i})$ for $i=1,2$;
$\nu_i$ and $\bar{\nu}_i$ are evaluated at the two end points of the two geodesics at the AdS boundary. Final results of disconnected geodesics are plotted as red graphs in Fig.\ref{fig:twocutDIS}.

The left graph of Fig.\ref{fig:twocutDIS} shows the increased amount of entanglement entropy as a function of the location of a fixed length interval at $t=0$ when $\rho=0.6$, corresponding to
 $q\simeq 5.3$. The blue and red curve correspond to the connected and disconnected geodesic. The latter, which is only allowed in the case (i) projection measurement, takes negative values near the origin. This is because due to the projection measurement removes large part of vacuum entanglement in this region. Since we always need to pick up the smaller contribution among the disconnected and connected
geodesic length, near the origin the disconnected one is favored. However, if we instead consider the setup of (ii) partial entangling of two CFTs, only connected one is allowed. In this case, the peak near the origin is clearly understood because the entangled pairs are expected to be localized around $|x|\leq l$.

The right graph of Fig.\ref{fig:twocutDIS} shows the time evolution for $A$ give by the interval $[-1/2,1/2]$.
   In the case (i) we observe the initial growth under the time evolution of the red curve (disconnected geodesic). This is common to the global quantum quenches \cite{GQ}. It gets saturated to a thermal value until $t\simeq q(\simeq 5.3)$ and after that it goes to zero. This is because the created entangled pairs all go out of the subsystem $A$  for $t>q$. In the case (ii), we start with a positive entanglement due to the partial entangling and it suddenly vanishes at $t\simeq q$ due to the same reason as that for (i).

Now let us focus on (ii) partial entangling of two CFTs and study the time evolution in more detail. This is plotted in the graphs of Fig.\ref{fig:twocutInt}. Both graphs
shows linear growth, saturations and linear decrease. This is clearly explained if we remember that at $t=0$ we created entangled pairs in the region $P$ homogeneously and that they will propagate in the left and right direction at the speed of light.
Indeed, the maximum value of $\Delta S_A$ in the second graph is very close to a half of entanglement entropy between two CFTs, which is explicitly computed as $S_1/2\simeq 19.3$ from the formula (\ref{entot}).

It will also be interesting to study the time evolution when the subsystem $A$ is given by a semi-infinite line.  We choose the subsystem $A$ to be an interval $[0,L]$ and we assume the late time and large size limit: $L\gg t\gg 1$ (remember that we set $p=1/2$). In this limit we find
\be
\nu_1\simeq \f{1}{2\pi t},\ \ \ \ti{\nu}_1\simeq 1-\f{1}{2\pi t},\ \ \ \
\nu_2\simeq \f{1}{2\pi(t+L)},\ \ \ \ti{\nu_2}\simeq \f{1}{2\pi (L-t)}.
\ee
Finally, this leads to the following estimation of entanglement entropy
\be
S_A(t)\simeq \f{c}{6}\log\left[\f{\s{2}t}{\beta}\sinh\left(\s{2}\pi\beta\right)\right]
+\f{c}{3}\log (L/a).
\ee
This logarithmic growth $\Delta S_A\sim \f{c}{6}\log t$ is the same as that found in
locally excited states \cite{NNT,CNT,Hat}, which is defined by exciting the CFT vacuum
by a primary field at a point \cite{Nozaki}. It is also intriguing to note that in the local quench
defined by attaching two semi-infinite lines, the entanglement entropy grows logarithmically with a doubled
coefficient $\Delta S_A\sim \f{c}{3}\log t$ \cite{CCL} (see \cite{Ugajin} for its gravity dual).

\section{Holographic Analogue of Quantum Teleportation}

In this section we would like to consider an analogue of quantum teleportation in quantum field theory
and present its holographic realization by employing our holographic local projections and entangling operations.

\subsection{Brief Review of Quantum Teleportation}

Let us start with a brief review of quantum teleportation of a single qubit.
Assume that Alice $A$ and Bob $B$ are sharing an EPR state
\be
|EPR\lb_{AB}=\f{1}{\s{2}}\left(|0\lb_A|0\lb_B+|1\lb_A|1\lb_B\right).
\ee
The physical distance between $A$ and $B$ can be very large. Alice can also access to another qubit
$V$, whose state is not known to either Alice nor Bob. We express the state of $V$ by
\be
|\psi_4\lb_V=\lambda_1|0\lb_V+\lambda_2|1\lb_V,  \label{vstate}
\ee
where $|\lambda_1|^2+|\lambda_2|^2=1$.

Now Alice wants to send the information of $V$ to Bob by LOCC. For this it is useful to note the following identity:
\ba
|\psi_4\lb_V\otimes |EPR\lb_{AB}=\f{1}{2}\sum_{k=1}^4 |\Psi_k\lb_{VA}\otimes |\psi_k\lb_{B},
\ea
where we defined
\ba
&& |\Psi_1\lb=\f{1}{\s{2}}\left(|0\lb|1\lb-|1\lb|0\lb\right),\no
&& |\Psi_2\lb=\f{1}{\s{2}}\left(|0\lb|1\lb+|1\lb|0\lb\right),\no
&& |\Psi_3\lb=\f{1}{\s{2}}\left(|0\lb|0\lb-|1\lb|1\lb\right),\no
&& |\Psi_4\lb=\f{1}{\s{2}}\left(|0\lb|0\lb+|1\lb|1\lb\right), \label{wwwqr}
\ea
and
\ba
&& |\psi_1\lb=\lambda_1|1\lb-\lambda_2|0\lb,\no
&& |\psi_2\lb=\lambda_1|1\lb+\lambda_2|0\lb,\no
&& |\psi_3\lb=\lambda_1|0\lb-\lambda_2|1\lb,\no
&& |\psi_4\lb=\lambda_1|0\lb+\lambda_2|1\lb.
\ea

Note that $|\Psi_k\lb$ $k=1,2,3,4$ are all orthogonal to each other and complete
\be
\sum_{k=1}^4|\Psi_k\lb\la \Psi_k|=1.
\ee
Therefore we can perform a projection measurement for the system $VA$ to distinguish these four different states. If Alice observes that the state $|\Psi_k\lb$ is realized, then this result is reported to Bob via a classical communication. After that Bob can act an unitary transformation $U_k$ on $B$ to reproduce the original state (\ref{vstate}) of $V$ as follows
\be
U_k |\psi_k\lb_B = |\psi_4\lb_B.
\ee
It is obvious how to choose $U_k$ as the four states $|\psi_k\lb$ are linear about $\lambda_1$ and$\lambda_2$.

In this way, Alice can teleport the quantum state of $V$ to Bob. This is called quantum teleportation \cite{Bennett}. Since the classical communication can not exceed the speed of light, quantum teleportation is consistent with the causality.

It is also straightforward to generalize the above construction for more than two dimensional states \cite{Bennett}. For $N$ dimensional states, we take its basis to be $|0\lb,|1\lb,\ddd,|N-1\lb$. The quantum teleportation can be done as follows: for any state $|\psi\lb_V$, we assume a maximally entangled state of Alice and Bob:
\be
|\psi\lb_V \otimes \f{1}{\s{N}}\sum_{j}|j\lb_A|j\lb_B.
\ee
We project the above state by the projection $|\Psi_{(n,m)}\lb_{VA}\la \Psi_{(n,m)}|_{VA}$, where
\be
|\Psi_{(n.m)}\lb_{VA}=\f{1}{\s{N}}\sum_{j} e^{\f{2\pi ijn}{N}}|j\lb_V|j+m\lb_A.
\ee
Then Alice send the result of measurement given by $(n,m)$ to Bob. Finally Bob perform the unitary transformation $U_{(n,m)}$ defined by
\be
U_{(n,m)}=\sum_{k}e^{\f{2\pi ikn}{N}}|k\lb\la k+m|,
\ee
to recover the original state as $|\psi\lb_B$.

\subsection{QFT Analogue}

Now we would like to explore an analogue of quantum teleportation in QFTs.\footnote{Refer to \cite{QTQFT,Hotta} for earlier studies of different modelings of quantum teleportation in QFTs.}
  To simplify our description, we focus on two dimensional QFTs. Though our setup is general, we will focus on two dimensional
CFTs soon later for a computational reason.

\begin{figure}
  \centering
  \includegraphics[width=8cm]{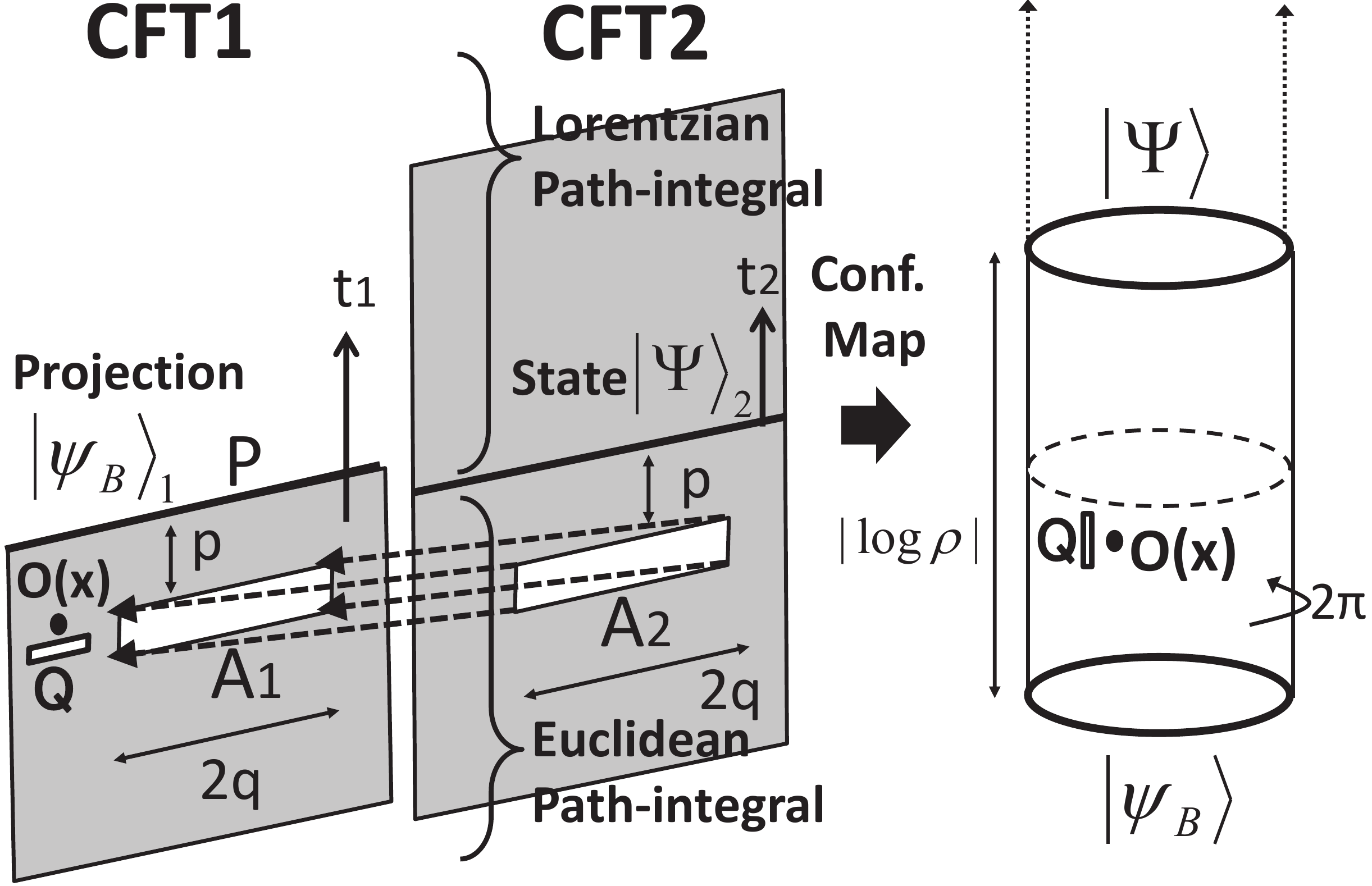}
  \caption{The path-integral description of quantum state under our quantum teleportation in CFT (left) and its conformal transformation into cylinder geometry (right). We glue the slit $A_1$ and $A_2$ together to realize the partial entangling operation. The projection measurement corresponds to putting a boundary in CFT$_1$. We assumed the projected region $P$ in CFT$_1$ is the whole space. If we add the slit $Q$ shown in the above pictures, then the point $x$ where the operator $O(x)$ is inserted gets disentangled with the other region, which looks more similar to the original quantum teleportation.}
\label{fig:QT}
  \end{figure}

Consider two identical QFTs, called QFT$_1$ and QFT$_2$, in two dimensions, each defined on an infinite line. We perform the partial entangling (see the left operation in Fig.\ref{fig:swap}) so that the two identical intervals $A_1$ in QFT$_1$ and $A_2$ in QFT$_2$ are entangled with each other. We choose the length of $A_1$ and $A_2$ to be $2p$ as before. Next we act a primary operator $O(x)$ localized at $x$ close to the interval $A_1$. Just after this, we perform a local projection measurement on each point in an interval $P$ which includes both $A_1$ and $x$. For a path-integral description of this procedure, refer to the left picture of Fig.\ref{fig:QT}. We expect we can recover the information of the operator $O(x)$ from the quantum state in QFT$_2$, which we call an analogue of quantum teleportation in QFTs.

For simplicity of our calculations, we assume that the two QFTs are conformal invariant and choose $P$ to be the total space in QFT$_{1}$. After the local projection measurement on $P$ in CFT$_1$, the state is projected to various quantum states $|\psi\lb_1$ which have the direct product structure in real space:
\be
|\psi\lb_1=\prod_{x\in R}|\psi(x)\lb_x.
\ee
A particular example of such a state is given in terms of a boundary state (or so called
Cardy state) $|B\lb$ as follows:
\be
 |\psi_b\lb_1={\cal N}_b\cdot e^{-p\cdot H}|B\lb_1,  \label{bdysk}
 \ee
 where $p$ is a small parameter which correspond to the UV cut off (i.e. lattice spacing) as the norm of a boundary state is divergent. The factor ${\cal N}_b$ is the normalization factor which guarantees $\la \psi_b|\psi_b\lb_1=1$. Note that an appropriate boundary state with this regularization has no real space entanglement if we regard $p$ as the lattice spacing \cite{MRTW}. General quantum states
 $|\psi\lb_1$ are obtained from (\ref{bdysk}) by acting the local unitary transformations
 \be
 |\psi\lb_1=\left(\prod_{x\in R} U_x\right)\cdot |\psi_b\lb_1.
 \ee
All such states $|\psi\lb_1$ have vanishing entanglement in real space decomposition and also are complete basis of all possible quantum states in CFT$_1$.

Our operations are summarized as follows. We start with the maximally entangled state at the Euclidean time $t_E=-p$ by gluing the intervals $A_1$ in CFT$_1$ and $A_2$ in CFT$_2$. Next we perform an Euclidean time evolution for both CFTs by a small period $p$ as a UV regularization. Then we project the state in CFT$_1$ by
$|\psi\lb_1 \la \psi|_1$. This leads to a pure state $|\Psi\lb_2$ at the same time $t_E=0$.

Now we can perform the conformal map (\ref{xcx}) so that the Euclidean path-integral $-\infty <t_E<0$
for CFT$_1$ is mapped to a cylinder with the width $-\f{1}{2}\log\rho$ and the circumference $2\pi$, where $\rho$ is a function of the ratio $q/p$ as depicted in Fig.\ref{fig:qprho}. In the same way, we can map the Euclidean path-integral $-\infty <t_E<0$ for CFT$_2$ into that on another cylinder. The gluing of $A_1$ and
$A_2$ is realized by attaching the two cylinders along each circle, leading to a longer cylinder
with the width $-\log\rho$. In this way, we end up with the cylinder depicted as the right picture in
Fig.\ref{fig:QT}. The boundary condition at its bottom is chosen to be the state $|\psi\lb$ obtained by the local projection measurement. The Euclidean path-integral over the cylinder leads to the state $|\Psi\lb_2$, which we wanted to compute, at its top boundary. Note that the length of the cylinder is estimated as $\f{\beta_H}{2}\equiv -\log\rho\simeq 2\pi \f{p}{q}\ll 1$ when $p/q\ll 1$. The insertion of the local operator $O(x)$ is in the middle of the cylinder. Thus by using this cylinder description, we find the final state $|\Psi\lb_2$ as follows
\be
|\Psi\lb_2={\cal N}_O\cdot e^{-\f{\beta_H}{4}H}\cdot O(x)\cdot e^{-\f{\beta_H}{4}H}|\psi\lb_2.
\label{fstate}
\ee

If we consider a linear combination of two operators $\lambda_1 O_1(x)+\lambda_2 O_2(x)$ with
$|\lambda_1|^2+|\lambda_2|^2=1$, the final state (\ref{fstate}) is also linear with respect to $\lambda_{1,2}$ i.e. we have
\be
|\Psi\lb_2={\cal N}_O\cdot e^{-\f{\beta_H}{4}H}\cdot (\lambda_1 O_1(x)+\lambda_2 O_2(x)) \cdot e^{-\f{\beta_H}{4}H}|\psi\lb_2.
\label{fstatee}
\ee
if the following conditions are satisfied:
\ba
&& \la\psi|e^{-\f{\beta_H}{4}H}O^\dagger_1e^{-\f{\beta_H}{2} H}O_1e^{-\f{\beta_H}{4}H}|\psi\lb=\la\psi|e^{-\f{\beta_H}{4}H}O^\dagger_2 e^{-\f{\beta_H}{2} H}O_2e^{-\f{\beta_H}{4}H}|\psi\lb=({\cal N}_O)^{-2},  \no
&& \la\psi|e^{-\f{\beta_H}{4}H}O^\dagger_1e^{-\f{\beta_H}{2} H}O_2e^{-\f{\beta_H}{4}H}|\psi\lb=\la\psi|e^{-\f{\beta_H}{4}H}O^\dagger_2 e^{-\f{\beta_H}{2} H}O_1e^{-\f{\beta_H}{4}H}|\psi\lb=0. \label{condq}
\ea
For example, if there is a $U(1)$ global symmetry, we can choose $O_1$ and $O_2$ to be positive and negative charged operators with appropriate normalizations to satisfy the above conditions by choosing $|\psi\lb$ to be arbitrary eigenstates of this $U(1)$ charge. In a 2d Dirac fermion theory, we can take $O_1=\psi$ and $O_2=\bar{\psi}$. We would like to argue these operations which start with the insertion of the operator
$\lambda_1 O_1(x)+\lambda_2 O_2(x)$ in CFT$_1$ and which finally lead to the state (\ref{fstatee}) in QFT$_2$, is an analogue of quantum teleportation in CFTs.

Let us compare the above procedure with the standard quantum teleportation. Following the idea of quantum teleportation, we started with a maximally entangled state on $A_1$ and $A_2$ by projecting the vacuum state with the operation ${\cal P}$ (\ref{pento}):
\be
|\Psi\lb_{12}={\cal P}|0\lb_1\otimes |0\lb_2 =\prod_{x\in A_1}\left[\sum_{n_x}|n_x\lb_{A_1} |n_x\lb_{A_2}\right]\otimes |\Psi(A^c_1\cup A^c_2)\lb_{12}
\ee
where $|0\lb_{1,2}$ are the vacuum states of CFT$_{1,2}$.  $|\Psi(A^c_1\cup A^c_2)\lb_{12}$
 is the state on the union of complements $A^c_{1,2}$ of $A_{1,2}$. Note that there is non-zero entanglement between $A^1_c$ and $A^2_c$.

Now we act the local operator at $x$ and we would like to teleport the state
\be
|\ti{\Psi}\lb=\ti{{\cal N}}_O(\lambda_1 O_1(x)+\lambda_2 O_2(x))|\Psi(A^c_1\cup A^c_2)\lb_{12}.
\label{starf}
\ee
Note that the normalization $\ti{{\cal N}}_O$ does not depend on $\lambda_1$ and $\lambda_2$
by assuming the U(1) charge conservation as in (\ref{condq}) owing to the fact that the state $|\Psi(A^c_1\cup A^c_2)\lb_{12}$ has zero charge. Therefore the map from this
 initial state (\ref{starf}) and the final state (\ref{fstatee}) is linear and may be regarded as a generalization of quantum teleportation. If we consider operators $O_i$ with different $U(1)$ charges, we can generalize the above analysis to linear combination $\sum_{i=1}^N O_i$ to get $N$ dimensional version of quantum teleportation.

However, one may notice that the state (\ref{starf}) is not purely defined as a state in CFT$_1$ because there is entanglement between $A^1_c$ and $A^2_c$. Actually this is the reason why our teleportation works even though the local measurement ${\cal P}$ projects to a state with no real space entanglement, as opposed to the projection to EPR states (or Bell states) $|\Psi_k\lb$ (\ref{wwwqr}) in the original quantum teleportation.

To make our setup closer to the original quantum teleportation, we can add a small slit $Q$ just below the operator insertion as in Fig.\ref{fig:QT}. This means that we start with a certain state at $x$ obtained by a projection ${\cal Q}$ onto a state $|\varphi\lb_Q$, which is not entangled with any other regions and then we act the operator on it. In other words, our initial state looks like
\ba
&& |\Psi'\lb_{12}={\cal P}O(x){\cal Q}|0\lb_1\otimes |0\lb_2  \no
&& \ \  =\ti{{\cal N'}}_O(\lambda_1 O_1(x)+\lambda_2 O_2(x))|\varphi\lb_Q\otimes \prod_{x\in A_1}\left[\sum_{n_x}|n_x\lb_{A_1} |n_x\lb_{A_2}\right]\otimes |\Psi(\ti{A}^c_1\cup A^c_2)\lb_{12}, \no \label{qtstate}
\ea
where $\ti{A}^c_1$ is defined by removing $Q$ from $A^c_1$.

Note that if we trace out the CFT$_1$ completely for the state (\ref{qtstate}), then we end up with the mixed density matrix for CFT$_2$ of the form:
\be
\rho_2=\mbox{Tr}_1\left[|\Psi'\lb_{12}\la \Psi'|_{12}\right]=\prod_{x\in A_2}\left[\sum_{n_x}|n_x\lb\la n_x|\right]_{A_2}\otimes \rho_{A^c_2},
\ee
where the information of the operator $\lambda_1 O_1(x)+\lambda_2 O_2(x)$ is completely missing.

However, by projecting the state (\ref{qtstate}) by a state $|\psi\lb_1$, we can extract the information of the operator. Indeed the final state in CFT$_2$ after the projection is written as
\be
|\Psi'\lb_2={\cal N'}_O\cdot e^{-\f{\beta_H}{4}H}\cdot (\lambda_1 \ti{O}_1(x)+\lambda_2 \ti{O}_2(x)) \cdot e^{-\f{\beta_H}{4}H}e^{itH}|\psi\lb_2,
\label{fstateee}
\ee
where $\ti{O}_1(x)$ and $\ti{O}_2(x)$ are dressed operator in the present of the slit boundary $Q$.
We also included a real time evolution by $t$. We apply an appropriate unitary transformation $U_\psi$, which depends on the state $|\psi\lb$ such that
\be
U \left({\cal N'}_O\cdot e^{-\f{\beta_H}{4}H}\cdot \ti{O}_i(x) \cdot e^{-\f{\beta_H}{4}H}|\psi\lb_2\right)
=\ti{{\cal N'}}_O O_i(x)|\varphi\lb_Q \otimes |\psi'(Q^c)\lb, \ \ (i=1,2)
\ee
for a certain state $|\psi'(Q^c)\lb$. The argument of linearity holds in a similar way as before by choosing the state $|\vp\lb_Q$ to be an eigenstate of $U(1)$ charge.

In this construction (\ref{fstateee}), if we take the limit $\beta_H\to 0$ and $t\to 0$, after the conformal transformation we find a large damping factor like $\lim_{y\to \infty}e^{-yH}$ because of the presence of boundary Q. Thus our teleportation fails in this limit. This is because the local projection measurement leads to a state with no real space entanglement and the mechanism of quantum teleportation does not work. Therefore we need to take $\beta_H$ or $t$ to be non-vanishing in this case.

\subsection{Holographic Interpretation}

We would like to consider holographic dual of quantum teleportation from CFT$_1$ to CFT$_2$ based on the previous setup. Let us start with an eternal BTZ black hole. Our entangled state between the two CFTs is conformally mapped into a cylinder as in the right picture in Fig.\ref{fig:QT}. Thus, if we do not perform any projection measurement, the entangled state is described by  a half of Euclidean BTZ geometry. Its Lorentzian time evolution is simply taken into account by Wick rotation, leading to an eternal Lorentzian BTZ black hole \cite{Mal}. In this geometry we cannot send any signal from CFT$_1$ to CFT$_2$ as there is no interaction between them in the CFT Hamiltonian. Holographically this is because the two boundaries are causally
disconnected.

Before we proceed, let us do an easy exercise. A ground state of a single CFT is dual to a pure AdS space. If we perform a local project measurement on the whole space at a time, then it ends up with a state $|\psi\lb$ with no real space entanglement, which is holographically dual to an empty spacetime. Thus, the pure AdS space only exists for $t<0$ and it is terminated by a boundary at $t=0$. This is the gravity dual of local project measurement of a single CFT. In this scenario, a collapse of wave function is equivalent to that of holographic spacetime.

Now consider the state (\ref{fstate}) just after the local projection measurement. The insertion of the local operator $O(x)$ gives a localized excitation near the AdS boundary region. In the version with the slit $Q$, we just need to replace $O(x)$ with a dressed operator $\ti{O}(x)$ as we mentioned. Since the projection measurement is interpreted as putting a boundary in the Euclidean path-integral as in Fig.\ref{fig:QT}, this is holographically dual to cutting out the upper-left wedge of BTZ black hole, i.e. shaded region in the left picture of Fig.\ref{fig:QT}. Following the prescription of AdS/BCFT
\cite{AdSBCFT}, we need to impose the condition of vanishing extrinsic curvature and this requires an introduction of backreactions. Eventually this leads to the gravity dual given by the right picture of Fig.\ref{fig:QT}, i.e. a half of an eternal BTZ black hole with the operator insertion. Note that the temperature of the black hole is now reduced by a factor two as is expected because the projection measurement reduces quantum entanglement between the two CFTs. As is clear in this holographic description, the information of the operator $O(x)$ originally inserted in CFT$_1$ is teleported to the CFT$_2$ through the Euclidean black hole. This is the basic mechanism of our holographic quantum teleportation.

Let us also comment on the classical communication. We need to send the result of local projection measurement in CFT$_1$ to CFT$_2$ via a classical communication. Actually, this can be done by swapping the information of projection measurement from CFT$_1$ to CFT$_2$. Note that this is still a classical communication as we send a direct product state.\footnote{We can insert a cut like $Q$ to make the swapped part of CFT$_2$ also have no real space entanglement.} The size of the geometry which involves this swapping procedure can be negligible in the gravity dual as there is no entanglement. Therefore we can regard it as a thin wire which can be included in the Euclidean space (i.e. the lower half parts) in Fig.\ref{fig:QT}.

The above holographic model of quantum teleportation is closely related to the one by Susskind \cite{Sk,Skk}, where we start with three copies of CFTs dual to three asymptotic regions of two AdS black holes. Both share the crucial property that the information is teleported through the Einstein-Rosen bridge.

\begin{figure}
  \centering
  \includegraphics[width=8cm]{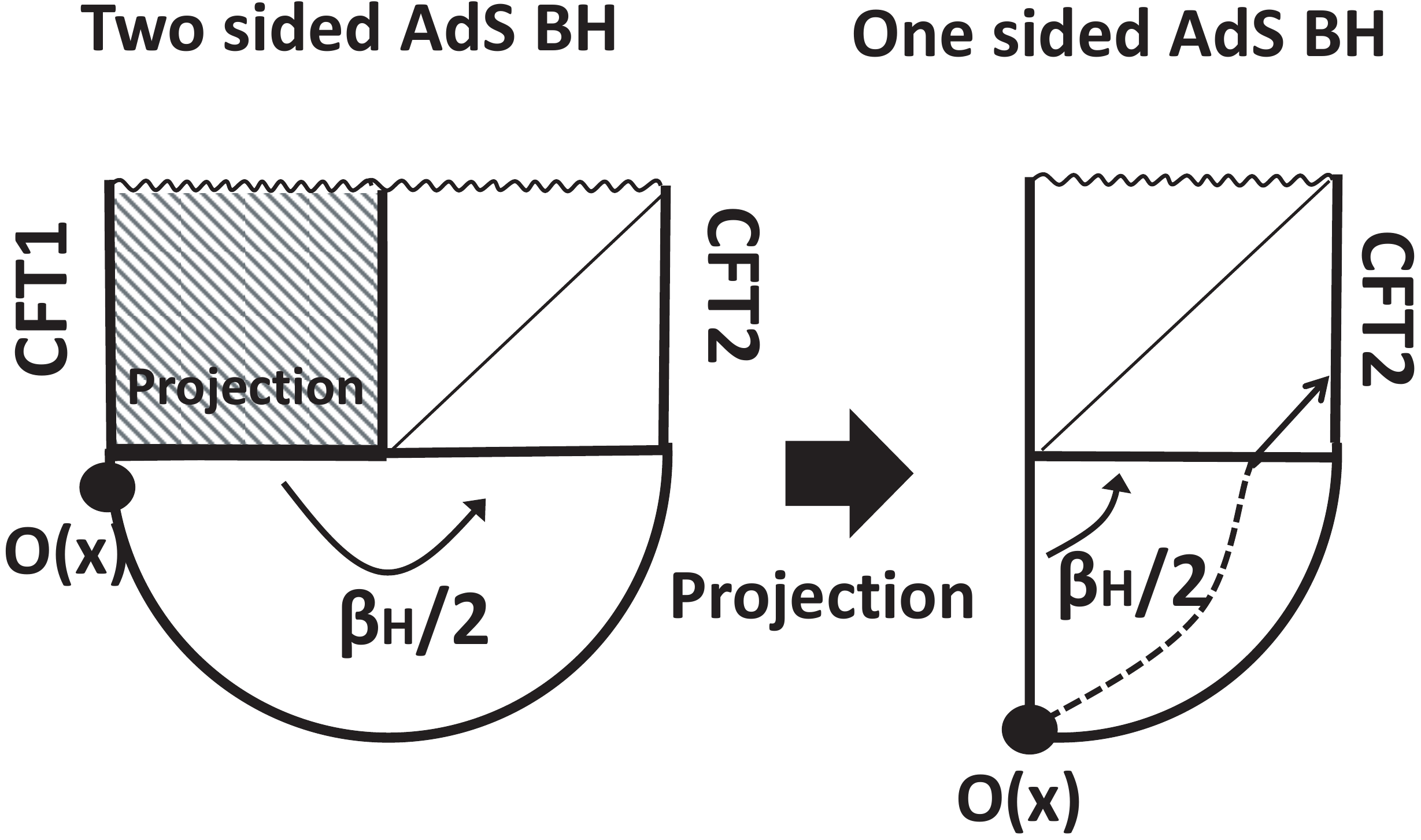}
  \caption{A holographic description of quantum teleportation.}
\label{fig:HQT}
  \end{figure}

\section{Conclusions and Discussions}

In this paper we formulated several important quantum operations in terms of quantum field theories, especially two dimensional conformal field theories (CFTs). First we considered local projection measurements. In a CFT, a class of states produced after local projection measurements are described by boundary states. Their holographic duals are given by removing some regions from the AdS space with backreactions taken into account using the AdS/BCFT prescription. We also consider quantum operations for two identical CFTs. We defined partial entangling of two CFTs by pasting two sheets in the path-integral along a slit. We also introduced swapping operations by exchanging two intervals.

The local projection measurement is conformally mapped to a path-integral on a cylinder. After this operation, the entanglement entropy is reduced. Later it grows linearly as in the case of quantum quenches and gets saturated for a while. Finally it decreases to the original value for the ground state. We find this behavior in both a free fermion CFT and a holographic CFT, though in the latter case we encounter a phase transition.

The partial entangling operation is described by a path-integral on a torus. The entanglement entropy between the two CFTs turned out to be proportional to the length of entangled region and thus follows the volume law.  The gravity dual of this state is given by a BTZ black hole. We also computed the time evolution of holographic entanglement entropy for an interval and we interpreted the results in terms of propagating entangled pairs.

On the other hand, the swapping operation corresponds to another torus with a different period and the entanglement entropy between the two CFTs is given by a twice of the familar logarithmic formula of ground state entanglement in 2d CFTs.

Finally we combined these quantum operations to give an analogue of quantum teleportation between two CFTs. We also present its holographic realization. The local projection measurement leads to a collapse of spacetime and this cuts out one of two boundaries of an eternal BTZ black hole. This allows us to send a signal from one CFT to another CFT.

One immediate future problem is higher dimensional generalizations. Even though in general we may not have any analytical control as gravity duals get more complicated in more than three dimensions, still we can think of simple settings. One obvious example is found by restricting two copies of Euclidean $d$ dimensional CFTs to the region $\Sigma_d$ specified by $t_E^2+\vec{x}^2\geq l^2$ (i.e. outside of a round disk) and gluing their boundaries $t_E^2+\vec{x}^2=l^2$ with each other. After an obvious conformal map, this geometry is equivalent to pasting two upper half planes along their boundaries to produce a full plane. A gravity dual of this background is given as follows. A CFT on $\Sigma_d$ is dual to a pure AdS$_{d+1}$
\be
ds^2=R^2\left(\f{d\eta^2+dt_E^2+d\vec{x}^2}{\eta^2}\right),
\ee
with a semi-sphere removed as
\be
t_E^2+\vec{x}^2+\eta^2\geq l^2. \label{region}
\ee
The entangling two CFTs are described by attaching the two boundaries (radius $l$ spheres).
After the holographic conformal map (\ref{mapz}), the glued geometry is simply equivalent to a full
geometry of a pure AdS$_{d+1}$. By using this map, we can understand the behavior of correlation functions and entanglement entropy etc.

The quantum operations discussed in this paper offer us a new method to study quantum information theoretical properties of CFTs. Indeed, we introduced a new quantity $\delta^B_A$ (\ref{ww}) using local projection measurement to probe some tripartite entanglement. It is intriguing to pursuit this direction to study multi-partite entanglement.

\section*{Acknowledgements}

  We are grateful to Hector Bombin, Yasunori Nomura, Eliezer Rabinovici, Yu Watanabe, Jie-Qiang Wu for useful conversations and especially Lenny Susskind for a valuable correspondence.
  TN, NS and KW are supported by Grant-in-Aid for JSPS Fellows No.14J02735, No.15J02740 and No.15J01358. TT is supported by the Simons Foundation and JSPS Grant-in-Aid for Scientific Research (B) No.25287058 and (A) No.16H02182. TT is also supported by World Premier International Research Center Initiative (WPI Initiative) from the Japan Ministry of Education, Culture, Sports, Science and Technology (MEXT).

\appendix

\section{Conventions of Theta Functions}\label{thetaf}

Here we present our conventions of $\theta$-functions. They are defined by
\ba
\eta(\tau)&=&q^{\f{1}{24}}\prod_{n=1}^{\infty}(1-q^n),\no
\theta_{1}(\nu,\tau)&=&2q^{\f18}\sin(\pi\nu)\prod_{n=1}^{\infty}(1-q^n)
(1-e^{2i\pi\nu}q^{n})(1-e^{-2i\pi\nu}q^{n}),\no
\theta_{2}(\nu,\tau)&=&2q^{\f18}\cos(\pi\nu)\prod_{n=1}^{\infty}(1-q^n)
(1+e^{2i\pi\nu}q^{n})(1+e^{-2i\pi\nu}q^{n}),\no
\theta_{3}(\nu,\tau)&=&\prod_{n=1}^{\infty}(1-q^n)
(1+e^{2i\pi\nu}q^{n-\f12})(1+e^{-2i\pi\nu}q^{n-\f12}),\no
\theta_{4}(\nu,\tau)&=&\prod_{n=1}^{\infty}(1-q^n)
(1-e^{2i\pi\nu}q^{n-\f12})(1-e^{-2i\pi\nu}q^{n-\f12}), \label{th}
\ea
where we set $q=e^{2i\pi\tau}$. Their modular transformations are summarized as follows
\ba
\eta(\tau)&=&(-i\tau)^{-\f12}\eta(-\f{1}{\tau}),\ \ \theta_{1}(\nu,\tau)
=i(-i\tau)^{-\f12}e^{-\pi i\f{\nu^2}{\tau}}\theta_{1}
(\nu/\tau,-\f{1}{\tau}), \no
\theta_{2}(\nu,\tau)&=&(-i\tau)^{-\f12}e^{-\pi i\f{\nu^2}{\tau}}
\theta_{4}(\nu/\tau,-\f{1}{\tau}), \ \ \theta_{3}(\nu,\tau)
=(-i\tau)^{-\f12}e^{-\pi i\f{\nu^2}{\tau}}\theta_{3}(\nu/\tau,-\f{1}{\tau})
, \no\theta_{4}(\nu,\tau)&=&(-i\tau)^{-\f12}e^{-\pi i\f{\nu^2}{\tau}}
\theta_{2}(\nu/\tau,-\f{1}{\tau}). \label{TF}
\ea

\section{Toy Analytical Example of Entangling Two CFTs}

\begin{figure}
  \centering
  \includegraphics[width=6.5cm]{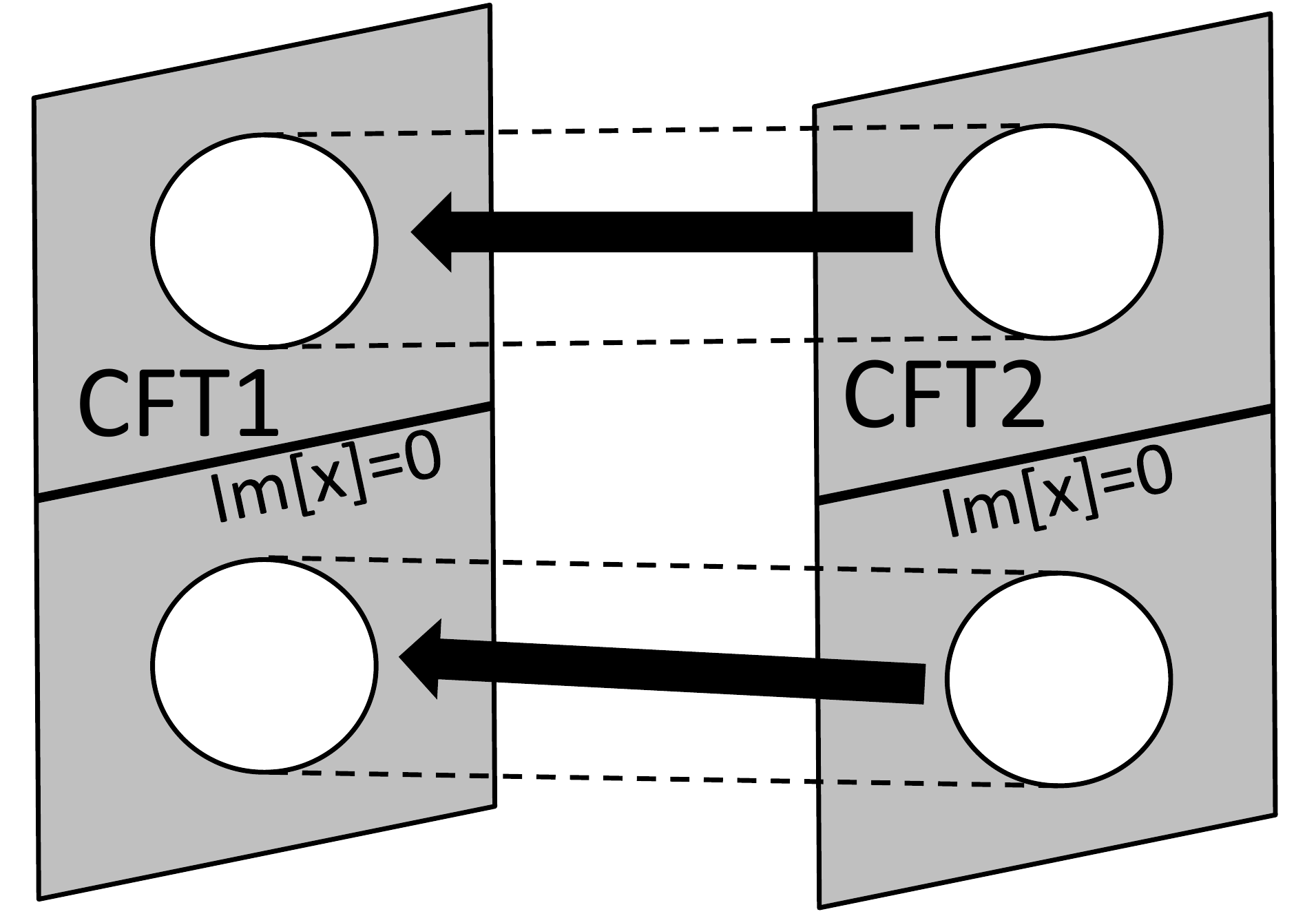}
\hspace{1cm}
  \includegraphics[width=7cm]{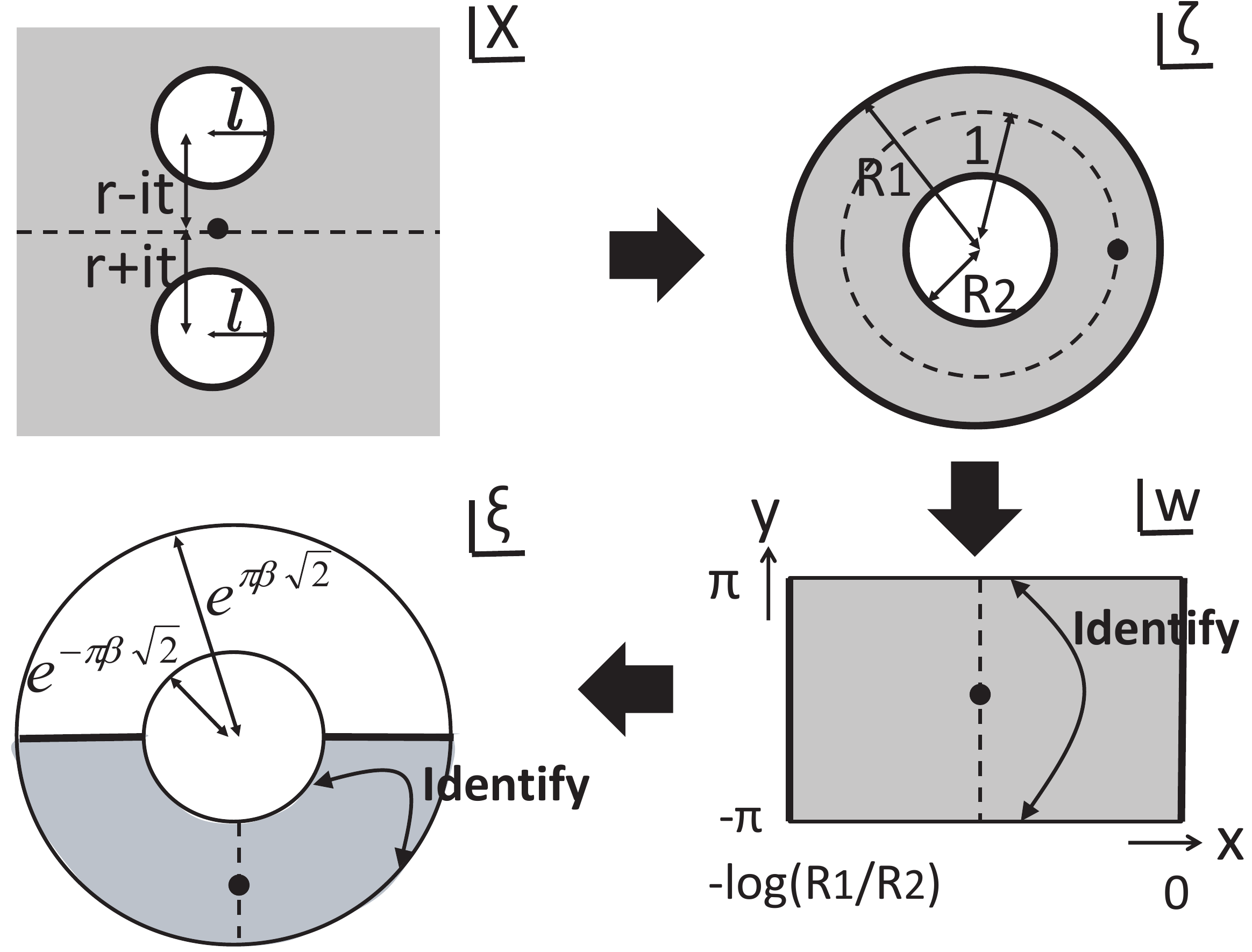}
  \caption{A simple way to entangle two CFTs is described by gluing two planes along two holes (left). Each of the two sheets can be mapped  into a cylinder in the chains of conformal transformations shown in the right picture.}
\label{fig:twocirmap}
  \end{figure}

Here we present an simple and analytical example which is analogous to the setup where we partially glue two CFTs. In the Euclidean path-integral, this model is defined by two complex planes with two circular holes attached with each other along the edges of the holes. Each sheet of the doubled planes is mapped into a cylinder as sketched in Fig.\ref{fig:twocirmap}. If we focus only this single sheet, the setup is analogous to the one with a projection measurement. We will describe the detail of this transformation and computations of holographic entanglement entropy below.

\subsection{Conformal Map of a Circle}

We define a complex plane with two holes (radius $l$) by
\be
|X+ir_2|\geq l,\ \ \ |X-ir_1|\geq l,  \label{antwc}
\ee
where $r_{1,2}(>l)$ describe the Euclidean time evolution. To describe
the real time evolution, we eventually perform the following analytical continuation:
\be
r_1=r-it,\ \ \ r_2=r+it,  \label{rexpv}
\ee
where $r$ is interpreted as the UV regularization parameter of the gluing procedure and $t$
is the real time coordinate.

We find that the following map
\ba
&& x=i\left(\f{a\zeta+b}{\zeta+1}\right), \no
&&  a=\f{r_2-r_1}{2}-\s{\f{(r_2+r_1)^2}{4}-l^2}, \no
&&  b=\f{r_2-r_1}{2}+\s{\f{(r_2+r_1)^2}{4}-l^2},
\ea
transforms the two circle $|x+ir_2|=l$ and $|x-ir_1|=l$ into a two circles centered at the origin
$|\zeta|=R_1$ and $|\zeta|=R_2$, where the two radii $R_1>R_2$ are
\ba
&& R_1=\f{l}{\f{r_1+r_2}{2}-\s{\f{(r_2+r_1)^2}{4}-l^2}},\no
&& R_2=\f{l}{\f{r_1+r_2}{2}+\s{\f{(r_2+r_1)^2}{4}-l^2}}. \label{radtim}
\ea

Using (\ref{rexpv}), we get
\ba
&& a=it-\s{r^2-l^2},\ \ \ b=it+\s{r^2-l^2},\no
&& R_1=\f{r+\s{r^2-l^2}}{l},\ \ \ R_2=\f{r-\s{r^2-l^2}}{l}.
\ea
In this way, the sheet is mapped into an annulus with the coordinate $\zeta$.

We act a further conformal transformation
\be
\zeta=R_2\cdot e^{\s{2}w},
\ee
with $w=\f{x+iy}{\s{2}}$. The annulus is now mapped into a cylinder
\be
-\pi\leq y \leq \pi, \ \ \ 0\leq x \leq \log\f{R_1}{R_2}. \label{onesi}
\ee

In a similar way, by the map $\zeta=R_2\cdot e^{-\s{2}w}$,
the same annulus is mapped into the cylinder with the range
\be
-\pi\leq y \leq \pi, \ \ \ -\log\f{R_1}{R_2} \leq x \leq 0.  \label{twosi}
\ee
If we paste the two of such cylinders along each edge, the total space is now described by a torus with the period $\tau=\tau_1+i\tau_2$:
\be
\tau_1=0,\ \ \ \tau_2=\f{\log\f{R_1}{R_2}}{\pi}.
\ee

Moreover it is useful to perform the following conformal transformation in order to conduct holographic computations:
\be
\xi=e^{2i\beta w}. \label{gggc}
\ee
All these maps are summarized in the right picture of Fig.\ref{fig:twocirmap}.

\subsection{Holographic Description}

Now we move on to the gravity dual of a holographic CFT on the previous torus geometry.
There are two phases depending on the length of the $x$- and $y$-circle, denoted by  $|C_x|$ and $|C_y|$, respectively:
\ba
&& \mbox{Thermal AdS Phase}:\ \ \tau_2=\f{|C_y|}{|C_x|}=\f{\log\f{R_1}{R_2}}{\pi}\leq 1,\label{ctad}\\
&& \mbox{BTZ BH phase}:\ \ \tau_2=\f{|C_y|}{|C_x|}=\f{\log\f{R_1}{R_2}}{\pi}\geq 1. \label{cbhp}
\ea
Note that in former (and latter) phase, the $y$-circle $C_y$ (and $x$-circle $C_x$) is contractible.
Their gravity dual of the former and latter phase can be simply obtained by assuming the Poincare AdS metric whose boundary give by the annulus in the $\zeta$ coordinate and $\xi$ coordinate, respectively,
where we impose the restriction and identification in the most symmetric way as in \cite{AdSBCFT}.

Since there is no black hole in the former phase, the entanglement entropy between the whole space in the first CFT and that in the second CFT is vanishing. Therefore we focus on the latter phase below.
The holographic dual of the phase (\ref{cbhp}) is given by the Euclidean BTZ black hole (\ref{btz}) which has the periodicity $x\sim x+\f{\s{2}\pi}{\beta}$. We also compactify $y$ such that
$y\sim y+2\pi$. Therefore we choose
\be
\beta=\f{\pi}{\s{2}\log\f{R_1}{R_2}}.
\ee
We can also map the gravity solution (\ref{btz}) into the Poincare AdS$_3$ (\ref{poi}) via the map
(\ref{mapz}) for the conformal transformation (\ref{gggc}). However note that we are actually considering the quotient of the Poincare AdS$_3$ (\ref{poi}) by the identification $y\sim y+2\pi$ in the  $w$ coordinate, which is equivalent to the identification: $(\eta,\xi,\bar{\xi})\sim e^{2\s{2}\pi\beta}(\eta,\xi,\bar{\xi})$. In other words, this is a solid torus defined by
\be
e^{-\s{2}\pi\beta}\leq \s{\f{\eta^2}{2}+|\xi|^2} \leq e^{\s{2}\pi\beta},
\ee
with the two half sphere boundaries identified.

\subsection{Entanglement Entropy between Two CFTs}

We can holographically compute the entanglement entropy $S_1$ for the first CFT when we trace out
the second one completely in our setup. In the BZT black hole phase (\ref{cbhp}), this entanglement entropy $S_1$ is given by the black hole entropy
\be
S_1=\f{\mbox{Horizon Length}}{4G_N}=\f{\pi R}{G_N}\cdot \f{\beta}{\s{2}}=\f{c\pi^2}{3}\cdot \f{1}{\log\f{R_2}{R_1}},
\ee
where $c$ is the central charge. Note also that we can justify this result in the limit $r\to l$
(i.e. $\tau_2\to 0$) without using the holography using the standard result of thermal entropy in the high temperature limit.

\begin{figure}
  \centering
  \includegraphics[width=6.5cm]{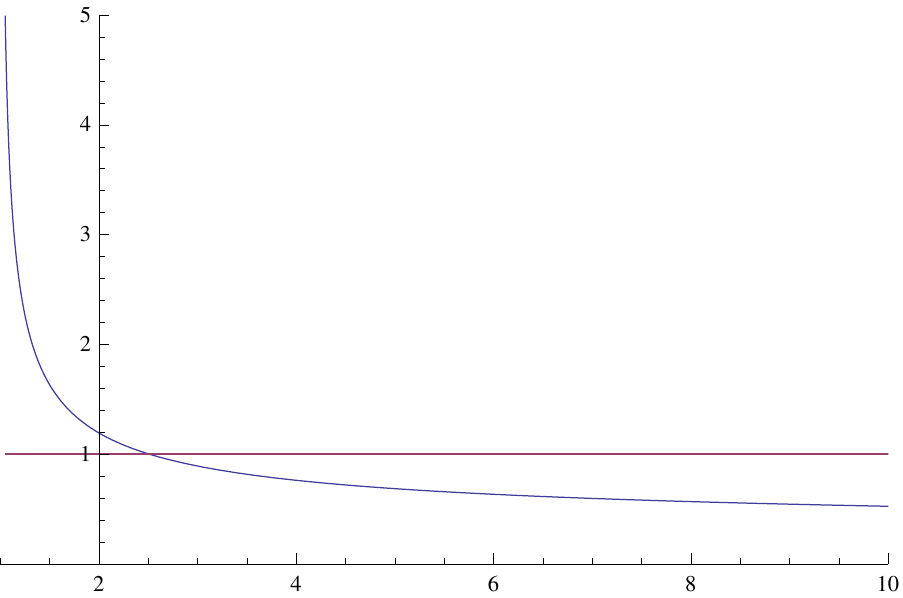}
  \caption{The plot of $\tau_2=\f{|C_y|}{|C_x|}$, which is proportional to entanglement entropy between the two CFTs, as a function of $r/l$.  The blue curve for $0<\gamma<1$ describes the behavior of the entanglement entropy. In AdS$_3/$CFT$_2$, at $\gamma=1$ there is a phase transition and the entropy becomes vanishing for $\gamma>1$.}
\label{fig:periodtwocircle}
  \end{figure}

In terms of the parameter $r$ and $l$, we find
\be
S_1=\f{2\pi^2 c}{3}\cdot \f{1}{\log\f{r+\s{r^2-l^2}}{l}}.
\ee
This is a monotonically decreasing function of the Euclidean time $r$ as we expect (refer to
Fig.\ref{fig:periodtwocircle}).
When $\log\f{r+\s{r^2-l^2}}{l}=\f{\pi}{2}$, the entropy $S_1$ suddenly changes to zero as we need to choose the thermal AdS phase (\ref{ctad}).

Since the radii (\ref{radtim}) only depends on $r_2+r_1$ and thus the resulting entanglement entropy
$S_A$ does not depend on the real time $t$ under the time evolution. This is because the time evolution is described by a unitary transformation which is a direct product with respect to the two CFTs.

It is intriguing to consider the limit $\delta\equiv r-l\ll l$. In this case $S_1$ behaves like
\be
S_1\sim c\s{\f{l}{\delta}}.  \label{entropci}
\ee

This result is actually consistent with our expectation. $\delta$ is interpreted as the UV cut off the
entangled pairs between the two CFTs along an interval. If the length of this interval is $\Delta x$, then the entanglement entropy is estimated as $S_B\sim c\cdot \f{\Delta x}{\delta}$ as follows from the
volume law of the maximally entangled state. The length $\Delta x$ is estimated by requiring the distance between the upper and lower boundary is of order $\delta$ namely,
\be
l-\s{l^2-(\Delta x)^2}\sim \delta,
\ee
which leads to $\Delta x\sim \s{l\delta}$. Thus we reproduce the behavior (\ref{entropci}).

\subsection{Time Evolution of Entanglement Entropy for an Interval in Two Entangling CFTs}

Finally we compute the time-dependent entanglement entropy $S_A$ when we choose $A$ to be an interval in the first CFT. The subsystem $A$ is describes as $[X_1,X_2]$ in the complex coordinate $X$. We choose
both $X_1$ and $X_2$ to be real. The gravity dual in the $X$ coordinate can be found by applying the previous chain of conformal transformations to the holographic map from (\ref{poi}) to (\ref{metm}).

To calculate the length of geodesic between the two points, we start with the Poincare AdS coordinate
$(\eta,\xi,\bar{\xi})$, where the holographic entanglement entropy is simply given by (\ref{heeint}).
The UV cut off $\eta=\ep_{1,2}$ should respect the original UV cut off $z=a$ in the $X$ coordinate.
Thus we find the following relation from (\ref{mapz})
\be
\f{\ep}{a}=\f{\s{2}\beta\s{r^2-l^2}}{\s{(x^2-t^2+r^2-l^2)^2+4t^2(r^2-l^2)}}\zeta^{-i\beta/\s{2}}
\bar{\zeta}^{i\beta/\s{2}},
\ee
where
\be
\zeta=\f{i\s{r^2-l^2}-x-t}{i\s{r^2-l^2}+x+t},\ \ \ \bar{\zeta}=\f{-i\s{r^2-l^2}-x+t}{-i\s{r^2-l^2}+x-t}.
\ee
Note that $\bar{\zeta}$ is not actual complex conjugate of $\zeta$ because we need to treat $t$ as an imaginal number due to the analytical continuation of the Euclidean time.

\begin{figure}
  \centering
  \includegraphics[width=6.5cm]{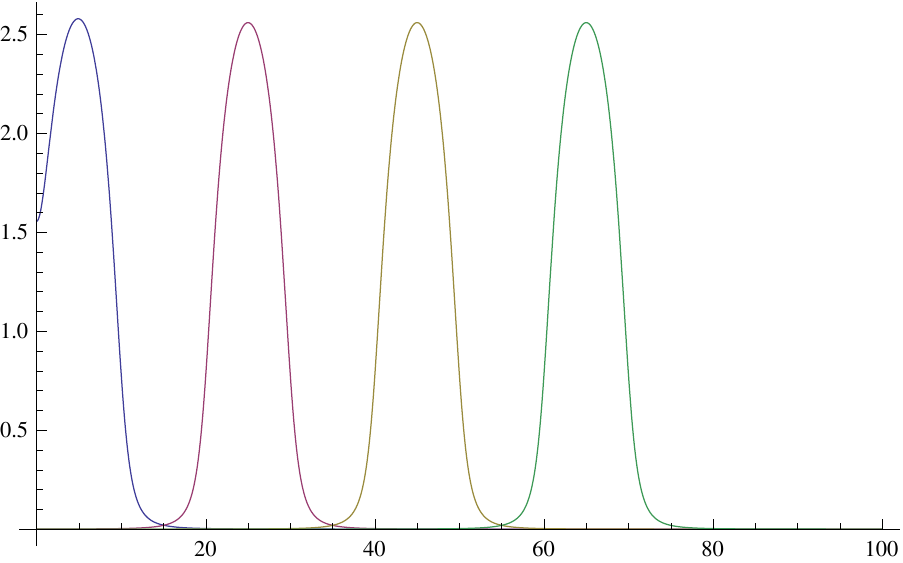}
  \hspace{1cm}
  \includegraphics[width=6.5cm]{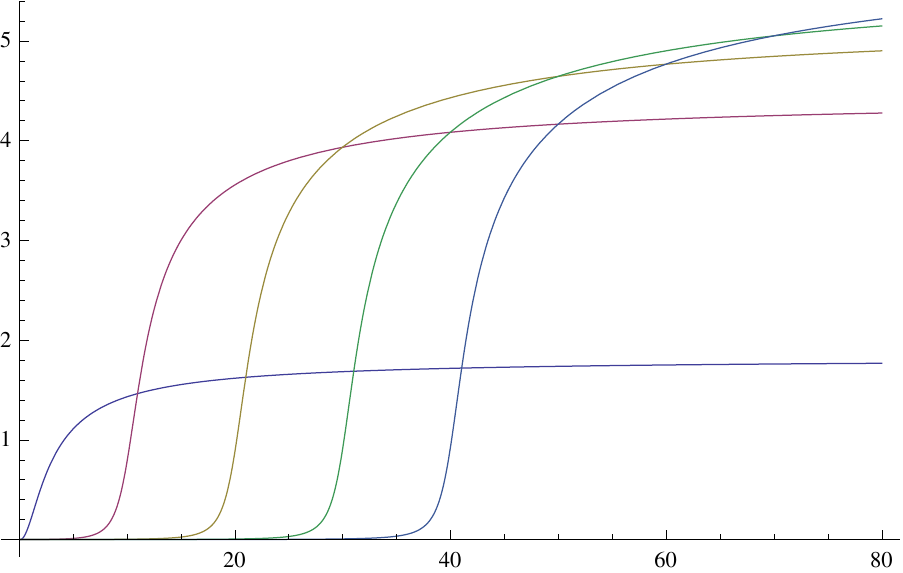}
  \caption{The behavior of holographic entanglement entropy with the vacuum contribution subtracted in the gravity dual of the two cuts geometry. We normalized the total value by choosing the central charge $c=6$ and assume the length parameters to be $r=2$ and $l=1$. In the left picture, we plotted $\Delta S_A$ as a function of the time $t$ for 4 different choices of interval subsystem $A$ i.e. $[0,10]$ (blue), $[20,30]$ (red), $[40,50]$ (yellow) and $[60,70]$ (green). In the right picture, we plotted $\Delta S_A$ as a function of $x$ for the interval subsystem $A$ given by $[0.1,x]$. The blue, red, yellow, green and blue curve corresponds to $t=0,10,20,30$ and $40$, respectively.}
\label{fig:twocircletime}
  \end{figure}

We presents some plots of holographic entanglement entropy in Fig.\ref{fig:twocircletime}. We subtracted the vacuum contribution. The peaks in the first plot is explained by noting that the entangled pairs between the two CFTs are locally generated around $X=0$ and propagate at the speed of light. When the entangled pairs come into the interval, $S_A$ gets increased. The second plot shows how the entropy grows as the size of interval increases at various times. Again we can confirm the light-like propagation of entangled pairs.

\subsection{Entanglement Entropy under Projection Measurement}

Consider a single sheet with two circular holes instead of the double sheet.
This is analogous to the projection measurement around the region near the holes. For this purpose we just need to restrict to the cylinder (\ref{onesi}) and ignore (\ref{twosi}).

It is obvious from our previous arguments that the gravity dual of this setup is obtained by cutting the previous solid torus into a half as in \cite{AdSBCFT}. Thus there are two possible types of geodesics: connected one and disconnected one as in Fig.\ref{fig:BTZ}. Notice that for the latter one, we need to carefully evaluate the location of end point of the geodesic as we did in (\ref{eet}).

\begin{figure}
  \centering
  \includegraphics[width=6.5cm]{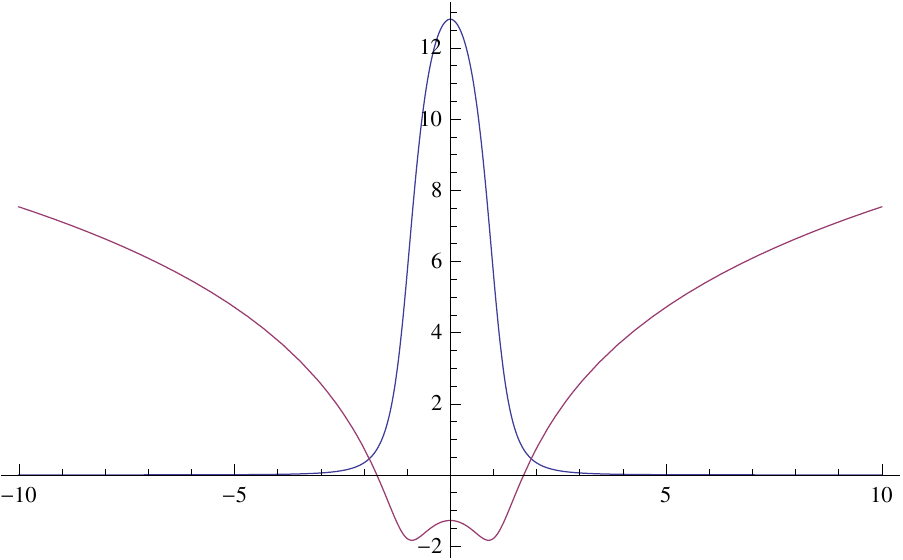}
  \hspace{1cm}
  \includegraphics[width=6.5cm]{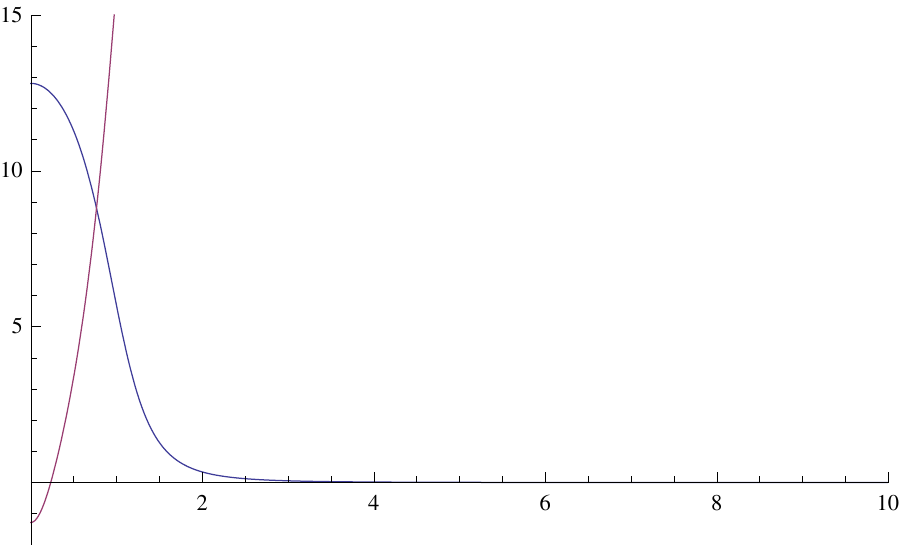}
  \caption{In the left picture, we showed $\Delta S_A$ as a function of $x$ for the interval subsystem $A$ defined by $[x-1,x+1]$. The blue and red graph correspond to the result from the connected and disconnected geodesic. In the right picture we plotted $\Delta S_A$
  as a function of time $t$ when we defined $A$ to be the interval $[-1,1]$. We normalized the total value by choosing the central charge $c=6$ and assume the length parameters to be $r=1.1$ and $l=1$.
  }
\label{fig:twocircleDIS}
  \end{figure}

We plotted both of them in Fig.\ref{fig:twocircleDIS}. The first graph shows the increased amount of entanglement entropy as a function of the location of a fixed length interval at $t=0$. The blue and red curve correspond to the connected and disconnected geodesic. The latter takes negative values near the origin. This is because due to the projection measurement removes large part of vacuum entanglement in this region. Since we always need to pick up the smaller contribution among the disconnected and connected
geodesic length, near the origin the disconnected one is favored. However, if we instead consider the previous setup of doubled CFTs, only connected one is allowed. In this case, the peak near the origin is clearly understood because the entangled pairs are expected to be localized around $|x|\leq l$. The second plot shows the time evolution for the interval $[-1,1]$. Initially, the growth of entropy is negative and increases to be slightly positive. After that, it switches to the connected geodesic solution and gets vanishing at late time as the system approaches to the vacuum state.

\section{HEE for Two Symmetric Intervals under Local Projection Measurements}

\begin{figure}
  \centering
  \includegraphics[width=7cm]{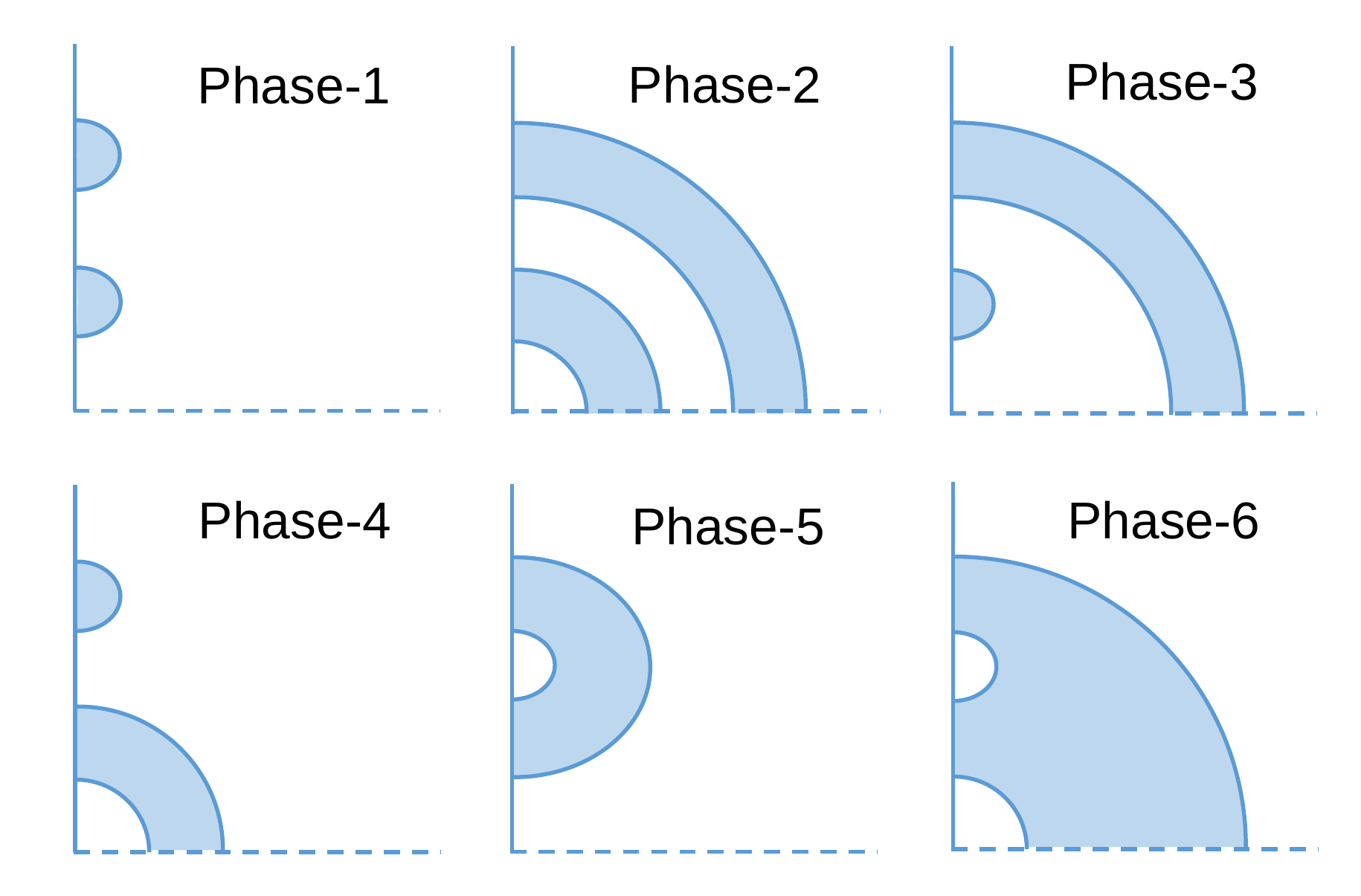}
  \hspace{1cm}
  \includegraphics[width=5cm]{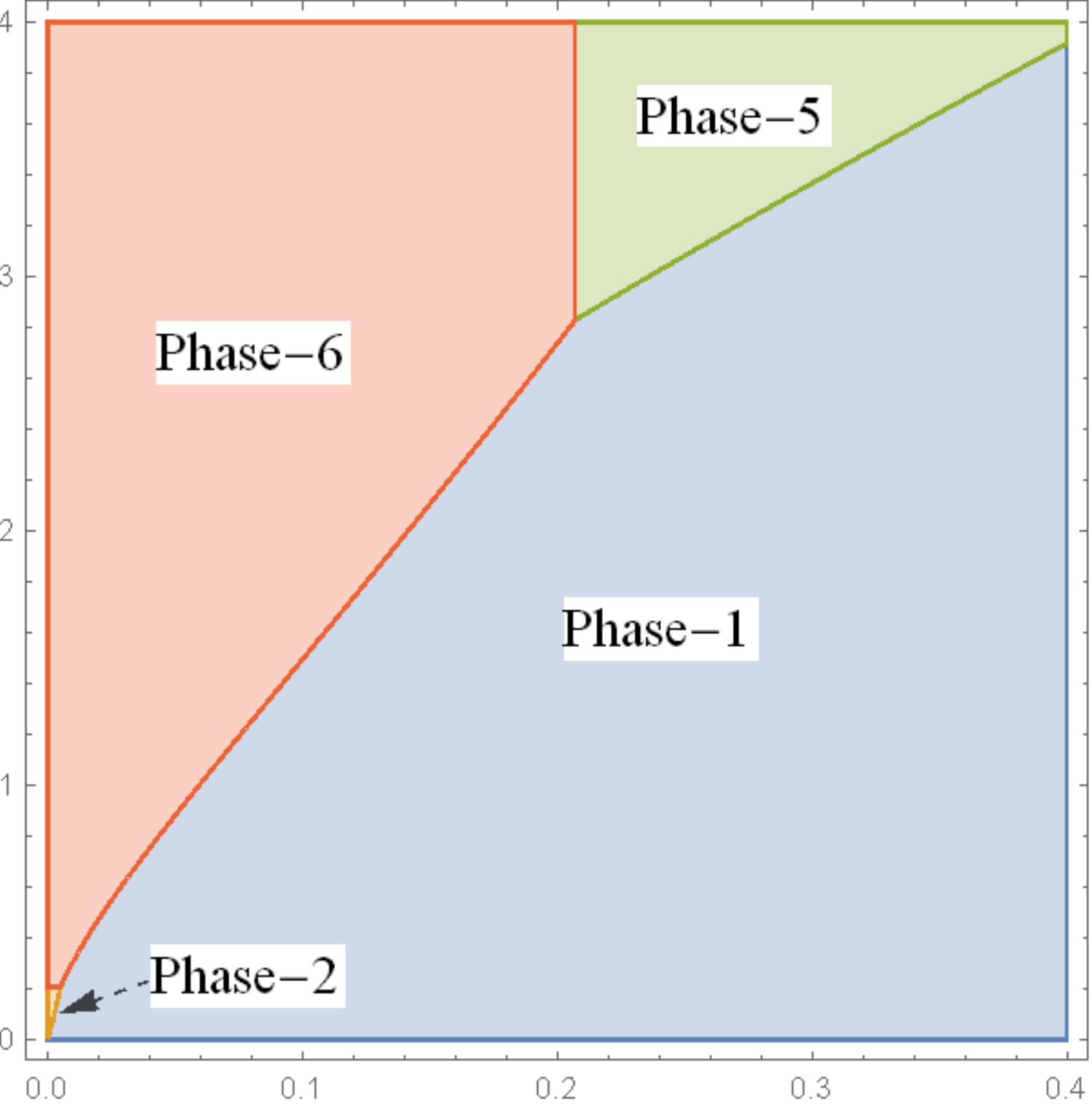}
  \caption{The left pictures show the six difference phases of geodesics. The right picture describes the phase diagram of the HEE $S_A$ in the $\f{l_1}{2q}$ -- $\f{l_2-l_1}{2q}$ plane.
We assigned colors to the phases :  blue to $S_A (1)$,  orange to $S_A (2)$ (tiny region), green to $S_A (5)$, and red to $S_A (6)$.}
  \label{fig:2int_HEEdia}
\end{figure}

Here we give a holographic calculation of $S_A$ when $A$ is a union of two (symmetric) intervals defined by
$q+l_1\leq x\leq  q+l_2$ and $-q-l_2\leq x\leq  -q-l_1$. In this case, we find there are six different phases, four out of them can be dominant depending on the values of parameters (see Fig.\ref{fig:2int_HEEdia}).

First, we consider phases which are symmetrical extensions of the phases in the one interval case (Phase-1 and -2).
Their contributions to HEE $S_{A}$ are
\begin{align}
S_A (1)
&= \f{c}{3} \log \left[\f{2(2q+l_1)l_1}{a q} \f{2(2q+l_2)l_2}{a q}
\f{ \left( \s{ \f{2q+l_1}{l_1}} - \s{ \f{2q+l_2}{l_2}}  \right)^2 }{ 4 \s{\f{2q+l_1}{l_1}} \s{\f{2q+l_2}{l_2}} }\right],
\end{align}
for Phase-1 (connected) and
\begin{align}
S_A (2)
&= \f{c}{3}\log \left[\f{2(2q+l_1)l_1}{a q} \f{2(2q+l_2)l_2}{a q}\right],
\end{align}
for Phase-2 (disconnected).

Next, we consider phases
where one interval is connected to the slit
and another interval is disconnected (Phase-3 and -4).
Their contributions are the same:
\begin{align}
S_A (3)&= \f{c}{3}\log \left[\f{2(2q+l_1)l_1}{a q} \f{2(2q+l_2)l_2}{a q}\right] + \f{c}{6} \log \left[\f{ \left( \s{ \f{2q+l_1}{l_1}} - \s{ \f{2q+l_2}{l_2}}  \right)^2 }{4 \s{ \f{2q+l_1}{l_1}} \s{ \f{2q+l_2}{l_2}}  }\right] \notag \\
&=S_A (4)=\f{1}{2} \left( S_A (1) + S_A (2) \right).
\end{align}
This means that $S_A (3) = S_A (4)$ cannot contribute to the HEE.

The remaining phases are 2 phases :
a phase where two intervals are connected each other (Phase-5)
and a phase where one of the edges of each interval is connected to the slit (Phase-6).
Their contributions to HEE $S_{A}$ are
\begin{align}
S_A (5)
&= \f{c}{6} \log \left[\f{2(2q+l_1)l_1}{a q } \f{2(2q+l_2)l_2}{a q} \left(\f{2q}{a} \right)^2\right],
\end{align}
for Phase-5 and
\begin{align}
S_A (6)
&= \f{c}{3} \log \left[\f{2(2q+l_1)l_1}{a q}\right] + \f{c}{6} \log \left[\f{2(2q+l_2)l_2}{a q} \f{2q}{a}\right],
\end{align}
for Phase-6.

In this way, $S_A$ has 4 phases: Phase-1,2,5,6.
$S_A(1)$ becomes the leading contribution for
\begin{align}
&\left( 0 < \f{l_2-l_1}{2q} < F_{12} \left( \f{l_1}{2q} \right) , 0 <\f{l_1}{2q} < \f{-7+5\sqrt{2}}{14} \right) \ , \notag \\
&\left( 0 < \f{l_2-l_1}{2q} < F_{16} \left( \f{l_1}{2q} \right) , \f{-7+5\sqrt{2}}{14} <\f{l_1}{2q} < \f{-1+\sqrt{2}}{2} \right) \ , \notag \\
&\left( 0 < \f{l_2-l_1}{2q} < F_{15} \left( \f{l_1}{2q} \right) , \f{ -1+\sqrt{2} }{2} < \f{l_1}{2q} \right) .
\end{align}

$S_A(2)$ becomes the leading contribution for
\begin{align}
\left( F_{12} \left( \f{l_1}{2q} \right)  < \f{l_2 - l_1}{2q} < \f{-1+\sqrt{2}}{2} - \f{l_1}{2q} , 0 < \f{l_1}{2q} < \f{-7+5 \sqrt{2}}{14}  \right) .
\end{align}

$S_A(5)$ becomes the leading contribution for
\begin{align}
\left( F_{15} \left( \f{l_1}{2q} \right)  < \f{l_2-l_1}{2q} , \f{ -1+\sqrt{2} }{2} < \f{l_1}{2q}  \right) .
\end{align}

$S_A(6)$  becomes the leading contribution for
\begin{align}
&\left( \f{-1+\sqrt{2}}{2} - \f{l_1}{2q} < \f{l_2 - l_1}{2q} , 0 < \f{l_1}{2q} <\f{-7+5 \sqrt{2}}{14}  \right) \ , \notag \\
&\left( F_{16} \left( \f{l_1}{2q} \right)  < \f{l_2 - l_1}{2q} ,  \f{-7+ 5 \sqrt{2}}{14} < \f{l_1}{2q} < \f{-1+\sqrt{2}}{2} \right) .
\end{align}
where
\begin{align}
F_{12} (x) &=
\f{x (1+x)}{\alpha - x} , \\
F_{15} (x) &=
1+ 2 x + 2 \sqrt{2 x (1+x)} , \\
F_{16} (x) &=
2 \sqrt{x (1+x)} \left[ 1+ 2 x + \sqrt{2 \left(2 x (1 + x) +\sqrt{x (1 + x)} \right)}  \right] .
\end{align}

The phase diagram is plotted in Fig.\ref{fig:2int_HEEdia}.

\end{document}